\documentclass{pasa}%

\usepackage{graphicx}

\title[CURVEPOPS II: SN with observed progenitors]{Supernova lightCURVE POPulation Synthesis II: Validation against supernovae with an observed progenitor. }

\author[Eldridge et al.]{J.J. Eldridge$^1$, N.-Y. Guo$^1$, N. Rodrigues$^{1}$,  E.R. Stanway$^{2}$, L. Xiao$^{3}$,\thanks{j.eldridge@auckland.ac.nz}
\affil{$^1$Department of Physics, Private Bag 92019, University of Auckland, Auckland 1010, New Zealand }%
\affil{$^2$Department of Physics, University of Warwick, Gibbet Hill Road, Coventry CV4 7AL, UK}
\affil{$^3$CAS Key Laboratory for Research in Galaxies and Cosmology, Department of Astronomy, University of Science and Technology of China, Hefei, 230026, China}
}%

\jid{PASA}
\doi{10.1017/pas.\the\year.xxx}
\jyear{\the\year}

\usepackage{aas_macros}
\usepackage{hyperref} 
\hypersetup{colorlinks,citecolor=blue,linkcolor=blue,urlcolor=blue}

\hypersetup{draft}

\begin{document}

\begin{frontmatter}
\maketitle

\begin{abstract}
We use the results of a supernova light-curve population synthesis to predict the range of possible supernova light curves arising from a population of single-star progenitors that lead to type IIP supernovae. We calculate multiple models varying the initial mass, explosion energy, nickel mass and nickel mixing and then compare these to type IIP supernovae with detailed light curve data and pre-explosion imaging progenitor constraints. Where a good fit is obtained to observations, we are able to achieve initial progenitor and nickel mass estimates from the supernova lightcurve that are comparable in precision to those obtained from progenitor imaging. For two of the eleven IIP supernovae considered our fits are poor, indicating that more progenitor models should be included in our synthesis or that our assumptions, regarding factors such as stellar mass loss rates or the rapid final stages of stellar evolution, may need to be revisited in certain cases. Using the results of our analysis we are able to show that most of the type IIP supernovae have an explosion energy of the order of log(E$_\mathrm{exp}$/ergs)=50.52$\pm$0.10 and that both the amount of nickel in the supernovae and the amount of mixing may have a dependence on initial progenitor mass.
\end{abstract}

\end{frontmatter}

\section{INTRODUCTION }
\label{sec:intro}

Stellar death-throes in supernovae are some of the most spectacular events that stars produce in the Universe. Attempting to understand these events is both a mature and constantly developing field. It is primarily driven by observations. New discoveries occur primarily when new observational techniques are employed, or when the peculiarities of an individual event attract attention and follow-up, leading to discovery of a new transient type. In comparison, theory may lag behind. Some rare transient types have been theoretically predicted before being observed, such as the kilonovae from the merger of two neutron stars, which appeared as theoretical predictions \citep{1998ApJ...507L..59L} and a tentative detection associated with a short gamma ray burst \citep{2013Natur.500..547T} before the electromagnetic radiation associated with GW\,170817 unambiguously confirmed their existence \citep{2017ApJ...848L..12A}.
However these have been deduced primarily from gravitational theory (Kilonovae, Tidal Disruption Events, etc). There has not yet been a clear case of stellar structure and evolution theory leading the way and predicting new supernovae types that were then observed.

The success of such theoretical predictions, verified by observation, in other areas indicates that we should redouble our theoretical work in understanding supernovae. In Eldridge et al. (2018, hereafter paper I), we introduced the supernova lightcurve population synthesis (CURVEPOPS) project with this goal. This comprises lightcurves derived from a large number of supernova progenitor models identified within the Binary Population and Spectral Synthesis project \citep[BPASS, ][]{2017PASA...34...58E} and exploded with the Supernova Explosion Code \citep[SNEC,][]{2015ApJ...814...63M}. By studying the synthetic lightcurves from a population of realistic progenitors drawn from populations including binary stars we were able to demonstrate that binary interactions are the main source of diversity of type II supernova lightcurves.

Here we take the next step which is to validate the CURVEPOPS models against observational data, and hence gain insight into the importance of explosion parameters such as explosion energy, nickel mass and amount of nickel mixing. The most effective way to perform this test is to consider supernovae for which observations of both the progenitor stars and the observed supernovae exist in the archive. Such studies have been carried out before for individual supernovae and large samples \citep[e.g.][]{2012ApJ...757...31B,2014AJ....148...68B,2017ApJ...838...28M,2018ApJ...858...15M}. Here we have focused our study on the sample of type IIP supernova progenitors that have detected progenitor stars as described in \citet{Smartt2015}. We have collated lightcurve data from the \textit{Open Supernova Catalogue} \footnote{\texttt{https://sne.space/}} and compared it to a large suite of SNEC explosion models derived from single-star progenitor stellar structures calculated as part of the BPASS project. We then constrain the nature of the progenitor star from fitting the supernova lightcurve and compare these inferences with those in the literature based on  fitting the colour and magnitude of the progenitor star in pre-explosion images. This allows us to gain insight into the accuracy and usefulness of our CURVEPOPS models. The SNEC inputs and outputs from this project have been made available at the BPASS website (\texttt{http://bpass.auckland.ac.nz}) and in the PASA datastore.

We note that there are many studies that have investigated the lightcurves of core-collapse supernovae \citep[e.g.][]{1994A&A...281L..89U,2005AstL...31..806U,2007A&A...461..233U,2008A&A...491..507U,2009A&A...506..829U,2010MNRAS.408..827D,2011MNRAS.414.2985D,2011ApJ...729...61B,2012ApJ...757...31B,2014MNRAS.440.1856D,2014AJ....148...68B,2015ApJ...814...63M,2016MNRAS.458.1618D,2016ApJ...829..109M,2017ApJ...838...28M,2018ApJ...858...15M,2018BAAA...60...23M,martinezbersten}. However what sets this study apart is the large number of SN models that we have created, aiming to model a number of supernovae simultaneously rather than attempting to produce a perfect fit for one SN alone. Furthermore we are attempting to determine how strong a link there is between the time evolution of the observed explosion and the properties of the progenitor star in a lightcurve-derived model. Our hope is that from this work we will be able to improve our CURVEPOPS lightcurves and use relations derived from this work in terms of final mass to explosion energy, nickel mass and nickel mixing to produce more realistic supernova lightcurve populations.

The structure of this paper is as follows, first we describe the creation of our grid of supernova models with BPASS and SNEC. Next we describe the observational sample of supernovae we employ in this project. We then outline the fitting method used to compare the models to observations. We present and discuss our results, before summarizing our conclusions.

\section{Creation of supernova simulations}

\begin{figure*}
\begin{center}
\includegraphics[width=\columnwidth]{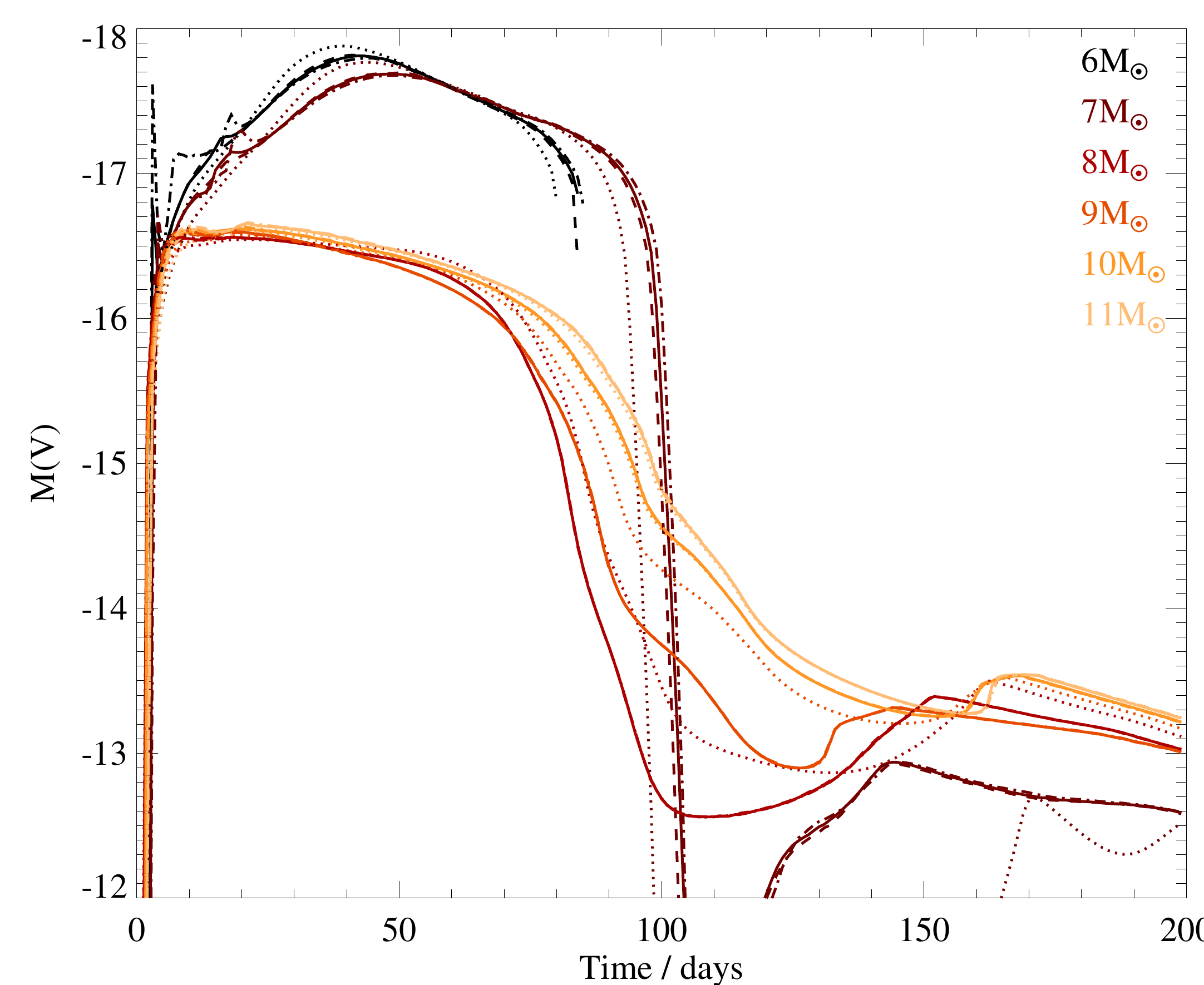}
\includegraphics[width=\columnwidth]{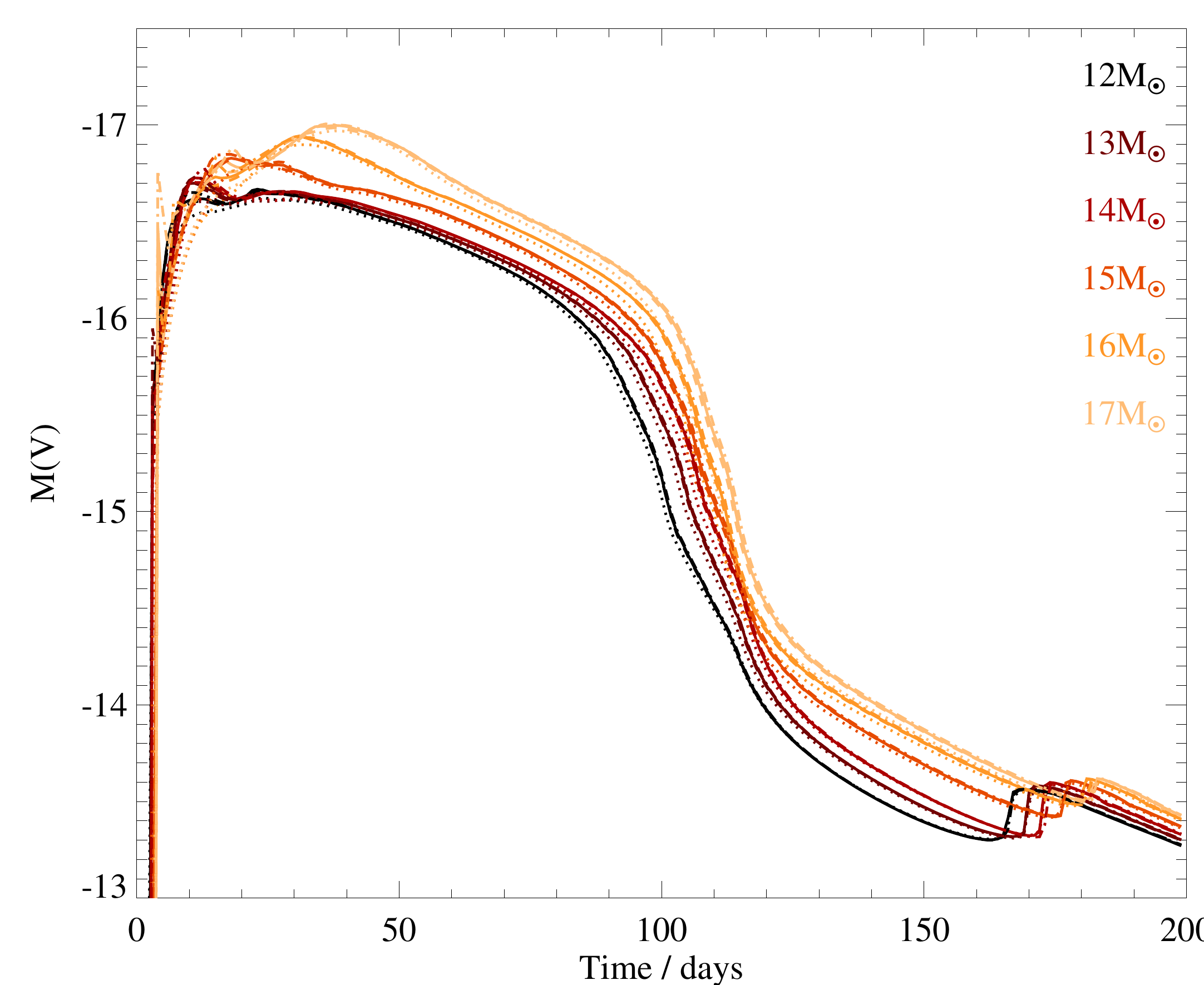}
\includegraphics[width=\columnwidth]{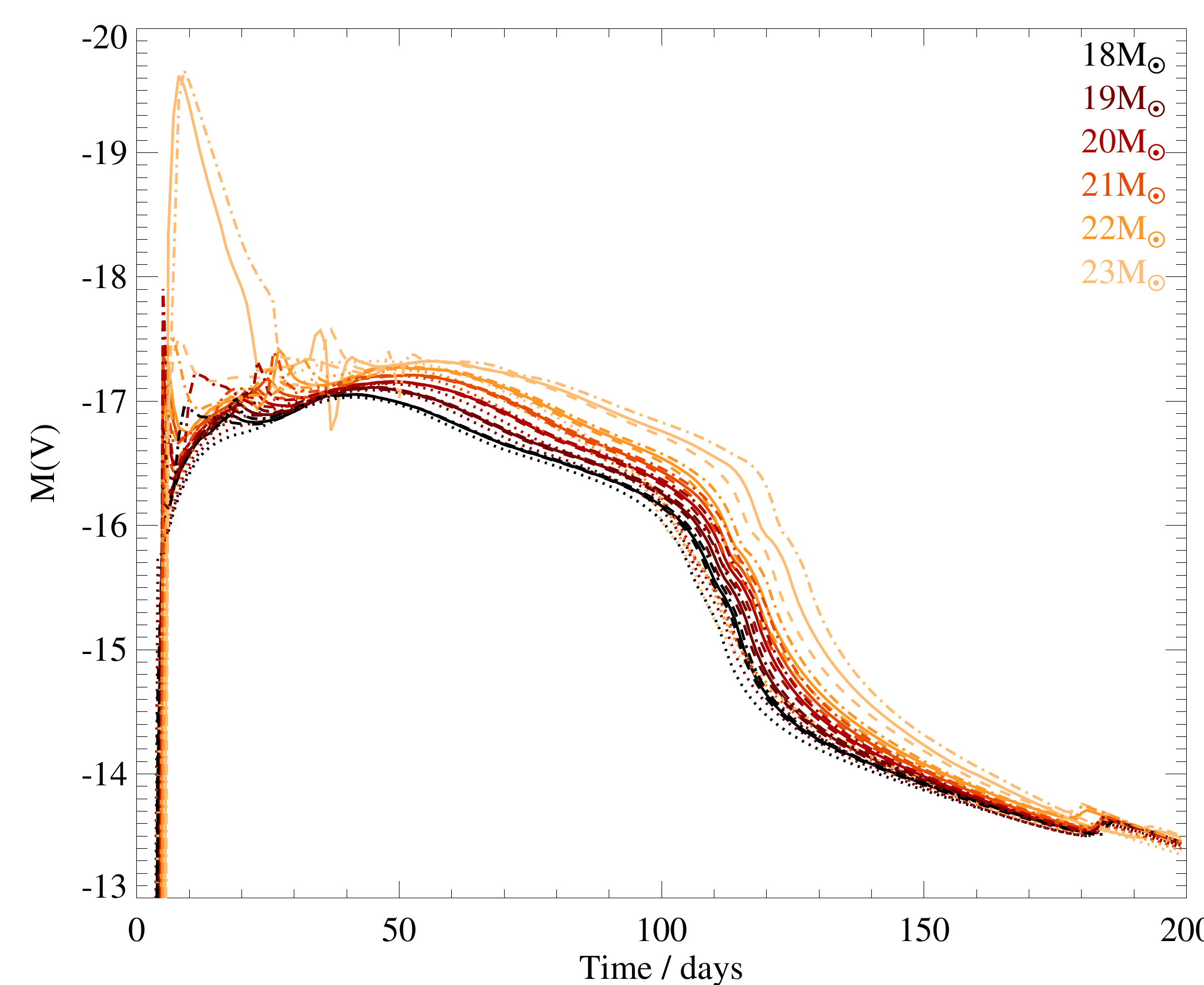}
\includegraphics[width=\columnwidth]{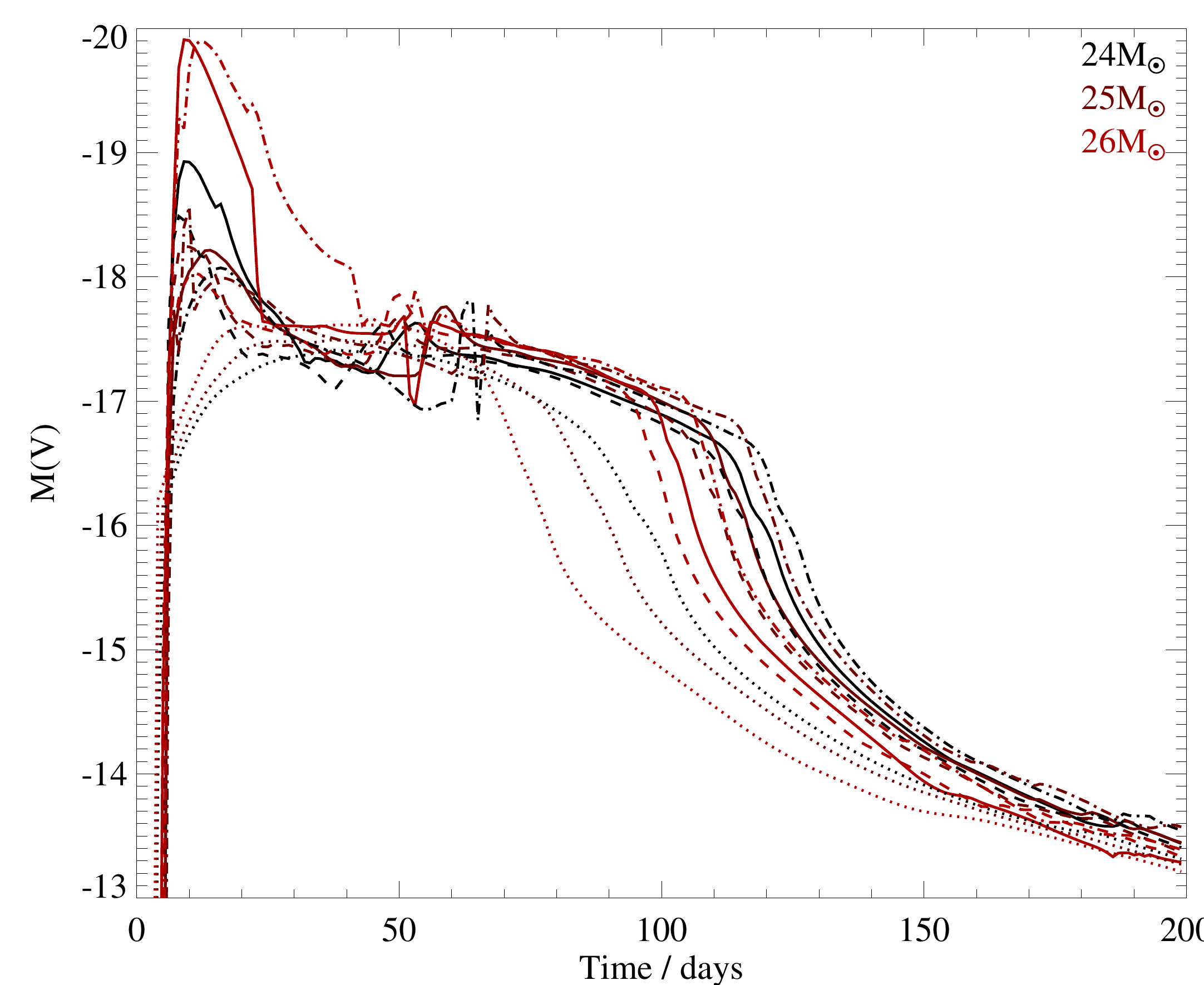}
\caption{Sample synthetic lightcurves, demonstrating how including the circumstellar material of the red supergiant's wind changes the lightcurve. The solid lines are for lightcurve models with the circumstellar material included as discussed in the text and the dotted lines assume no material surrounding the progenitor star. The dashed line is where the circumstellar material density has been reduced by a factor of 2. While the dash-dotted line is where the circumstellar material density has been increased by a factor of two. The explosion energy, nickel mass and mixing are kept constant (10$^{50.5}$erg~s$^{-1}$, 10$^{-1.5}$M$_{\odot}$ and mid-strength mixing). Figures for each initial mass on its own are included in Appendix D.}
\label{fig:examplecsmmodels}
\end{center}
\end{figure*}

\begin{figure*}
\begin{center}
\includegraphics[width=\columnwidth]{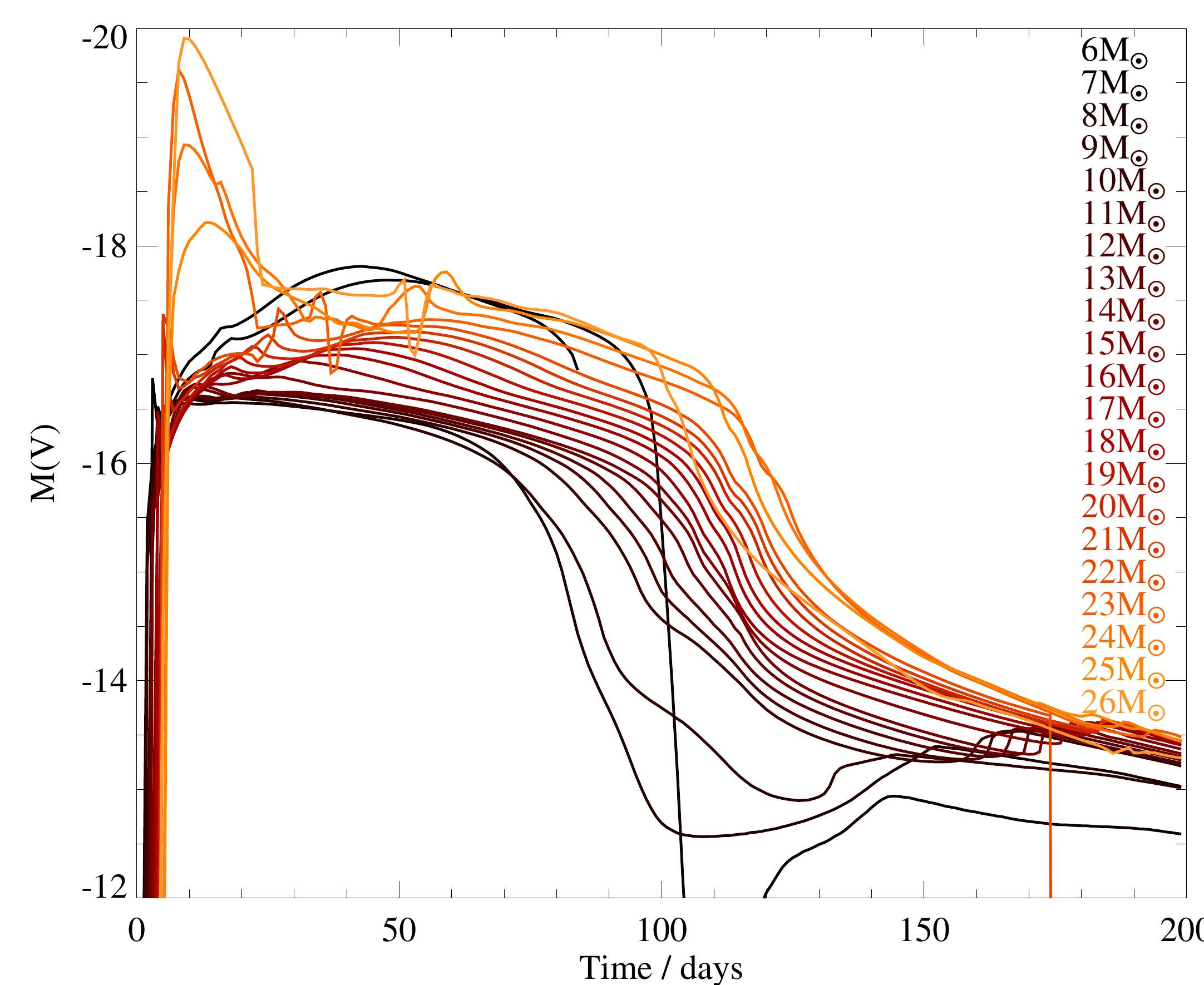}
\includegraphics[width=\columnwidth]{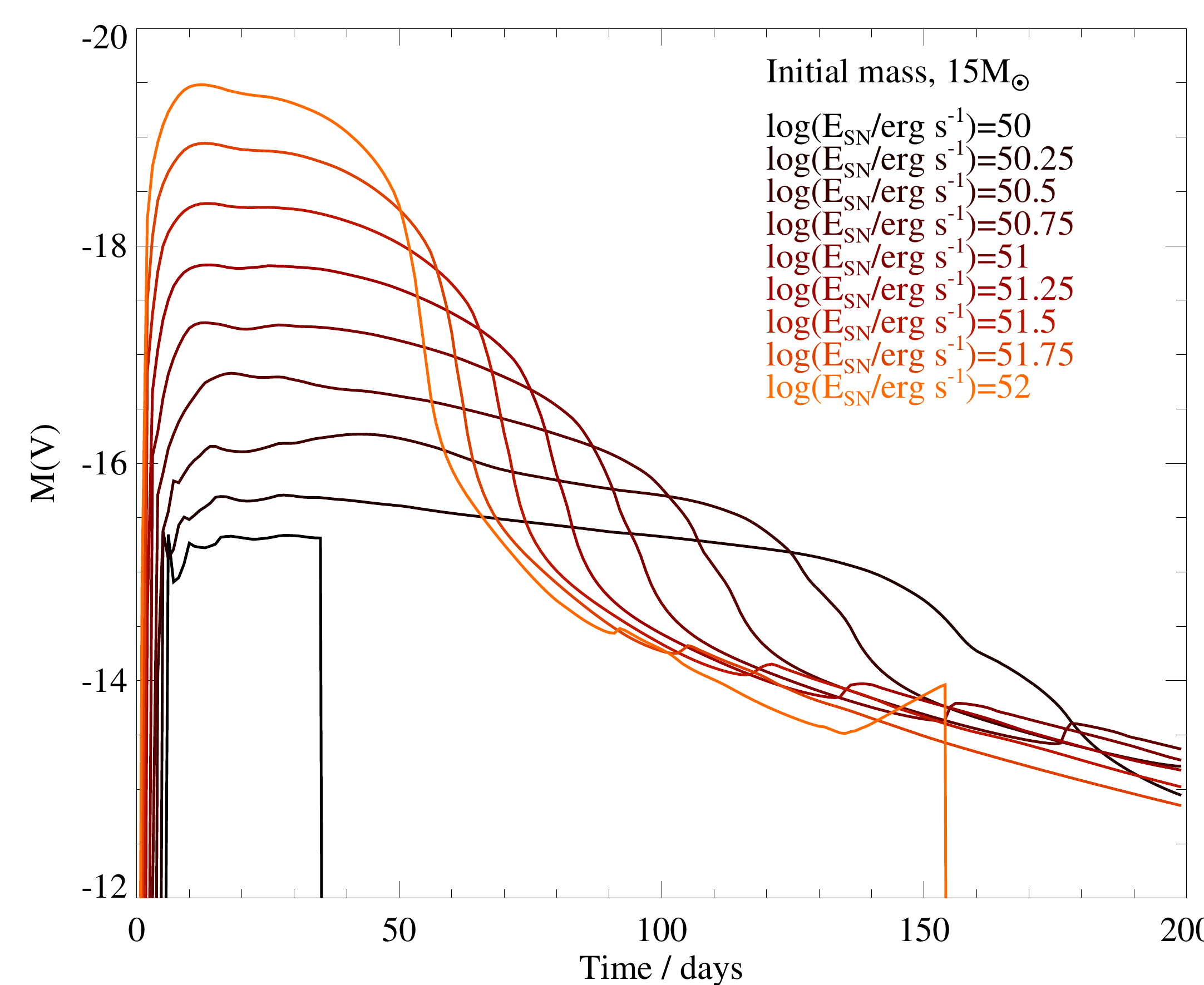}
\includegraphics[width=\columnwidth]{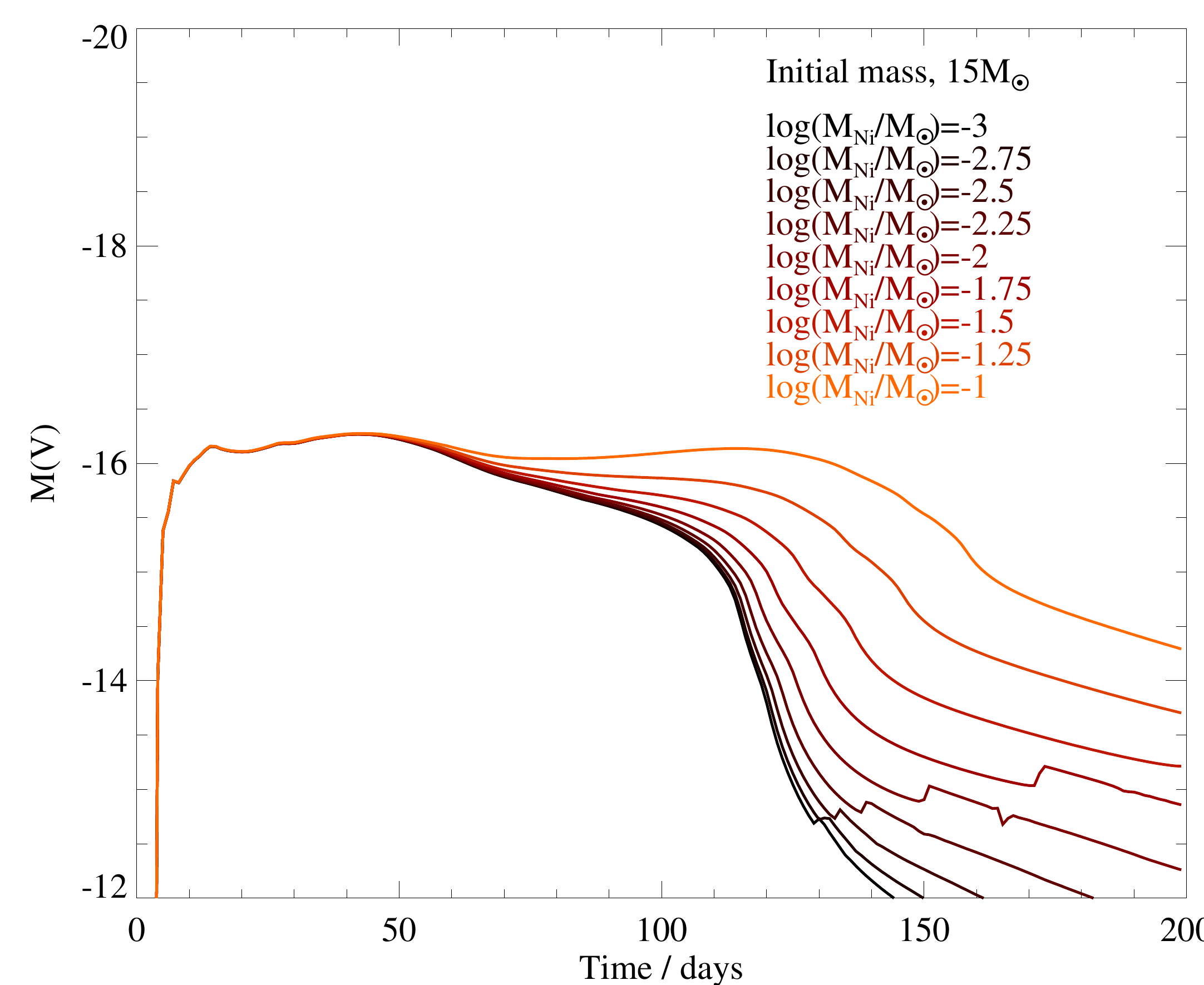}
\includegraphics[width=\columnwidth]{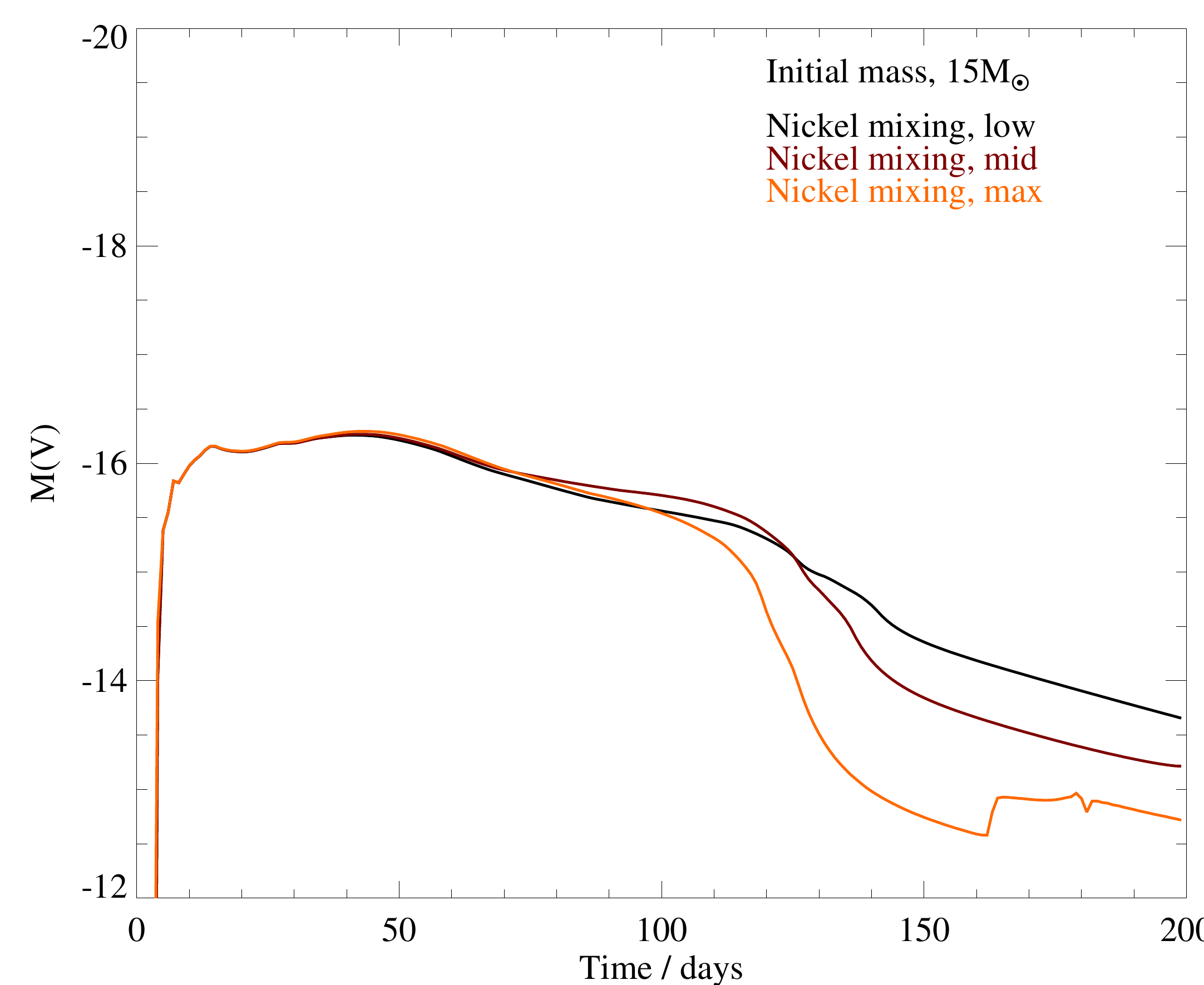}
\caption{Sample synthetic lightcurves, demonstrating how the explosion parameters change the lightcurve. The upper left panel shows how changing the initial mass of star varies the lightcurve when the stellar structure, explosion energy, nickel mass and mixing are kept constant (10$^{50.5}$erg~s$^{-1}$, 10$^{-1.5}$M$_{\odot}$ and mid mixing). In the other panels one of the explosion parameters are varied while the stellar structure and other parameters are kept constant: upper right - explosion energy, lower left - nickel mass and lower right - nickel mixing.}
\label{fig:examplemodels}
\end{center}
\end{figure*}

We use the v2 single-star models from the Binary Population and Spectral
Synthesis code \textsc{BPASS} \citep{2017PASA...34...58E}. The models
are calculated using a version of the Cambridge STARS code that has been adapted to
follow binary evolution \citep[see][for full details]{2017PASA...34...58E}. The results of the code are available from the website, \texttt{bpass.auckland.ac.nz}. We summarize the most important details here. 

The models are calculated from the zero-age main-sequence up to
the end of core carbon burning. At this point the models stop since the STARS code is unable to compute further due to the increasing complexity and time resolution of late time evolution. We assume that our models are close enough to the time of core-collapse that the parameters we use for our lightcurve models
will not vary significantly. We have used a model calculated from the Modules for Experiments in Stellar Astrophysics \citep[MESA-r10398,][]{Paxton2011, Paxton2013, Paxton2015, Paxton2018} stellar evolution code for a typical SN\,IIP progenitor -- a single-star with M=15.6\,M$_\odot$ -- to explore how different later models might be if evolution is taken closer to core-collapse. The results of this analysis are presented in Appendix C. In brief, the outer core structure changes little while the density of the central core increases; the greatest changes are within the region that is assumed to form the remnant. This may change the late time evolution of our supernova models and so our late-time lightcurves should be assumed to be subject to increased uncertainty. If we compare our models to other supernova models \citep[e.g.][]{2013MNRAS.433.1745D,2019A&A...625A...9D,2019arXiv190309114G} we find that they are generally similar in shape. Differences are greatest at the end of the plateau phase where the extra structure and mixing within the model will have the greatest impact on the resultant lightcurve.

We use models with a metallicity mass fraction of $Z=0.014$ which is close to estimates of massive stars in the Solar neighborhood \citep{2012A&A...539A.143N}. We use every integer initial mass model from 5 to 26\,M$_{\odot}$ as these all have sufficiently massive hydrogen envelopes to give rise to a long lasting plateau in their lightcurves. We do not use the binary models because, as shown in paper I, type IIP supernovae mostly arise from progenitors that have not experienced significant binary interactions. While some stars in binaries may experience interactions and provide type IIP progenitors with interior structures different to single stars, if we were to explode all the binary progenitors from paper I with the range of explosion parameters below this would require calculation of 154,791 supernova simulations. Doing so is beyond the scope of the current pilot study.

We next format the stellar structure models, taken at their last time step, for input into the SuperNova Explosion Code, SNEC. This is an open-source code that is available online from \texttt{https://stellarcollapse.org/SNEC} and which has been applied to modelling various aspects of supernovae \citep[e.g.][]{2015ApJ...814...63M,2016ApJ...829..109M,2017ApJ...838...28M,2018ApJ...858...15M}. We only include the composition variables that we have within the BPASS stellar evolution code which are hydrogen, helium, carbon, nitrogen, oxygen, neon, magnesium, silicon and iron. We also include a nickel-56 variable but leave this blank, inputting the nickel within the explosion parameters of the SNEC model. We have made all these input files available on the PASA datastore as well as on the BPASS website (\texttt{http://bpass.auckland.ac.nz}).

We extend the surface layers of the progenitor models to include the stellar progenitor's wind. Recent studies have shown how including the red supergiant's wind can have important impact on the early lightcurve of the progenitor star \citep{2016ApJ...829..109M,2017ApJ...838...28M,2018ApJ...858...15M,2018MNRAS.476.2840M}. To incorporate this into our progenitor models we determine the mass-loss rate from the stellar model and calculate the wind velocity using the method outlined in \citet{2006MNRAS.367..186E}. The mass-loss rates used are those of \citet{1988A&AS...72..259D}. We do not vary the wind parameters as in \citet{2018MNRAS.476.2840M} but leave them fixed so that the nature of the circumstellar medium is linked to the initial mass of the progenitor. We attach the outer most stellar mesh point to the wind assuming a beta wind velocity law as used by \citet{2018MNRAS.476.2840M} with $\beta=5$ which is typical of cool red supergiants \citep{1985A&A...147..103S,2016MNRAS.455..112G} although we note this value might not be correct for {\it all} red supergiants \citep{2017Natur.548..310O}.  Varying this parameter will make small changes to the very early lightcurve and future observations, with better sampling of the early time evolution, will be required to determine the best value to use. We have run supernova models with values of $\beta$ between 2 and 6, and these are shown in Appendix D. We find that the effects of varying $\beta$ are somewhat degenerate with the total density of the circumstellar material density assumed. The effect of the material however is restricted to the early lightcurve for most stellar masses, with an effect on the late time plateau duration for only the most massive stars. Given that most of the progenitors are expected to have initial masses below about 20\,M$_\odot$, at which the full range of $\beta$ explored changes the plateau length by $<10$\,days, we do not expect a significant impact on our results. Nonetheless it is clear that this area should be considered in future studies.

We extend the stellar wind out to approximately 10000 times the radius of the star which is typically shorter than a parsec. We show examples of these lightcurves compared to those without the stellar wind included in Figure \ref{fig:examplecsmmodels}. We find inclusion of this causes brighter phases in our early lightcurves. This is most significant for the most massive stars above 20M$_{\odot}$. There are also some spurious jumps in the nickel tail of our model as varying shells of material collide at late times. We suggest that this is an artifact of our one-dimensional simulations and that this behaviour would be smoothed out in a 3-dimensions.

{The mass-loss rates of red supergiants are uncertain as there is a question as to whether the rates of  \citet{1988A&AS...72..259D} are correct or not. Two studies, \citet{2011A&A...526A.156M} and \citet{2018MNRAS.475...55B}, found that while \citet{1988A&AS...72..259D} have the correct scale of mass-loss rates there are significant departures away from the predicted rates for some stars. \citet{2011A&A...526A.156M} found that red supergiants varied by up to a factor of four away from the \citet{1988A&AS...72..259D} predictions. While \citet{2018MNRAS.475...55B} found more significant deviations, especially that  \citet{1988A&AS...72..259D} rates may be overestimating the mass lost during the red supergiant phase significantly (we note in their study that the STARS models, effectively identical to BPASS models, are closest to their estimated mass losses of all the stellar evolution models they consider). In addition to these studies there is evidence from radio observations of supernovae that the mass-loss rates we assume are similar to those of typical red supergiants \citep{2006ApJ...641.1029C}. In light of this uncertainty, in addition to varying the $\beta$ parameter, we explore this further by testing how varying the circumstellar material density affects our model lightcurves. We show models with double or half the wind density of our fiducial models in  Figure \ref{fig:examplecsmmodels} and Appendix D. Again we find that, for the majority of the lightcurves, varying the wind density by a factor of two causes minimal changes to the resulting light curves. As noted before, the effects are also in degenerate with the assumed value of the wind acceleration parameter $\beta$.}

The largest impact of varying the circumstellar density is again observed in the early lightcurves, and particularly for the most massive stars. For many of the observed lightcurves available in the literature, the early evolution is not well sampled or is simply unobserved, leaving models largely unconstrained. The uncertainties in the winds of RSGs, while important to consider, should thus not alter our interpretations to any significant degree. We also note that the most massive stars leading to core collapse supernovae, for which the wind effects are greatest, are also more likely to have massive binary partners and interact, leading to stripped-envelope supernovae. By not accounting for binary star progenitors in these models we are also ignoring variations in the circumstellar medium that would be caused by binary interactions which may increase or decrease apparent mass-loss rates for a progenitor. Therefore rather than varying the circumstellar medium in our fitting grid based on mass, we link the assumed circumstellar medium density to the progenitor at the point of explosion in line with our current best understanding.

Within SNEC, as well as the progenitor structure, other parameters must be specified. These are the explosion energy, nickel mass, amount of nickel mixing and excised mass. In paper I we assumed that the first three parameters were constant while the excised mass was determined by the progenitor structure. Here we compute multiple models with varying explosion energy, nickel mass and nickel mixing. We still derive the excised mass to be the remnant mass computed as described in \citet{2017PASA...34...58E}. Our grid of values is as follows,
\begin{enumerate}
\item The explosion energy varies from $\log(E_{\rm SN}/{\rm erg \, s^{-1}}=50$ to 52 in steps of 0.25~dex.
\item The nickel mass varies from  $\log(M_{\rm Ni}/M_{\odot})=-3$ to -1 in steps of 0.25~dex.
\item The nickel mixing is determined by setting the nickel boundary mass, out to which the nickel is mixed from the excised mass. We set this bound to one of three values: low, mid or max. These are determined as follows,
\begin{enumerate}
\item $M_{\rm Ni\,boundary\,low}=M_{\rm excised}+0.1M_{\rm ejecta}$ 
\item $M_{\rm Ni\,boundary\,mid}=M_{\rm excised}+0.5M_{\rm ejecta}$ 
\item $M_{\rm Ni\,boundary\,max}=M_{\rm excised}+0.9M_{\rm ejecta}$ 

\end{enumerate}
Where $M_{\rm excised}$ is the mass which collapses into the compact remnant, and $M_{\rm ejecta}$ is the mass of material ejected by the supernova explosion. This therefore assumes significant mixing of nickel into the envelope in all cases, but more mixing in some cases than others.
\end{enumerate}
Covering all of the parameters of initial mass, explosion energy, nickel mass and nickel mixing requires 5346 SNEC models to be run. For each observed supernova we then use this grid of these models to find the parameters that best match its lightcurve.

We show in Figure \ref{fig:examplemodels} a sample of some of explosion models. Here we have shown how varying each of the explosion parameters varies certain aspects of the light curves. In these plots the base-line model is an initially 15\,M$_{\odot}$ progenitor, with an explosion energy of $10^{50.5}\,{\rm erg \,s^{-1}}$, a nickel mass of $10^{-1.5}\,M_{\odot}$ and the mid strength nickel mixing. 

In the panel with varying initial masses we see that most models have strong plateaus in their light curves. However, above 20\,M$_{\odot}$ we see early strong rises in the light curve due to denser winds around our progenitor stars. The plateau phase for these lightcurves is also extended due to the circumstellar medium. The exact shape of this rise will vary on the density of our assumed wind and also the value of $\beta$ assumed for the wind acceleration as shown by \citet{2018MNRAS.476.2840M}. For the lowest mass progenitors we see a late time rising of the light curves due to growing interaction between the supernova ejecta and our model circumstellar medium. In both case these features rely on our assumption that the surrounding wind is directly linked to the nature of the progenitor star as prescribed by \citet{1988A&AS...72..259D} rather than allowing the mass-loss rates to vary arbitrarily. We stress that this initial work is largely exploratory and a full future study will involve models of varying metallicities, which will have different circumstellar environments, and also include progenitors that are the result of binary interactions. These will be very different to those from single-star evolution alone.

In the other panels we see first that varying the explosion energy changes the length and brightness of the plateau. Second, varying the nickel mass changes the late time evolution of the plateau as well as the brightness of the nickel tail. Finally changing the strength of nickel mixing has a complex impact: if the mixing is low then more nickel remains in the ejecta to power the late time light curve, while stronger mixing allows more of the nickel decay energy to escape without contributing to the V-band magnitude in the nickel tail.

We note that some light-curve tracks end abruptly. This is because of the known problem with SNEC that once a significant fraction of nickel is outside the photosphere the code stops predicting broad-band magnitudes for the supernova. There are also apparent bumps and short term features in some of the model light curves. This is the result of our circumstellar medium as well as the structure of our stellar progenitors. How realistic these features are is difficult to determine, but  since most only occur late in the nickel tail we expect they will have only a small impact on our fitting here.

\section{Observational sample}

\begin{figure*}
\begin{center}
\includegraphics[width=\linewidth]{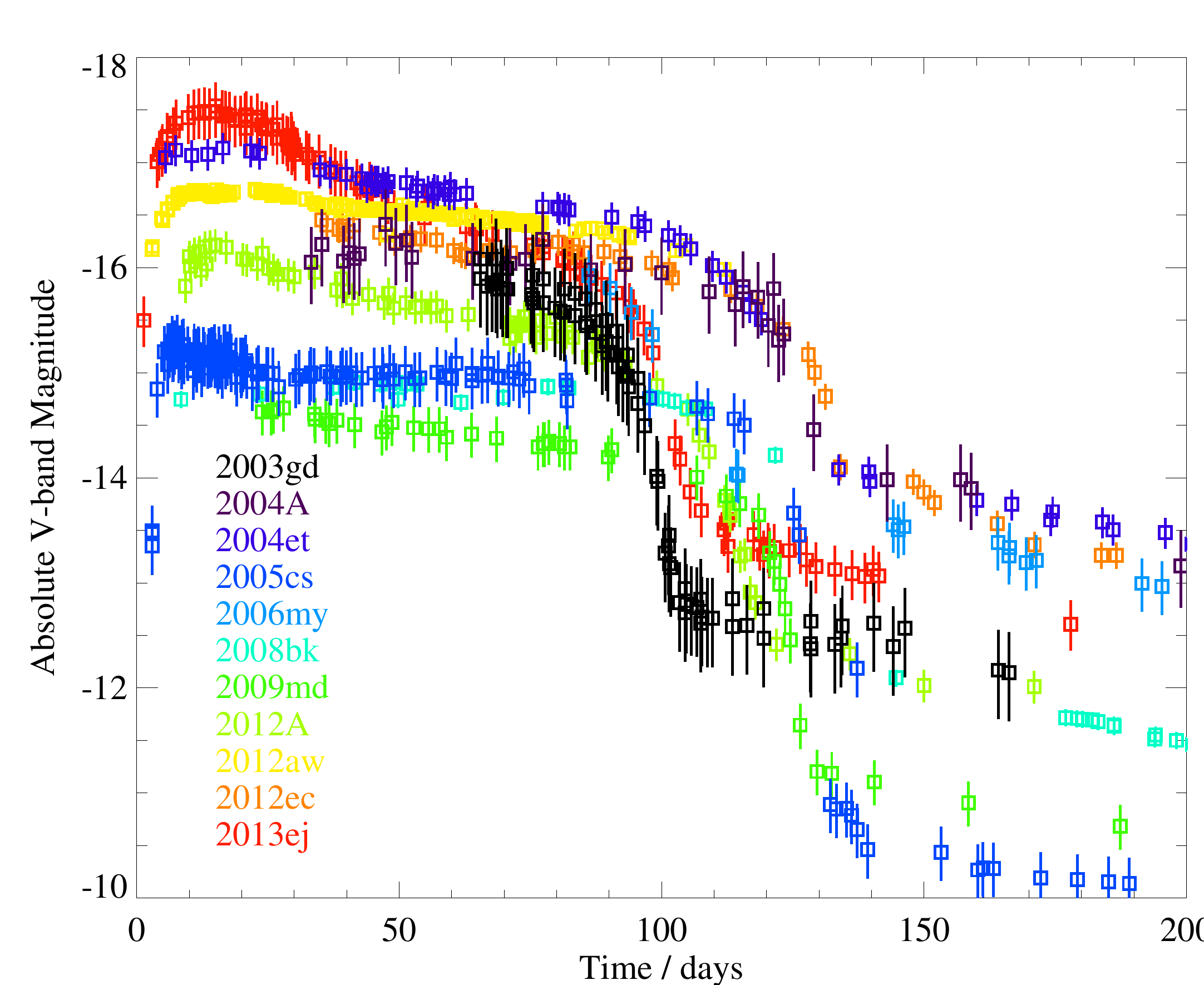}
\caption{Light curves of observed supernova in V-band absolute magnitude. SN host details and sources of data are given in Table \ref{tab:SNlist}.}
\label{fig:Tb1SNePopulation}
\end{center}
\end{figure*}


This work only considers Type II-P supernovae with observed progenitors as listed in Table 1 of \cite{Smartt2015}. These are supernovae with direct progenitor detection and are all within a distance of approximately 20~Mpc. We list these in Table \ref{tab:SNlist} along their host galaxy identification, assumed distance, foreground Galactic extinction for the host galaxy and the source of the photometric data we have used to reconstruct their lightcuves.

Distances were taken from cross-referencing literature, NED \footnote{The NASA/IPAC Extragalactic Database (NED) is operated by the Jet Propulsion Laboratory, California Institute of Technology, under contract with the National Aeronautics and Space Administration.} and SIMBAD \citep{SIMBAD} values for the host galaxies. Where discrepancies exist values from the paper containing progenitor measurements were prioritized. Error calculations were performed using the {\fontfamily{lmtt}\selectfont uncertainties} package of Python.

\begin{table*}
\caption{List of Supernovae used in project}
\label{tab:SNlist}
\begin{center}		
\begin{tabular}{cccc p{9cm}}
\hline\hline
Supernova& Host Galaxy & Distance (Mpc) & $A_{\rm V}$ & Source\\
\hline
SN2003gd & M74 & 9.30 $\pm$ 1.80$^{2}$ & 0.192
& \cite{Ni2008bk} \cite{2016AJ....151...33G} \cite{2014MNRAS.442..844F} \cite{SN2003gdDistance} \cite{2003PASP..115.1289V}$^{11}$ \\
SN2004A&NGC6207 & 20.3 $\pm$ 3.40$^{3}$ &0.042 & \cite{Ni2004A} \cite{TsvetkovD.Yu.2008Poot}$^{11}$\\
SN2004et&NGC6946 &5.70 $\pm$ 0.39$^{8}$&  0.938
&\cite{SN2006myData}\\
SN2005cs & 	M51  & 8.40 $\pm$ 1.00$^{3}$ & 0.096 &\cite{2009MNRAS.394.2266P}$^{11}$\\
SN2006my & 	NGC4651	& 22.3 $\pm$ 2.60$^{5}$ &0.073 &\cite{SN2006myData}\\
SN2008bk&NGC7793 & 3.44 $\pm$ 0.13$^{9}$ & 0.053& \cite{Ni2008bk} \cite{SN2008bkDistance}$^{11}$\\
SN2009md & NGC3389& 21.3 $\pm$ 2.10$^{4}$ & 0.074&\cite{SN2009mdData}\\
SN2012A &NGC3239 &	9.80 $\pm$ 0.70$^{6}$ & 0.088
&\cite{SN2012AProgPaper}\\
SN2012aw & M95  & 9.82 $\pm$ 0.20$^{1}$
& 0.076	& \cite{Sn2012awDistance}\\
SN2012ec&NGC1084&17.3 $\pm$ 1.00$^{10}$& 0.073 & \cite{2015AA...579A..40S}$^{11}$\\
SN2013ej&NGC628 &9.10 $\pm$ 1.00$^{7}$ & 0.192 & \cite{2016MNRAS.461.2003Y} \cite{2015ApJ...807...59H}$^{11}$\\
\hline\hline
\end{tabular}
\end{center}
\tabnote{$^1$The NASA/IPAC Extragalactic Database (NED) is operated by the Jet Propulsion Laboratory, California Institute of Technology, under contract with the National Aeronautics and Space Administration.}
\tabnote{$^2$\cite{SN2003gdDistance} $^3$\cite{SN2005csProgPaper}  $^4$\cite{SN2009mdData}  $^5$  \cite{SN2006myData}  $^6$ \cite{SN2012AProgPaper}  $^7$ \cite{SN2013ejProg}  $^8$ \cite{SN2004etProg}  $^9$ \cite{SN2008bkDistance}  $^{10}$ \cite{SN2012ecProg}  $^{11}$Through Open Supernova Catalog: \cite{OSC}}
\end{table*}
	
While there are some supernovae with observations from multiple filters in the source data sets, we have focused on the V band as there was plentiful data for each supernovae. In addition, U and B band are more strongly affected by extinction, and emission lines such as H$\alpha$ may effect the R band magnitudes.

We show our sample lightcurves in Figure \ref{fig:Tb1SNePopulation}, after correction for distance and extinction. Each supernova in our sample has a clear plateau phase that lasts from approximately 75 to 125 days. There is a range of plateau luminosities of about 2 magnitudes and a broader range in the nickel tail of 4 magnitudes.

\section{Lightcurve Fitting}

\subsection{Preparing the synthetic lightcurves}

All the synthetic lightcurves were visually inspected, since for some of the models the input stellar models have a small number of meshpoints that introduced oscillatory behaviour in the simulated lightcurves, where the luminosity of the lightcurve would vary by about a magnitude timestep by timestep, which was set to 6 hours. Such oscillations would begin/disappear and the mean behaviour however was still that of a plateau lightcuve. We expect these were numerical in nature due to the sparse number of mesh points in some of our progenitor models, which causes numerical problems with the integration when the timestep was too high. To remove these spurious sequences we smoothed these parts of the light curve with the following simple algorithm:
\begin{enumerate}
\item Mask all model points where the neighbouring value differs from the previous by more than 0.1 magnitude.
\item Take the resulting array and mask again if the $i+2$ value differs from $i$ by more than 0.1 magnitude.
\item Compress the resulting array and repeat step 1.
\end{enumerate}
The algorithm works as the time domain is very densely sampled and plotting of model light curves show that normally, consecutive model points should not differ from previous by more than 0.1 magnitude. The result removes most oscillations with a few that remained which were manually removed from our sample. These were mainly those with low explosion energies. Oscillatory behaviour also exists above 20$M_\odot$, however the light curves at higher masses exhibit different shapes in the early plateau phase so it is unclear whether the behaviour is purely from model computation. Thus, smoothing was only applied to models up to 20$M_\odot$.

\subsection{Fitting Procedure}
As the first magnitude measurement of the SNe is not necessarily the explosion date, a search through literature was performed for all SNe to find the minimum and maximum possible explosion dates. The minimum possible date is taken as the date of SNe discovery, and the maximum is the last reported non-discovery. If the difference between minimum and maximum is larger than 10 days, then the explosion epochs tried are found by linearly splitting the range into 10 intervals and trying those 10 values. For SNe with no maximum possible explosion date, the range is set to 100 days. The best-fit date is then found and a second run testing best-fit date $\pm20$ days in integer days is performed.\par 

The model is linearly extended at the tail to the maximum time of the observed data points by using {\fontfamily{lmtt}\selectfont numpy.polyfit()} on the last 15 model magnitude values to allow better fit of the radioactive-decay tail.\par 


A minimum $\chi^2$ method was employed to find the best fitting model. The photometric uncertainty associated with each observed data point is assumed to be Gaussian distributed around the measured value $y_{\mathrm{obs}}$, and the probability $\mathrm{P}$ for a match between the data and model is calculated as
\begin{equation}
\chi^2 = \Sigma_{\rm mod} \left(\frac{y_{\mathrm{mod}}-y_{\mathrm{obs}}}{\sigma_{\mathrm{tot}}}\right)^2
\end{equation}
where $\sigma_{\mathrm{tot}}$ is the quadrature-combined error from magnitude measurement and the error of the model, estimated at 0.25 mag as half of the magnitude difference between models of successive explosion energies. $y_{\mathrm{mod}}$ is the model value at the same time $t$ as the data measurement. Because the model is time-wise densely sampled, the model value at $t$ is found by linear interpolation between model values using Python's {\fontfamily{lmtt}\selectfont numpy.interp()} function.\par 

We then identify the model that has the minimum $\chi^2$ and estimate the uncertainty in each of the parameters as the range of values of each fit parameter for which $\Delta \chi^2<5.89$. Best fit and optimal $\chi^2$ values for each supernova are given in the appendix. The parameters we derive for each supernovae are the initial mass, nickel mass, explosion energy and amount of mixing. For the nickel mixing we introduced a numerical value, $X$, for the mixing of 0.1 for low mixing, 0.5 for intermediate mixing and 0.9 for max mixing. This thus represents the amount of nickel mixing into the ejecta mass as a fraction of the ejecta mass. We report these values in Table \ref{tab:freefittingresults} and \ref{tab:progconstrainedfittingresults} and include figures showing the distribution of $\chi^2$ over these parameters in the Appendix. When the uncertainties on our fits were less the spacing of our rather course grids in the parameters we assumed a minimum uncertainty for all our fit values taken from half the grid spacing of our models. For example, 0.13 (rounded up from 0.125) for our uncertainty in the explosion energy and 0.5 for the error in the initial mass.

\section{Results}
Two sets of fitting were undertaken, one where the fitting code was free to run over the entire grid of models with masses from 6 to 26$M_\odot$, and the other where the initial masses were constrained to vary over the $\pm$1\,$\sigma$ mass range deduced from progenitor observations in \cite{Smartt2015}. These two cases evaluate first what parameters can be derived from the lightcurve alone, and second whether a quantitative difference would be found in these fits in the rare cases where progenitor imaging provides additional constraints. 
We perform a quality assessment of the best fitting lightcurves (see Appendix) into broad categories, assigning a flag value:  A - robust fit with $\chi^2 <\frac{1}{2}N_{\rm obs}$, B - good fit but with some features not captured and $\chi^2_{\rm min}\sim N_{\rm obs}$, or C - poor fit with $\chi^2_{\rm min}>N_{\rm obs}$. In the quantitative analysis that follows, poor quality class C fits are omitted from analysis, although they are shown in the figures for information. Only supernovae 2005cs and 2006my fall into this category due to poor end of plateau matching and a limited number of observed points for 2006my.

The derived parameters of the best-fitting lightcurve model for each supernova are shown in Tables \ref{tab:freefittingresults} for fitting across the full, unconstrained mass range, and \ref{tab:progconstrainedfittingresults} for mass-constrained fits. 
The observational data used, together with the best fit model and the model closest to the fit parameters in each case are shown in Figures \ref{fig:freefit2013ej}-\ref{fig:constrained2003gd} in the appendices.

We again point out that our derived parameters are model dependent and have an inherent uncertainty that results from the assumptions put into SNEC. Most notably is the use of simple bolometric corrections to obtain broad-band magnitudes rather than the more complex method such as those in \citet{2013MNRAS.433.1745D}. However checking against the results of other studies enables us to have confidence in the derived parameters even if the models have limitations. As an extra test we have compared the photosphere velocity given by SNEC in our best fitting models to the velocities of some spectral lines observed. We find that we predict values of the correct order although underpredict the velocities at early times. This is again complex as we do not calculate spectra for our models and so precise comparisons are difficult.

\subsection{Validation against Progenitor Fitting}

We test the validity and precision of our approach by comparing the progenitor parameters we derive from lightcurve fitting alone against those determined by Smartt (2015) based on direct analysis of progenitors directed in pre-explosion imaging. 

\begin{figure*}
\begin{center}
\includegraphics[width=0.9\columnwidth]{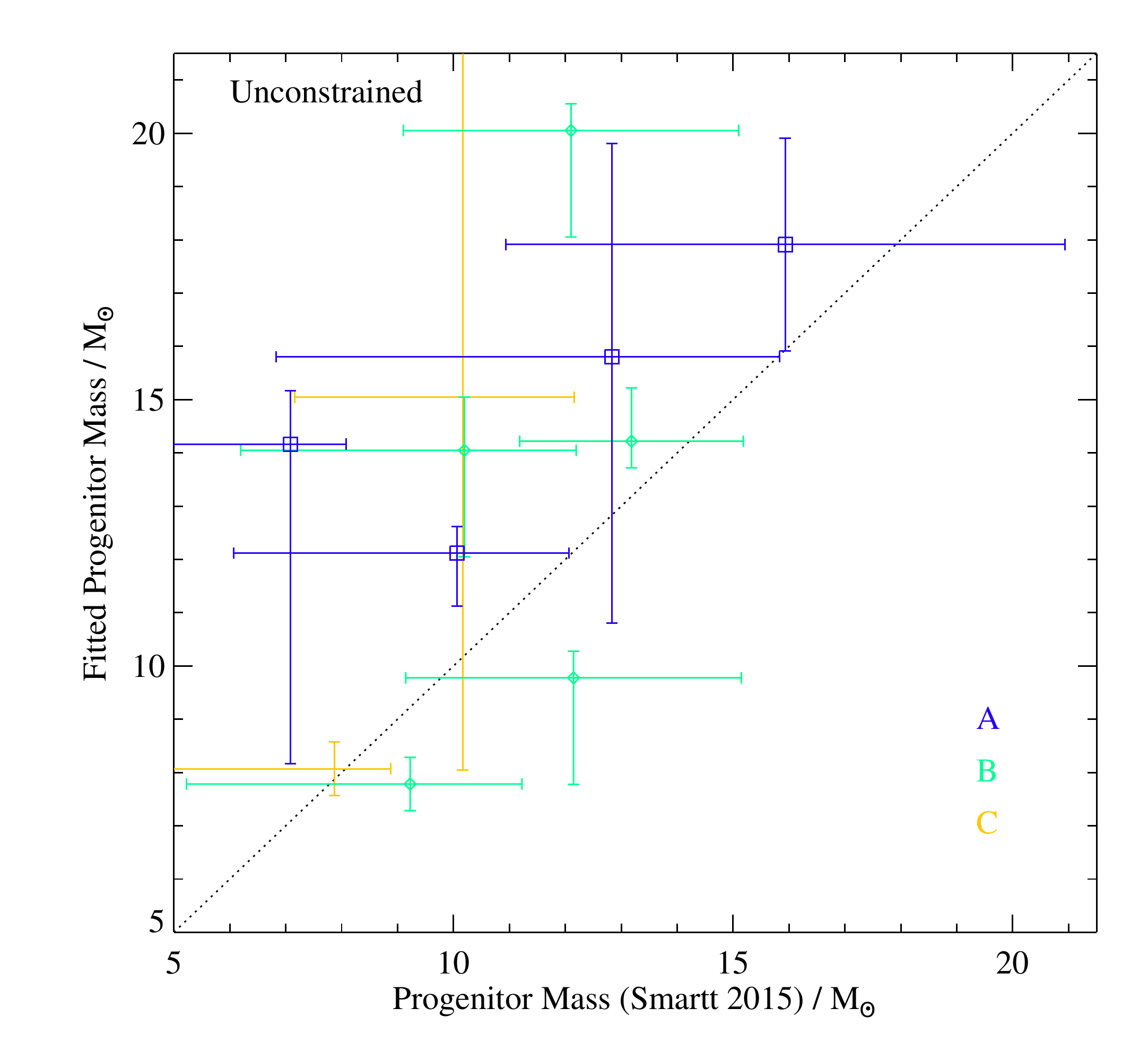}
\includegraphics[width=0.9\columnwidth]{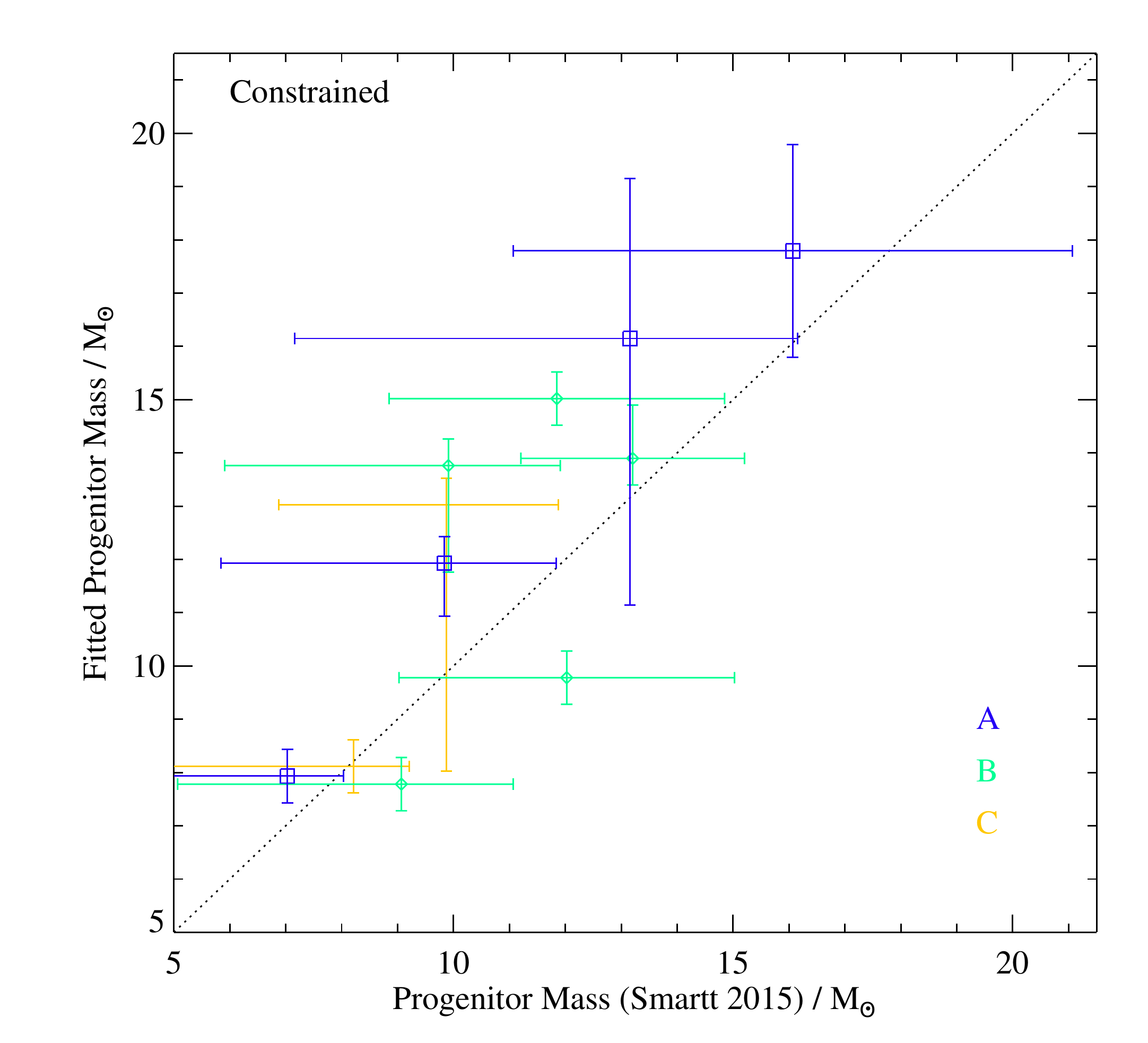}
\caption{A comparison of the progenitor initial mass derived from lightcurve fitting to that derived from analysis of progenitor observations. We show cases where the range of permitted values for lightcurve fitting is constrained by the range found by the progenitor fitting of Smartt (2015) and where it unconstrained (i.e. where the two model fits are independent). Symbols and colours indicate reliable fits (classified A, blue squares), reasonable fits (B, green diamonds) and poor (C, yellow points). Small offsets are applied to models with the same integer masses for clarity.}
\label{fig:initial_mass}
\end{center}
\end{figure*}

\begin{figure*}
\begin{center}
\includegraphics[width=0.9\columnwidth]{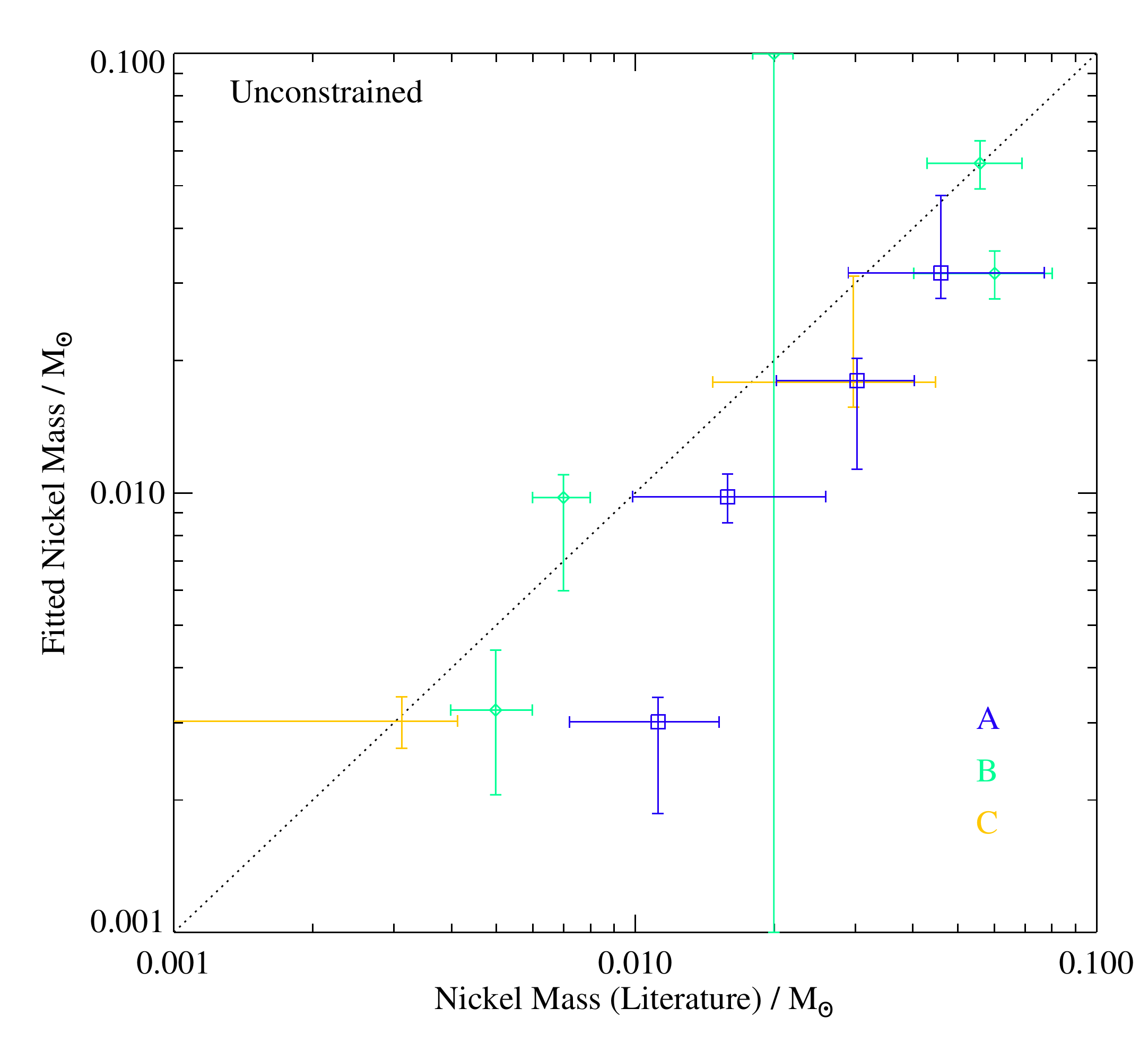}
\includegraphics[width=0.9\columnwidth]{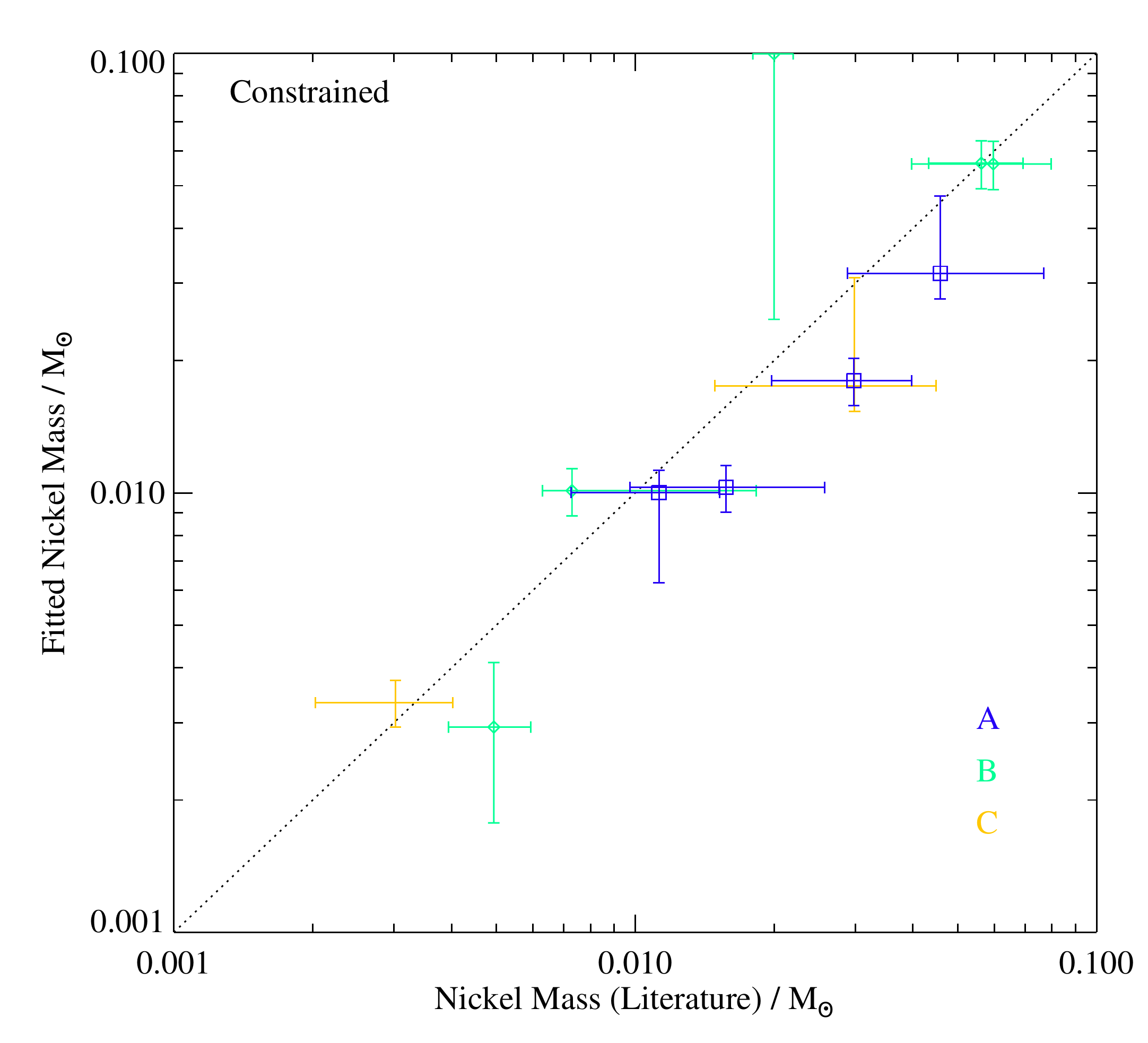}
\caption{A comparison of the progenitor Nickel-56 mass derived from lightcurve fitting to that derived from analysis of progenitor observations, as described in figure \ref{fig:initial_mass}.}
\label{fig:nickel_mass}
\end{center}
\end{figure*}

In Figures \ref{fig:initial_mass} and \ref{fig:nickel_mass} we demonstrate the results of our two approaches - fits in which the progenitor parameters are unconstrained across our model grid, and fits in which the progenitor properties are permitted only to vary within the 1\,$\sigma$ uncertainty range associated with the Smartt (2015) progenitor mass values. The former demonstrates the power of CURVEPOPS lightcurve fitting to yield an independent estimate of the parameters. The latter indicates the added value that lightcurve constraints can yield to reduce the range of uncertainties on parameters already estimated from progenitor observations.

As the figures demonstrate, the ability of CURVEPOPS to recover progenitor constraints {\it without any dependence on pre-explosion imaging} is impressive. Our unconstrained fits and those by Smartt (2015) are entirely consistent, given the still rather large uncertainties on each parameter. There is a tendency for our more robust fits (classified A) to yield slightly higher progenitor initial masses (by an average of 2.8\,M$_\odot$) although this is reduced (to 1.3\,M$_\odot$) if class A and B fits are considered. 

In Tables \ref{tab:freefittingresults} and \ref{tab:progconstrainedfittingresults} we also compare our progenitors to the estimates from analysis of the surrounding stellar population from \citet{2017MNRAS.469.2202M}, the similar study using SNEC of \citet{2018ApJ...858...15M} and updated progenitor masses from \citet{2018MNRAS.474.2116D}. We see that generally there is agreement consistent with the quoted uncertainties, however for certain supernovae there is some disagreement. For supernova 2004A the stellar population age is less than that which we derive and that inferred from the progenitor detection which could indicate that the progenitor was a runaway star and is not associated with the surrounding stellar population. The reverse is true for supernova 2009md where the stellar population inferred mass is too high but the light curve mass and pre-explosion mass match well.

Supernova 2012aw has the greatest mis-match as the \citet{2018ApJ...858...15M} mass is significantly higher than that we estimate. Our fit does have regions that do not match the light curve despite the generally good (class A) fit and the agreement of our mass with the progenitor detections. For supernova 2012ec the reverse is true and we predict a higher initial mass than \citet{2018ApJ...858...15M}. We are uncertain why we achieve such a different fit. As the $\chi^2$ parameter space indicates, there are a number of degeneracies in fitting to the light curves and the results can be dependent on the progenitor models.

A final verification of our fitting can be gained by comparing our masses to those estimated from late-time nebular spectra of some of our supernova sample by \citet{2015MNRAS.448.2482J}. The initial mass constraints they suggest are that SN\,2006my was less than 12M$_{\odot}$, SN\,2012A was 12M$_{\odot}$, while supernovae 2004et, 2012aw and 2012ec were all in the range between 12 to 15M$_{\odot}$. Again these broadly agree with our derived masses and suggests an accurate future method to estimate progenitor masses.

It is also notable that when we constrain the permitted values to lie within the 68\% confidence interval of the progenitor detection model, we are able to substantially reduce the uncertainty on both the initial and nickel masses, with our inferred uncertainties typically dominated by our sampling of these parameters rather than by the lightcurve data.

%
%

\begin{table*}
\caption{Reference and Free Fitted Parameters}
\label{tab:freefittingresults}
\begin{center}
\setlength\tabcolsep{2pt}
\begin{tabular}{lccccccccccc}
\hline\hline
 &  & \multicolumn{5}{c}{ \hspace*{1cm}Initial Mass / M$_\odot$} \hspace*{1cm}& \multicolumn{3}{c}{ } & & \\
 SN &Fit &  &   &Morozova & Davies \& & This & \multicolumn{3}{c}{ \hspace*{0.3cm}$^{56}$Ni Mass / $10^{-3}$\,M$_\odot$ \hspace*{0.15cm}} & \hspace*{0.15cm}log(Explosion \hspace*{0.15cm}  &  $^{56}$Ni Mixing\\
Name& Quality &  Smartt & Maund& et al.&  Beasor&Work  & \multicolumn{2}{c}{  Literature} & This Work &  Energy / ergs) & parameter, $X$\\
\hline
2003gd & A &  $ 7_{-1}^{+4}$ &5--14        & --           & 6.4$^{+0.6}_{-0.4}$        & 14$^{+1}_{-6}$  &  $16_{-6}^{+10 }$ &${[3]}$ &       10.0$^{+3.3}_{-2.5}$  &  50.75$^{+0.13}_{-0.38}$ &  0.9$^{+0.1}_{-0.3}$  \\
2004A  & A &  $13_{-3}^{+6}$ & 7--10       & --            & 12.7$^{+1.6}_{-1.5}$      & 16$^{+4}_{-5}$ &   $46_{-17}^{+31}$ &${[4]}$ &    31.6$^{+68.4}_{-7.9}$ &  50.5$^{+0.38}_{-0.13}$ &  0.5$\pm$0.5  \\
2004et & B &  $12\pm3$       &17$\pm$2     &16.5$^{+5.5}_{-1.5}$ & 10.7$^{+0.9}_{-0.8}$ & 20$^{+0.5}_{-2}$ &   $60\pm20$     &${[5]}$ &   31.6$^{+10.5}_{-7.9}$    &  50.75$\pm$0.13 &  0.5$\pm$0.3  \\
2005cs & C &  $ 8_{-1}^{+4}$ &7.9$\pm$0.5  &9.5$^{+2.5}_{-0.5}$ & 7.1$^{+0.5}_{-0.5}$&  8.0$\pm$0.5 &   $3_{-1}^{+1}$       &${[5]}$ &    3.2$^{+1.1}_{-0.8}$ &  50.25$\pm$0.13 &  0.1$^{+0.3}_{-0.1}$  \\
2006my & C &  $10_{-2}^{+3}$ & --          & --                & 13.9$^{+2.9}_{-3.0}$  & 15$^{+11}_{-7}$ &   $30\pm{15}$     &${[5]}$ &   17.7$^{+82.2}_{-4.4}$    &  50.75$^{1.13}_{-0.63}$ &  0.1$^{+0.9}_{-0.1}$  \\
2008bk & B &  $12\pm3$       &11$\pm$0.8   & --                 & 8.3$^{+0.6}_{-0.6}$   & 10.0$^{+0.5}_{-2}$ &   $7\pm1$     &${[6]}$ &   10.0$^{+3.3}_{-5.8}$ &  50.00$\pm$0.13 &  0.9$^{+0.1}_{-0.3}$  \\
2009md & B &  $ 9_{-2}^{+4}$ &13$\pm$1     & --                  & 8.0$^{+1.9}_{-1.5}$  &  8.0$\pm$0.5 &   $5\pm1$          &${[7]}$ &    3.2$^{+4.3}_{-1.8}$ &  50.00$\pm$0.13 &  0.9$^{+0.1}_{-0.7}$  \\
2012A  & A &  $10_{-2}^{+4}$ & --          &9.5$^{+4.5}_{-0.5}$& 8.6$^{+0.9}_{-0.8}$  & 12$^{+0.5}_{-1}$ &   $11\pm4$       &${[8]}$ &    3.2$^{+1.1}_{-1.8}$  &  50.50$\pm$0.13 &  0.9$^{+0.1}_{-0.7}$  \\
2012aw & B &  $13\pm2$       &13.5$\pm$1   &20$^{+3}_{-1}$     & 13.0$^{+1.9}_{-2.0}$   & 14$^{+1}_{-0.5}$ &   $56\pm13$   &${[9]}$ &     56.2$^{+18.8}_{-14.1}$ &  50.75$\pm$0.13 &  0.5$\pm$0.3  \\
2012ec & A &  $16\pm5$       &16-27        &10.5$^{+7.5}_{-1.5}$& 16.8$^{+1.4}_{-1.3}$& 18$\pm$2 &   $30\pm10$             &${[10]}$&     17.8$^{+5.9}_{-10.3}$    &  50.50$\pm$0.13 &  0.5$\pm$0.5  \\
2013ej & B &  $10_{-2}^{+4}$ &14$\pm$1.5   &13$^{+5.5}_{-3}$ & 9.8$^{+0.8}_{-0.7}$ &    14$^{+1}_{-2}$ &   $20\pm2$        &${[11]}$&     100$\pm$90           &  51.00$\pm$0.13 &  0.9$^{+0.1}_{-0.7}$  \\
\hline\hline
\end{tabular}
\end{center}
\tabnote{$^1$From progenitor observations and modelling, as in \cite{Smartt2015}}
\tabnote{$^3$\cite{Ni2003gd}  $^4$\cite{Ni2004A}  $^5$\cite{deathofmassivestars}  $^6$\cite{Ni2008bk}  $^7$\cite{SN2009mdData}  $^8$\cite{SN2012AProgPaper}  $^9$\cite{Ni2012aw}  $^{10}$\cite{Ni2012ec}  $^{11}$\cite{Ni2013ej}}
\end{table*}


\begin{table*}
\caption{Reference and Progenitor Constrained Fitting Parameters}
\label{tab:progconstrainedfittingresults}
\begin{center}
\setlength\tabcolsep{2pt}
\begin{tabular}{lccccccccccc}
\hline\hline
 &  & \multicolumn{5}{c}{ \hspace*{1cm}Initial Mass / M$_\odot$} \hspace*{1cm}& \multicolumn{3}{c}{ } & & \\
SN  &Fit &   &  & Morozova & Davies \& & This & \multicolumn{3}{c}{ \hspace*{0.3cm}$^{56}$Ni Mass / $10^{-3}$\,M$_\odot$ \hspace*{0.15cm}} & \hspace*{0.15cm}log(Explosion \hspace*{0.15cm}  &  $^{56}$Ni Mixing\\
Name& Quality & Smartt & Maund& et al.&Beasor &Work  & \multicolumn{2}{c}{  Literature} & This Work &  Energy / ergs) & parameter, $X$\\
\hline
2003gd & A &  $ 7_{-1}^{+4}$ & 5--14       & --           & 6.4$^{+0.6}_{-0.4}$      &   8$\pm$0.5 &   $16_{-6}^{+10 }$      &${[3]}$&  10.0$^{+3.3}_{-2.5}$  &  50.50$^{+0.38}_{-0.13}$ &  0.9$^{+0.1}_{-0.7}$  \\
2004A  & A &  $13_{-3}^{+6}$ & 7--10       & --           & 12.7$^{+1.6}_{-1.5}$      &  16$^{+3}_{-5}$ &   $46_{-17}^{+31}$ &${[4]}$&  31.6$^{+68.4}_{-7.9}$    &  50.50$\pm$0.13 &  0.5$\pm0.3$  \\
2004et & B &  $12\pm3$       & 17$\pm$2    &16.5$^{+5.5}_{-1.5}$ & 10.7$^{+0.9}_{-0.8}$ &  15$\pm$0.5 &   $60\pm20$          &${[5]}$&  56.2$^{+18.8}_{-14.1}$    &  50.75$\pm$0.13 &  0.5$\pm0.3$  \\
2005cs & C &  $ 8_{-1}^{+4}$ &7.9$\pm$0.5  &9.5$^{+2.5}_{-0.5}$ & 7.1$^{+0.5}_{-0.5}$ &  8$\pm$0.5 &   $3_{-1}^{+1}$         &${[5]}$&  3.2$^{+1.1}_{-0.8}$  &  50.25$\pm$0.13 &  0.1$^{+0.3}_{-0.1}$  \\
2006my & C &  $10_{-2}^{+3}$ & --          & --                & 13.9$^{+2.9}_{-3.0}$ &   13$^{+0.5}_{-5}$ &   $30\pm{15}$   &${[5]}$&  17.8$^{+82.2}_{-4.4}$  &  50.50$^{+0.5}_{-0.38}$ &  0.1$^{+0.9}_{-0.1}$  \\
2008bk & B &  $12\pm3$       &11$\pm$0.8   & --               & 8.3$^{+0.6}_{-0.6}$  &   10$\pm$0.5 &   $7\pm1$              &${[6]}$&  10.0$^{+3.3}_{-2.5}$  &  50.00$\pm$0.13 &  0.9$^{+0.1}_{-0.9}$  \\
2009md & B &  $ 9_{-2}^{+4}$ &13$\pm$1     & --                & 8.0$^{+1.9}_{-1.5}$ &    8$\pm$0.5 &   $5\pm1$              &${[7]}$&  3.16$^{+4.3}_{-1.8}$ &  50.00$\pm$0.13 &  0.9$^{+0.1}_{-0.9}$  \\
2012A  & A &  $10_{-2}^{+4}$ & --          &9.5$^{+4.5}_{-0.5}$& 8.6$^{+0.9}_{-0.8}$ &    12$^{+0.5}_{1}$ &   $11\pm4$       &${[8]}$&  10.0$^{+3.3}_{-5.8}$  &  50.50$\pm$0.13 &  0.9$^{+0.1}_{-0.9}$  \\
2012aw & B &  $13\pm2$       &13.5$\pm$1   &20$^{+3}_{-1}$     & 13.0$^{+1.9}_{-2.0}$ &   14$^{+1}_{-0.5}$ &   $56\pm13$     &${[9]}$&  56.2$^{+18.8}_{-14.1}$    &  50.75$\pm$0.13 &  0.5$\pm0.3$  \\
2012ec & A &  $16\pm5$       &16-27        &10.5$^{+7.5}_{-1.5}$ & 16.8$^{+1.4}_{-1.3}$ &    18$\pm$2 &   $30\pm10$          &${[10]}$& 17.8$^{+5.9}_{-4.4}$    & 50.50$\pm$0.13 &  0.5$^{+0.3}_{-0.5}$  \\
2013ej & B &  $10_{-2}^{+4}$ &14$\pm$1.5   &13$^{+5.5}_{-3}$  & 9.8$^{+0.8}_{-0.7}$&    14$^{0.5}_{-2}$ &   $20\pm2$         &${[11]}$&  100$^{+82.2}_{-82.2}$    & 51.00$\pm$0.13 &  0.9$^{+0.1}_{-0.7}$  \\
\hline\hline
\end{tabular}
\end{center}
\tabnote{$^1$From progenitor observations and modelling, as in \cite{Smartt2015}}
\tabnote{$^2$0: low $0.9M_r+0.1M_\ast$, 1 : mid $0.5M_r+0.5M_\ast$, 2: max $0.1M_r+0.9M_\ast$ where $M_r$ is mass of the remnant and $M_\ast$ is the mass of the ejecta.}
\tabnote{$^3$\cite{Ni2003gd}  $^4$\cite{Ni2004A}  $^5$\cite{deathofmassivestars}  $^6$\cite{Ni2008bk}  $^7$\cite{SN2009mdData}  $^8$\cite{SN2012AProgPaper}  $^9$\cite{Ni2012aw}  $^{10}$\cite{Ni2012ec}  $^{11}$\cite{Ni2013ej}}
\end{table*}

\subsection{Additional Explosion Parameters}

In addition to the progenitor parameters, CURVEPOPS analysis involves fitting over explosion parameters which are not explored in Smartt 2015. These represent added value from this approach, which cannot be derived from pre-explosion progenitor imaging.

\subsubsection{Explosion Energy}

\begin{figure*}
\begin{center}
\includegraphics[width=0.9\columnwidth]{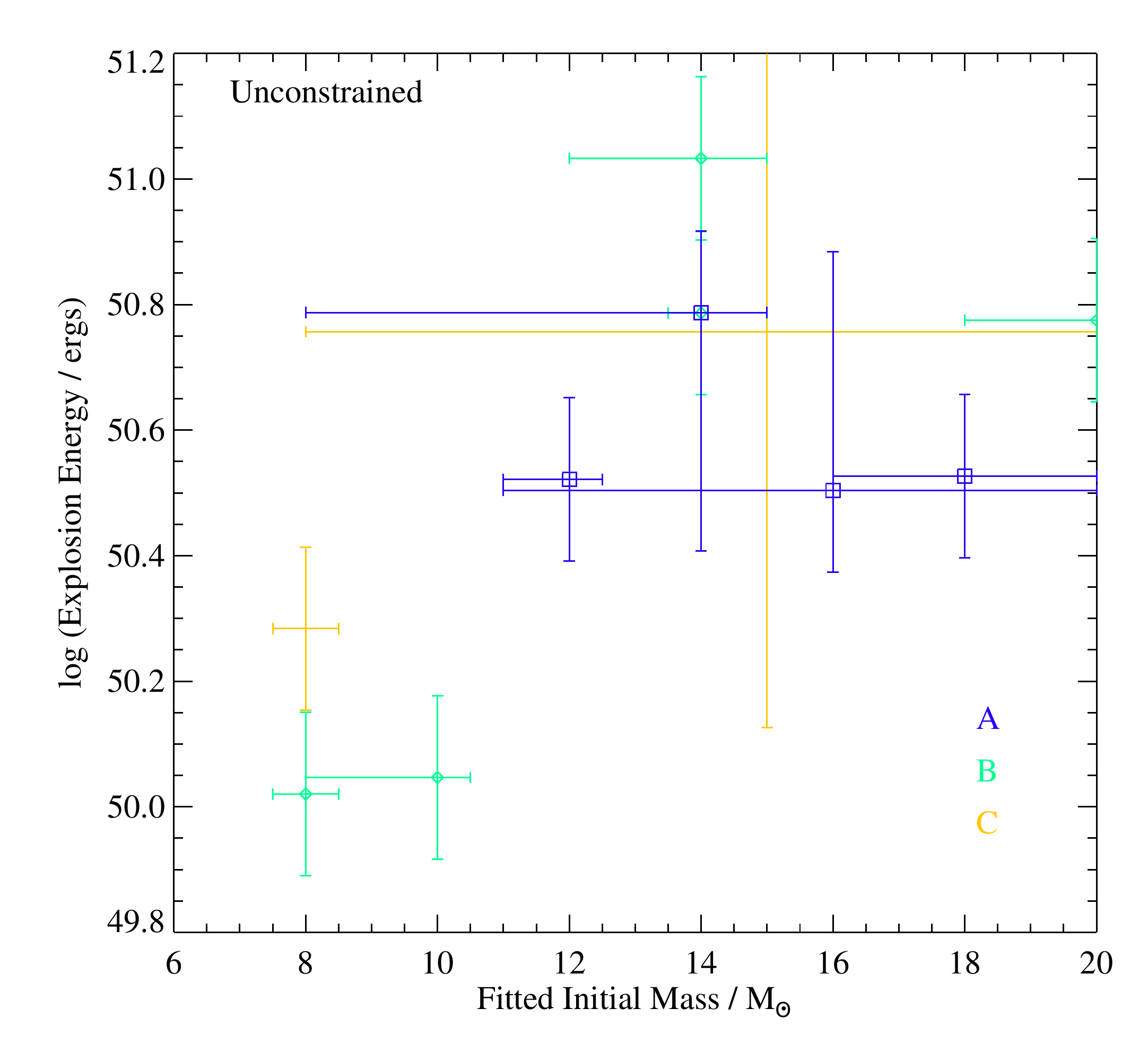}
\includegraphics[width=0.9\columnwidth]{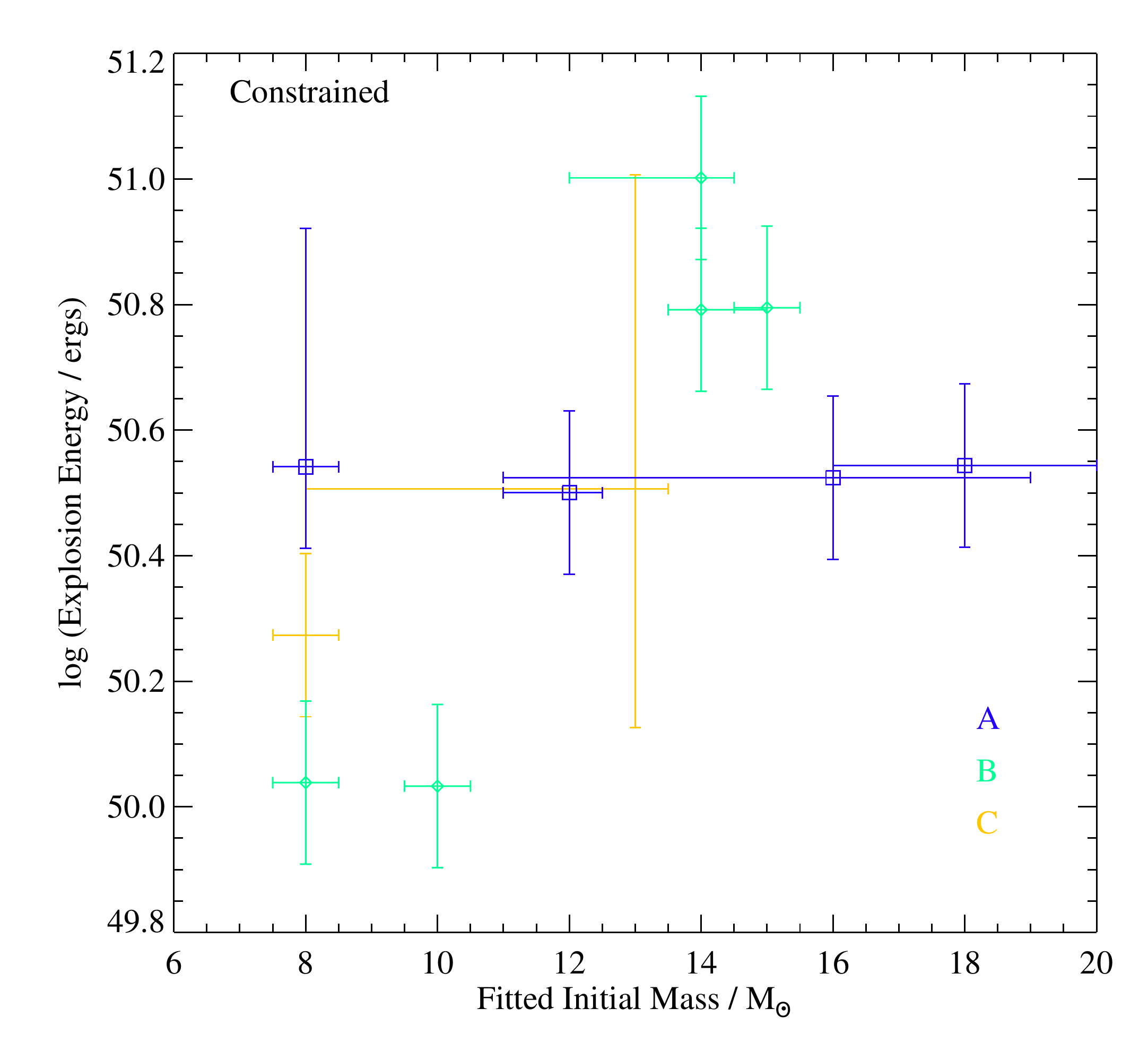}
\caption{The explosion energy derived from lightcurve fitting and its dependence on initial mass, with symbols as described in figure \ref{fig:initial_mass}.}
\label{fig:mass_energy}
\end{center}
\end{figure*}

In figure \ref{fig:mass_energy} we investigate the dependence of assumed supernova explosion energy on the progenitor initial mass for the best fitting CURVEPOPS model.

We find no strong dependence of energy on mass, with a mean log(E$_\mathrm{exp}$/ergs)=50.52$\pm$0.10 for our most robust, class A lightcurve fits. When our slightly less-good, class B fits are included, we see some evidence for an increased spread of explosion energies particularly for low progenitor initial masses. It will be interesting to see if this trend is robust in larger samples of lightcurves.

\subsubsection{Nickel Mixing}

\begin{figure*}
\begin{center}
\includegraphics[width=0.9\columnwidth]{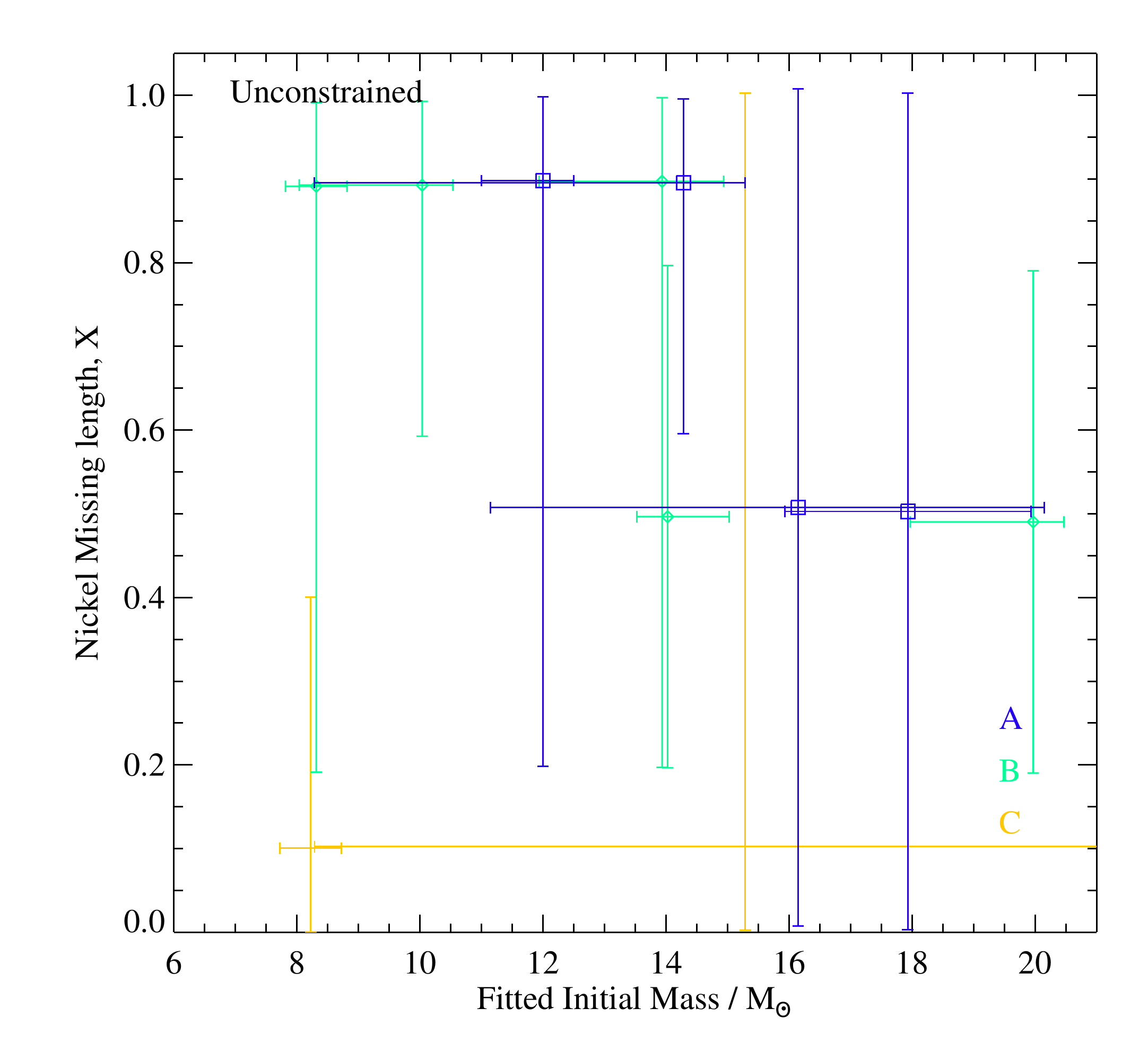}
\includegraphics[width=0.9\columnwidth]{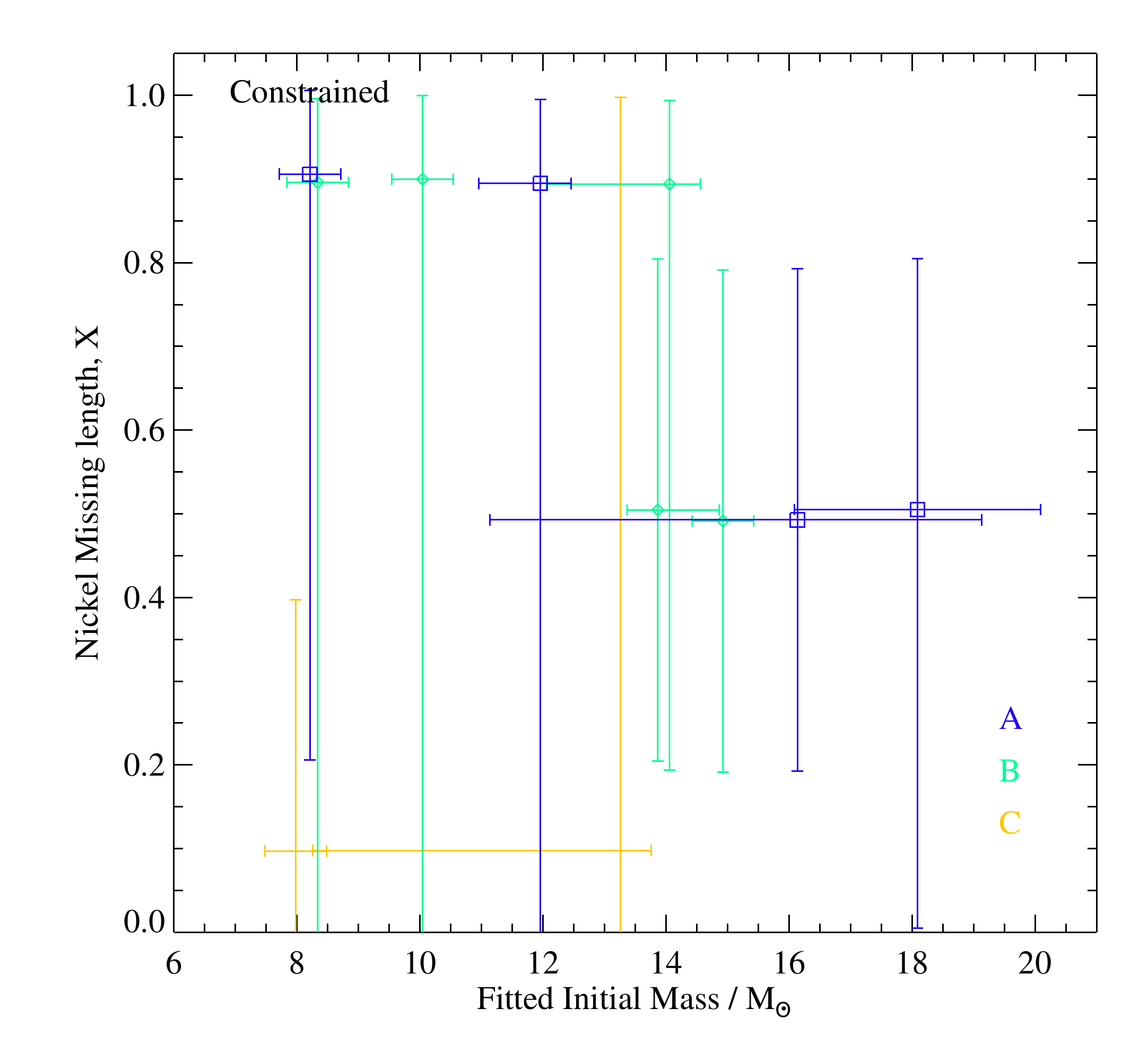}
\caption{The nickel mixing length parameter derived from lightcurve fitting and its dependence on initial mass, with symbols as described in figure \ref{fig:initial_mass}.}
\label{fig:mass_mlength}
\end{center}
\end{figure*}

In figure \ref{fig:mass_mlength} we investigate the dependence of nickel mixing length on the progenitor initial mass for the best fitting CURVEPOPS model. As explained above, we have calculated models with a nickel mixing mass coordinate defined as $M_{\rm excised}+X\,M_{\rm ejecta}$ where the parameter $X$=[0.1, 0.5, 0.9] with larger values indicating more mixing. We interpolate between these to find the best fitting value.

%

Unfortunately, the extremely large uncertainties on our inferred Nickel mixing lengths makes definite conclusions from this analysis impossible. We note that our very coarse grid (with only three samples) is likely inflating these uncertainties, although there may well be underlying degeneracies in the fit parameters, as demonstrated by the projections of $\chi^2$ parameter space for individual fits illustrated in the Appendices. We do perhaps see a hint that lower mixing lengths might be favoured for high mass progenitors, but further work with a much finer grid of models will be required to verify this and explore its origin further.

\subsubsection{Explosion Epoch}

During CURVEPOPS fitting, an explosion epoch must be derived, particularly for those events in which the pre- and early post-explosion light curve is poorly sampled. Initial 
constraints on the SNe explosion epochs are determined by searching the literature for the last reported non-detection of each object to act as the upper limit on its age and the supernova discovery date to be the lower limit. These are refined in the literature as described in section 4.2. The details of constraints, along with fitted explosion epochs, are summarized in Table \ref{tab:ExplosionEpoch}.

\begin{table*}
\caption{Constraints and Fitted Results of Explosion Epochs in MJD}
\label{tab:ExplosionEpoch}
\begin{center}		
\begin{tabular}{ccc p{4.5cm} cc}
\hline\hline
& {Minimum} & {Maximum } &  &{Free} & {Progenitor}\\
& {Explosion} &{ Explosion } &  &{Fitted } & {Constrained}\\
Supernova& { Date} & { Date} & \multicolumn{1}{c}{Source} & { Result} & {Result}\\
\hline
SN2003gd & 52667.6 & 52802.32 & \cite{2003IAUC.8150....2E} \cite{2003IAUC.8163....3G} & 52868 $^{+1}_{-14}$    & 52885 $^{+1}_{-9}$\\
SN2004A & - & 53012.9&          \cite{SN2005csProgPaper}                              & 53042 $^{+11}_{-15}$     & 53042$^{+11}_{-15}$\\
SN2004et& 53270 & 53275        & \cite{2004IAUC.8413....1Z}                           & 53283 $^{+2}_{-1}$       & 53285$^{+1}_{-1}$\\
SN2005cs & 53516 & 53548.933 &  \cite{2005IAUC.8553....1M}                            & 53420 $^{+1}_{-1}$      & 53420$^{+1}_{-1}$\\
SN2006my & 53914 & 54047.32 &    \cite{2006CBET..727....1N}                           & 53856 $^{+20}_{-20}$     & 53854 $^{+21}_{-18}$\\
SN2008bk& 54467.742 & 54550.15&    \cite{Ni2008bk}                                    & 54556 $^{+6}_{-3}$     & 54556  $^{+6}_{-3}$\\
SN2009md & 55154 & 55174.31 &     \cite{SN2009mdData}                                 & 55198 $^{+1}_{-2}$     & 55198 $^{+1}_{-2}$\\
SN2012A & 55924 & 55933.39 &      \cite{2012CBET.2974....2L}                          & 55946 $^{+1}_{-4}$    & 55946$^{+1}_{-5}$\\
SN2012aw & 56000.77 & 56002.4 &  \cite{2013MNRAS.433.1871B}                          & 55804$^{+1}_{-1}$      & 55804 $^{+1}_{-1}$\\
SN2012ec & - & 56150.039 &       \cite{SN2012ecProg}                                 & 56189 $^{+7}_{-9}$      & 56189$^{+7}_{-9}$\\
SN2013ej& 56496.04 & 56496.625    &\cite{Ni2013ej}                                   & 56501 $^{+1}_{-1}$      & 56501 $^{+1}_{-1}$\\
\hline\hline
\end{tabular}
\end{center}
\end{table*}



\section{DISCUSSION}

A first important question is to evaluate how well our fitting method has recovered the initial and nickel masses of the SNe we have studied. We can see in Figures \ref{fig:initial_mass} and \ref{fig:nickel_mass}  that in general the both of these masses are consistent with those from pre-explosion mass constraints from the literature. For the initial mass we find that at first glance there is little difference between the fits that are constrained by the progenitor mass from pre-explosion images and those that are not. A closer look reveals that the scatter is less in the latter case. This is probably due to the degeneracy in SN light curves between mass and energy. As for the nickel mass we again reproduce the literature values, however for some of the greater values the scatter is larger. 

We have also compared our masses to those from estimating the age of co-eval stars surrounding the progenitor site as estimated by \citet{2017MNRAS.469.2202M}. Again here we see that we are at least consistent with these masses in most cases. Other indirect measurements of a few of the progenitor mass are also possible, for example \citet{2018arXiv180501213X}, and these in general also agree with the masses we derive. 

While our derived masses are comparable to those of \citet{Smartt2015} in general we derive higher masses. This is agreement with others such as \citet{2018ApJ...858...15M} with a similar lightcurve fitting method and \citet{2018MNRAS.474.2116D} who re-evaluated the pre-supernova luminosity of the detected progenitor stars.

We note that \citet{2019A&A...625A...9D} and \citet{2019arXiv190309114G} have pointed out the difficulty of deriving initial masses of supernova progenitors from the lightcurves. We do find our results do have large uncertainties in some cases. However our results are dependent on for example our matching of the circumstellar environment to the stellar progenitor model as well as other caveats. At best our estimates give a relative estimate of whether the progenitors are more or less massive, similar to the broad scheme suggested by \citet{2006ApJ...641.1029C}. 

For several of the supernovae in our sample, 2004et, 2005cs, 2008bk, 2009md, 2012aw and 2013ej the fits have some part that does not match the observed light curves. We suspect this is due to the limited scope of this study which only uses single-star progenitor models. We expect that if we were to also explode progenitor models that had undergone binary interactions we should be able to have a greater variety of light curves (see paper I for example). The other parameter we have not varied is initial metallicity of the progenitors. Here we have only used one metallicity, but varying this will change in subtle ways the progenitor structure and also the density of the circumstellar medium, by changing the stellar wind mass-loss rates and the wind velocity. The alternative would be to adopt a similar approach to \citet{2018ApJ...858...15M} and instead compute models over a range of circumstellar medium parameters rather than those from the progenitor model. Their work does suggest in the cases where we achieve a poor fit we may be underestimating the amount of circumstellar material. Both these approaches could allow us to find a better fit, however the computational demand of calculating synthetic light curves from more progenitor models or more circumstellar environments, while still varying the explosion parameters as here, for each progenitor is extreme. We have demonstrated here that the CURVEPOPS concept is useful and can thus now begin to undertake this mammoth set of numerical calculations. 

\citet{2018ApJ...858...15M} found good fits to a number of the same supernova lightcurves by varying the amount of circumstellar material around each of their progenitor models. In each case the amount required was significant, of the order of a few times 0.1M$_{\odot}$. Here we have also included the circumstellar medium and achieve a similar early brightening of our theoretical lightcurves. In contrast to \citet{2018ApJ...858...15M} we achieved this with a mass-loss rate and wind velocity determined from the progenitor model. In addition, our use of the the wind acceleration model of \citet{2018MNRAS.476.2840M} means that the density of the circumstellar medium is higher close to the star before decreasing out to the density expected for a freely expanding wind. This produced similar high densities as found by \citet{2018ApJ...858...15M}. This suggests how the circumstellar medium is modelled does not matter so much as the mass of material that is close to the progenitor star upon explosion. Future investigations of progenitors at different metallicities and thus different wind parameters will enable us to understand the importance of the circumstellar medium to a greater degree. 

\begin{figure*}
\begin{center}
\includegraphics[width=0.9\columnwidth]{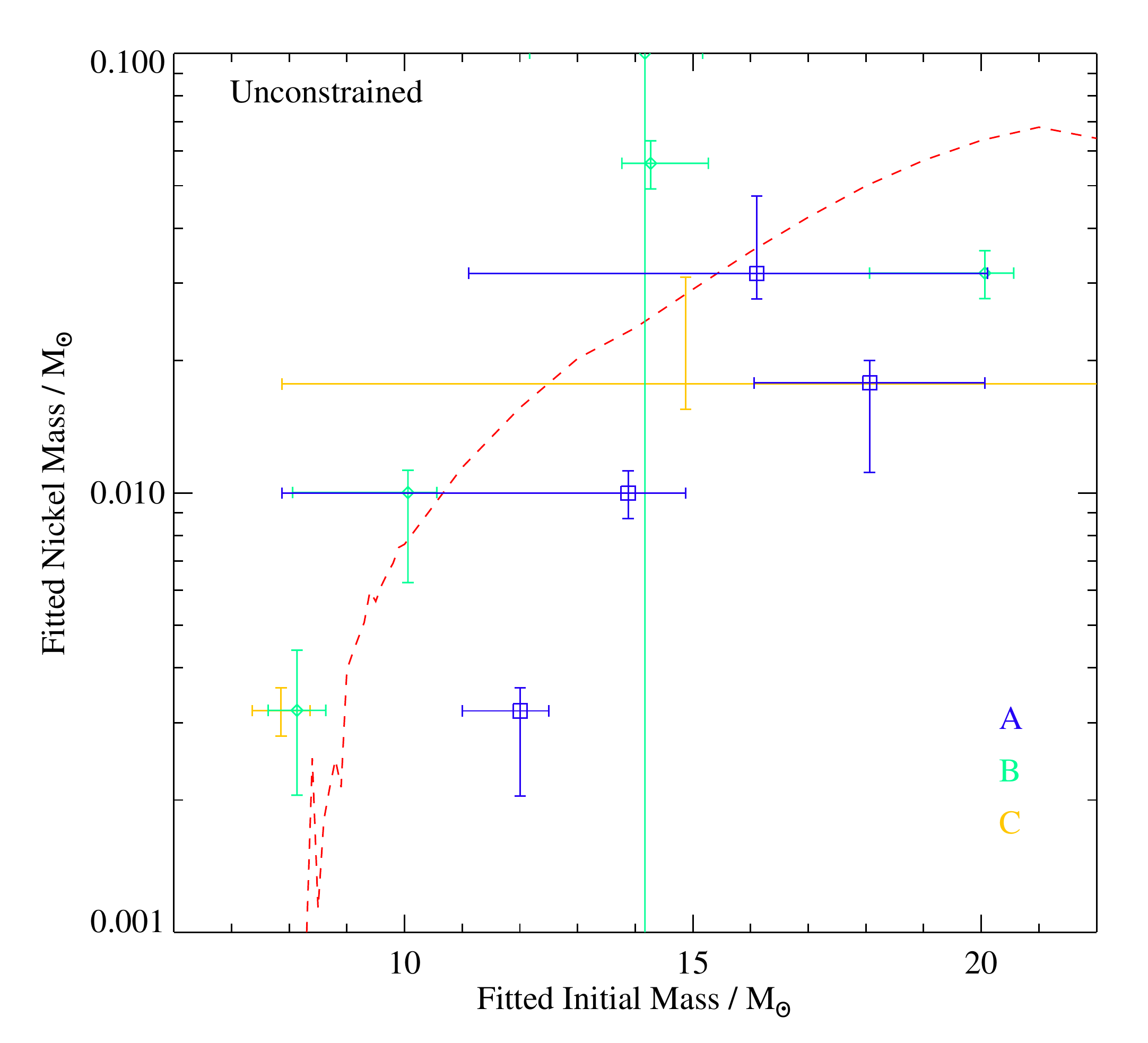}
\includegraphics[width=0.9\columnwidth]{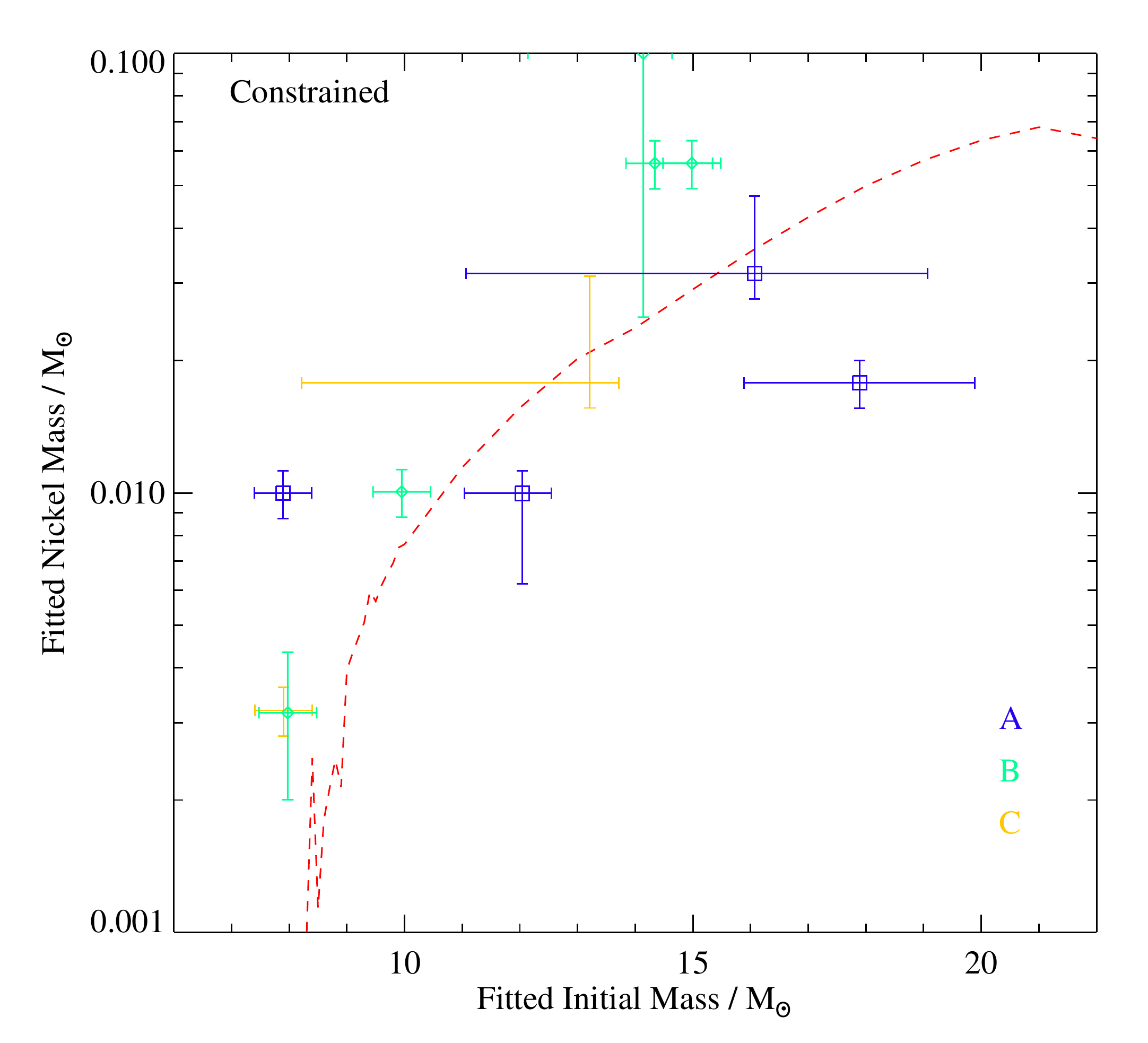}
\caption{The nickel mass derived from lightcurve fitting and its dependence on fitted initial mass, with symbols as described in figure \ref{fig:initial_mass}. Lines show a simple linear relation between the logarithms of initial and nickel mass (dotted) and also a model in which the nickel mass is proportional to the amount of mass in the CO core before collapse which is not removed into a compact remnant (dashed line, see text).}
\label{fig:mass_nimass}
\end{center}
\end{figure*}

Smartt et al (2009) proposed that the evolution of explosively-synthesised nickel mass with progenitor initial mass was mediated by the amount of oxygen in the star at the supernova epoch, and specifically considered the ratio between oxygen mass and carbon-oxygen (CO) core mass at explosion. In Figure \ref{fig:mass_nimass} we consider the same parameters based on light curve fitting alone. In common with Smartt et al, we see a trend to lower nickel masses for supernovae with lower initial progenitor masses. The only outliers from this trend are supernova for which the light curve fit is particularly poor and the inferred parameters unreliable.  Overplotted on the figure we show a plausible relation between these quantities. 
The dashed line indicates the size of the carbon-oxygen mantle that surrounds the forming compact remnant, scaled to match the data at 15\,M$_\odot$. This appears to track the datapoints (subject to the substantial uncertainties) suggesting that this may be an important parameter in determining the nickel mass.

\section{CONCLUSION}

This article is the second in the CURVEPOPS series and again shows the utility of calculating large grids of synthetic supernova light curves for comparison to observed supernovae. While not every fit to an observed supernova was good, in general the population of results can be used to find various relationships between initial progenitor mass and explosion energy, nickel masses or nickel mixing. 

We stress that there are limitations to our study, which are primarily caused by computational limits on the number of models considered. The only solution is to calculate more models to allow for the full diversity of stellar structure and circumstellar environment around each star. We estimate that to calculate all the necessary synthetic light curves requires would be approximately 25 million supernova simulations which would take approximately 120 million CPU hours. This is certainly possible but even with significant computing resources it still takes time and is beyond the scope of this article.

From considering the fits together we find that,
\begin{enumerate}
\item The typical explosion energy to be input into SNEC is log(E$_\mathrm{exp}$/ergs)=50.52$\pm$0.10.
\item We find a relation between nickel mass and initial mass which may track the size of the carbon-oxygen core at core collapse.
\item We find suggestions of a weak dependence of nickel mixing on initial mass with less mixing when there is a more massive ejecta and initial progenitor.
\item It is possible to achieve strong constraints on the progenitors of type IIP supernovae from the light curves alone.
\item As found by \citet{2018ApJ...858...15M} and \citet{2018MNRAS.476.2840M} it is important to include the circumstellar material surrounding the progenior stars to correctly model type IIP supernova lightcurves. However exactly how to include this in the lightcurve modelling requires further study.
\item Good fits are not possible for every observed supernovae which suggests that there are other factors at play in specifying the shape of type IIP light curves. These include the initial stellar metallicity and binary interactions.
\end{enumerate}

In summary, we can produce progenitor constraints independent of progenitor observations and rivaling them in quality and, where progenitor imaging exists, we add additional data and tighten the uncertainties on key parameters. Finally the synthetic light curves and SNEC input files are freely available from the BPASS website and PASA data store as a resource for the community to use. This data will be continually added to as more simulations are computed.

\begin{acknowledgements}
This work would not have been possible without use of the
NeSI Pan Cluster, part of the NeSI high-performance computing
facilities. New Zealand's national facilities are provided
by the NZ eScience Infrastructure and funded jointly
by NeSI's collaborator institutions and through the Ministry
of Business, Innovation \& Employment's Research Infrastructure programme. URL: \texttt{https://www.nesi.org.nz.}.
JJE acknowledges support from the
University of Auckland and thanks the OCIW Distinguished Visitor Program at the Carnegie Observatories, which funded JJE's visit to gain advice from Anthony L. Piro on using SNEC and including the circumstellar media around the progenitors.
ERS acknowledges support from the University of Warwick. NYG and NR both acknowledges support from University of Auckland's Summer Scholarships.  LX acknowledges support from the China Postdoctoral Science Foundation (grant No.2018M642524). LX acknowledge the grant from the National Key R\&D Program of China (2016YFA0400702), the National Natural Science Foundation of China (No. 11673020 and No. 11421303), and the National Thousand Young Talents Program of China.

\end{acknowledgements}

\bibliographystyle{pasa-mnras}
\bibliography{ReportReferences}

\newpage

\begin{appendix}

\section{Results from a free fit of the light curves}
Here we present the best fitting V-band magnitude lightcurves for the supernovae when a free fit across the full range of modelled initial masses is allowed as well as corner plots showing how the $\chi^2$ varies with the 5 parameters we fit. We also include plots showing how $\chi^2$ only depends on one parameter at a time as shown by the solid black line while the horizontal dashed lines show the $\Delta \chi^2$ for the 1$\sigma$, 2$\sigma$ and 3$\sigma$. In the bottom figure of the lightcurves we plot the best fitting model (black line) along with the lightcurves at are within the 1$\sigma$ uncertainty in grey while the observations are shown in red. 
\begin{figure*}[h]
\begin{center}
\includegraphics[width=2\columnwidth]{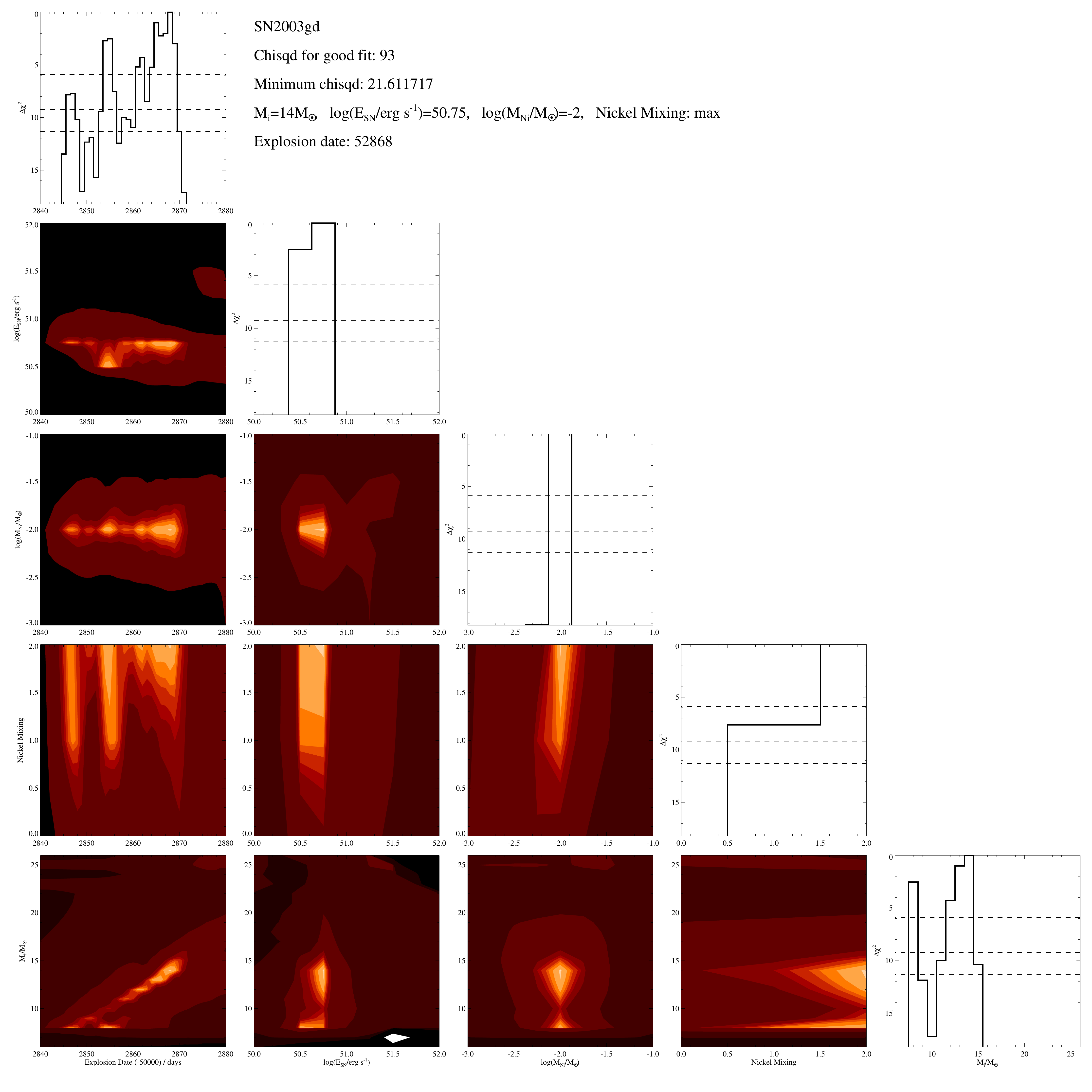}
\includegraphics[width=\columnwidth]{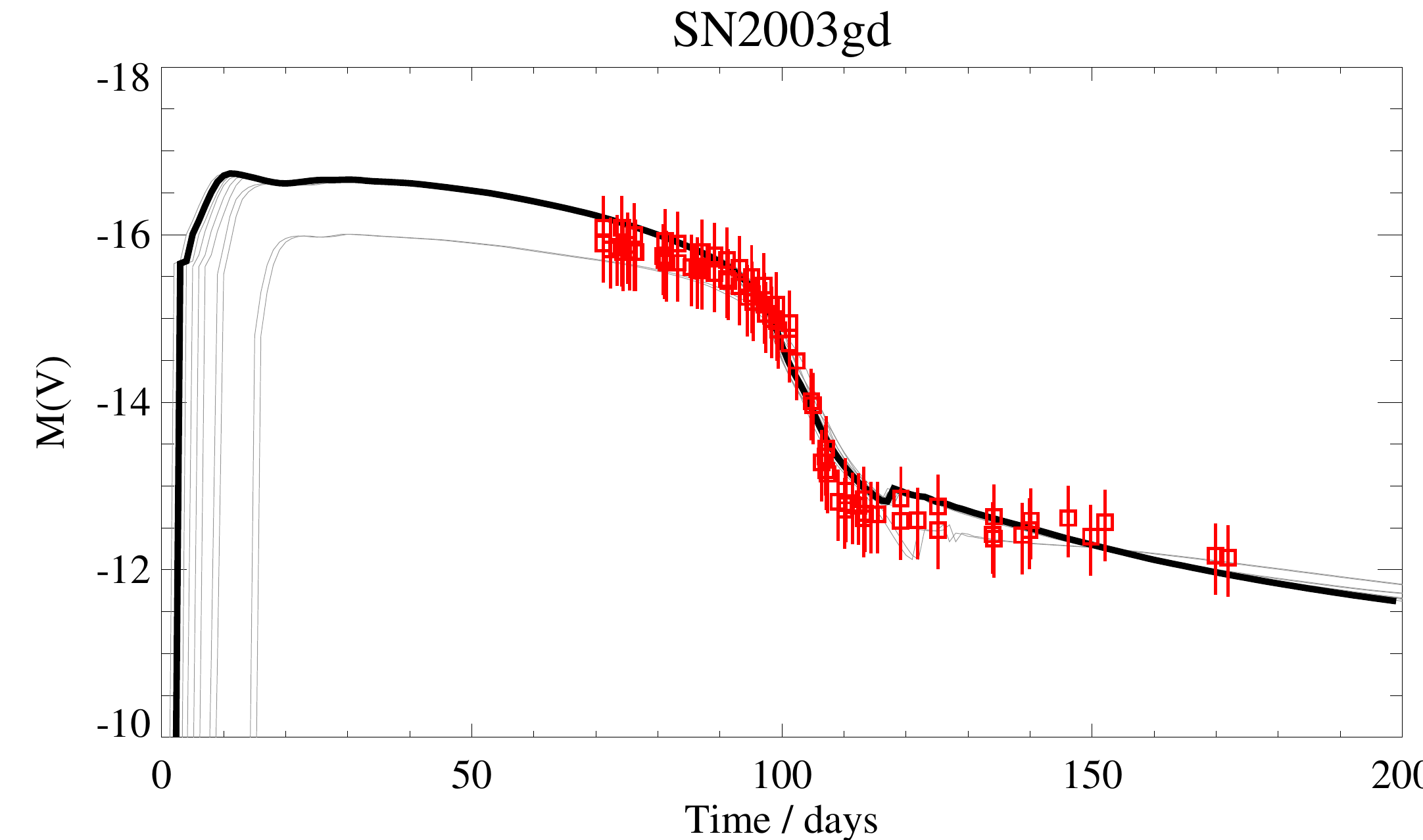}
\caption{SN2003gd Free-Fit, corner plots showing how $\chi^2$ varies over the 5 parameters as well a plot comparing the observed lightcurves to the matching theoretical models.}
\label{fig:freefit2003gd}
\end{center}
\end{figure*}

\begin{figure*}[!h]
\begin{center}
\includegraphics[width=2\columnwidth]{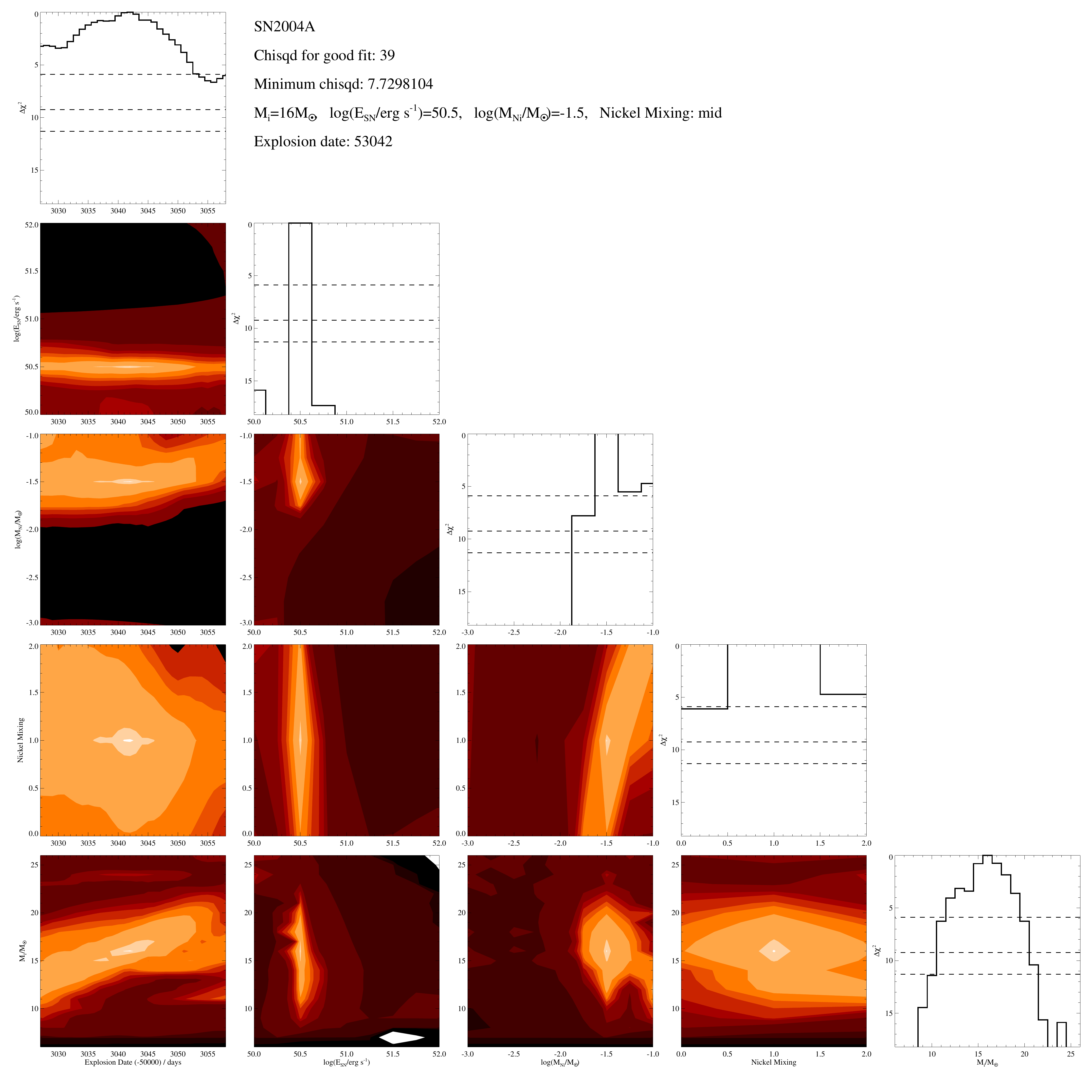}
\includegraphics[width=\columnwidth]{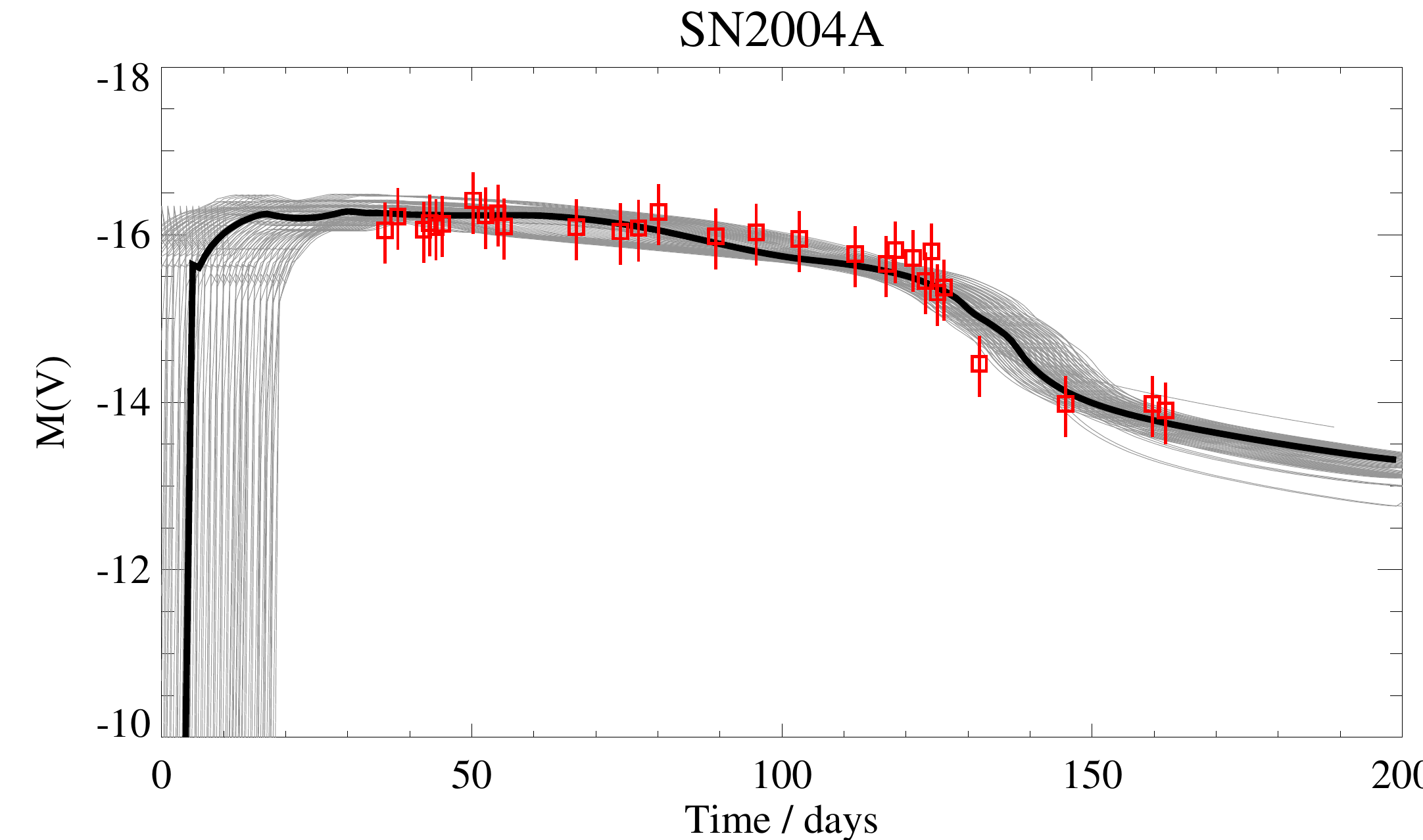}
\caption{SN2004A Free-Fit, as in Figure \ref{fig:freefit2003gd}}
\label{fig:freefit2004A}
\end{center}
\end{figure*}

\begin{figure*}[h]
\begin{center}
\includegraphics[width=2\columnwidth]{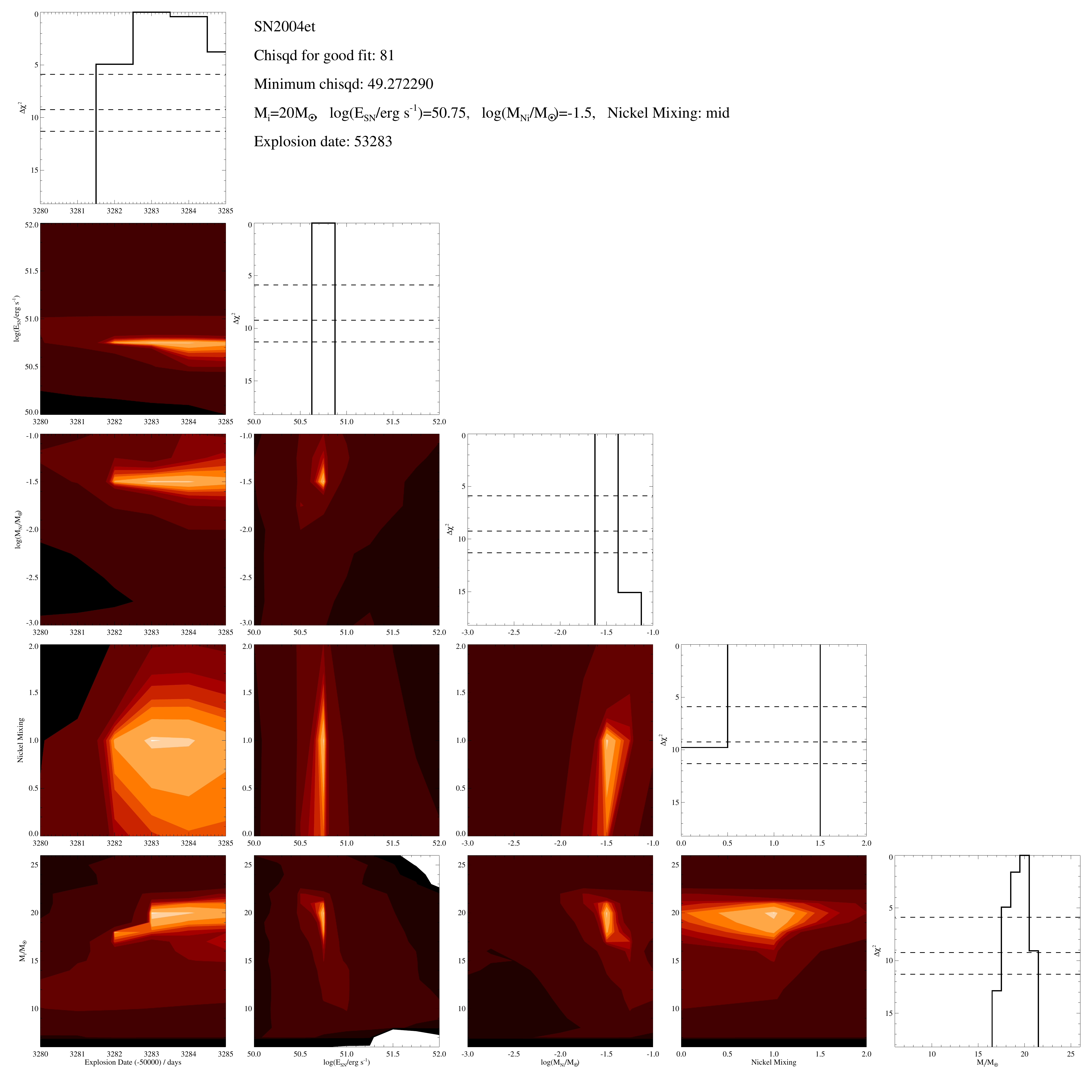}
\includegraphics[width=\columnwidth]{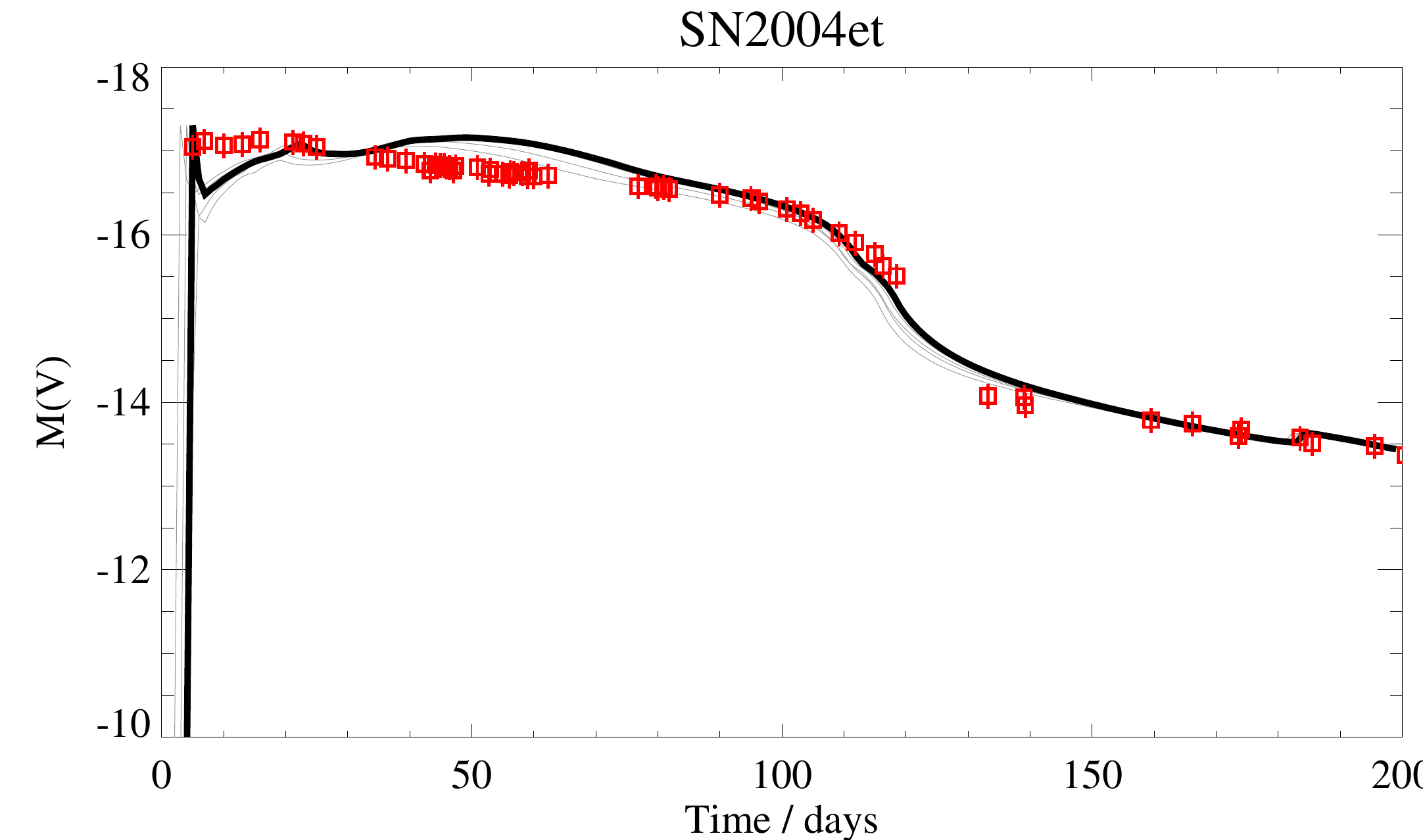}
\caption{SN2004et Free-Fit, as in Figure \ref{fig:freefit2003gd}}
\label{fig:freefit2004et}
\end{center}
\end{figure*}

\begin{figure*}[!h]
\begin{center}
\includegraphics[width=2\columnwidth]{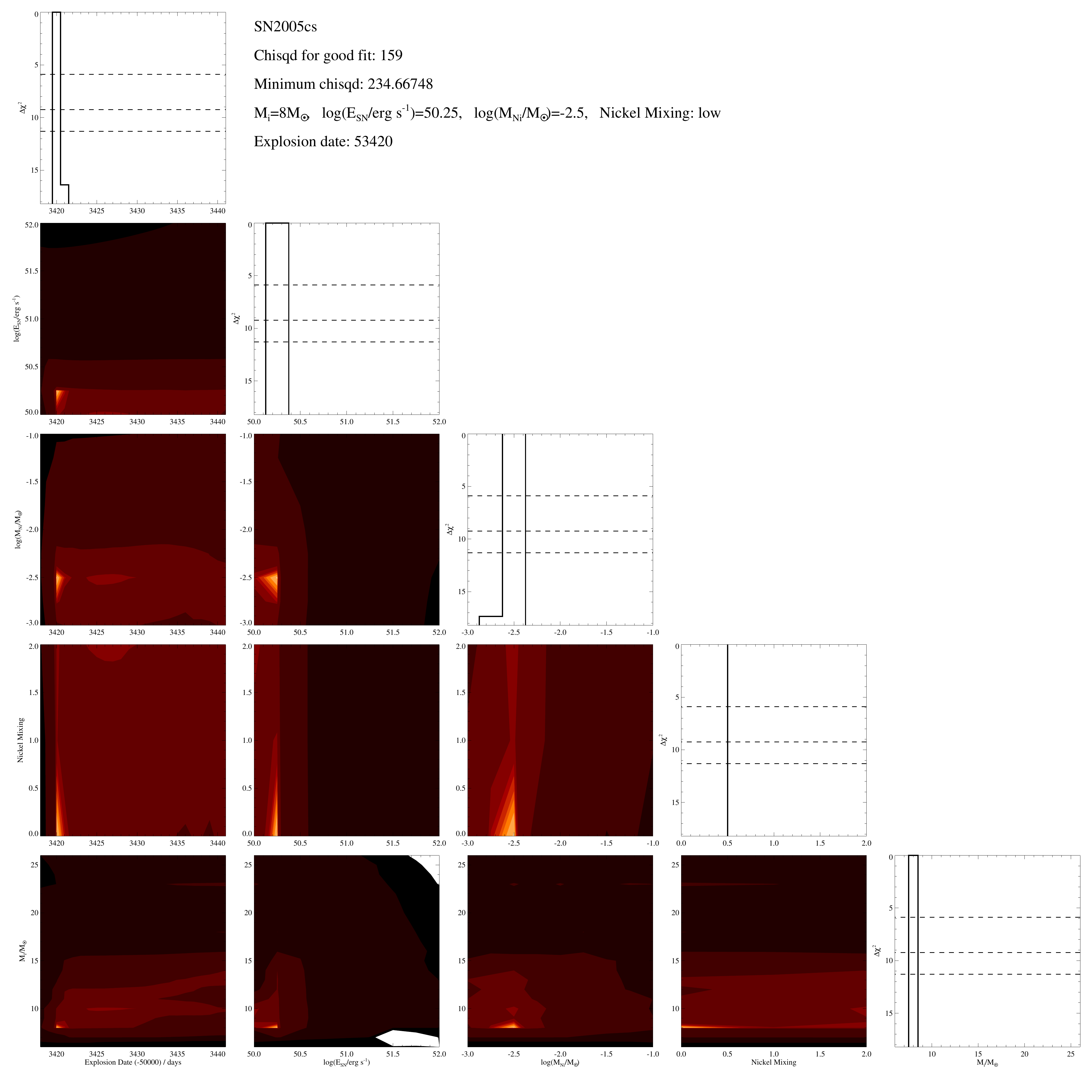}
\includegraphics[width=\columnwidth]{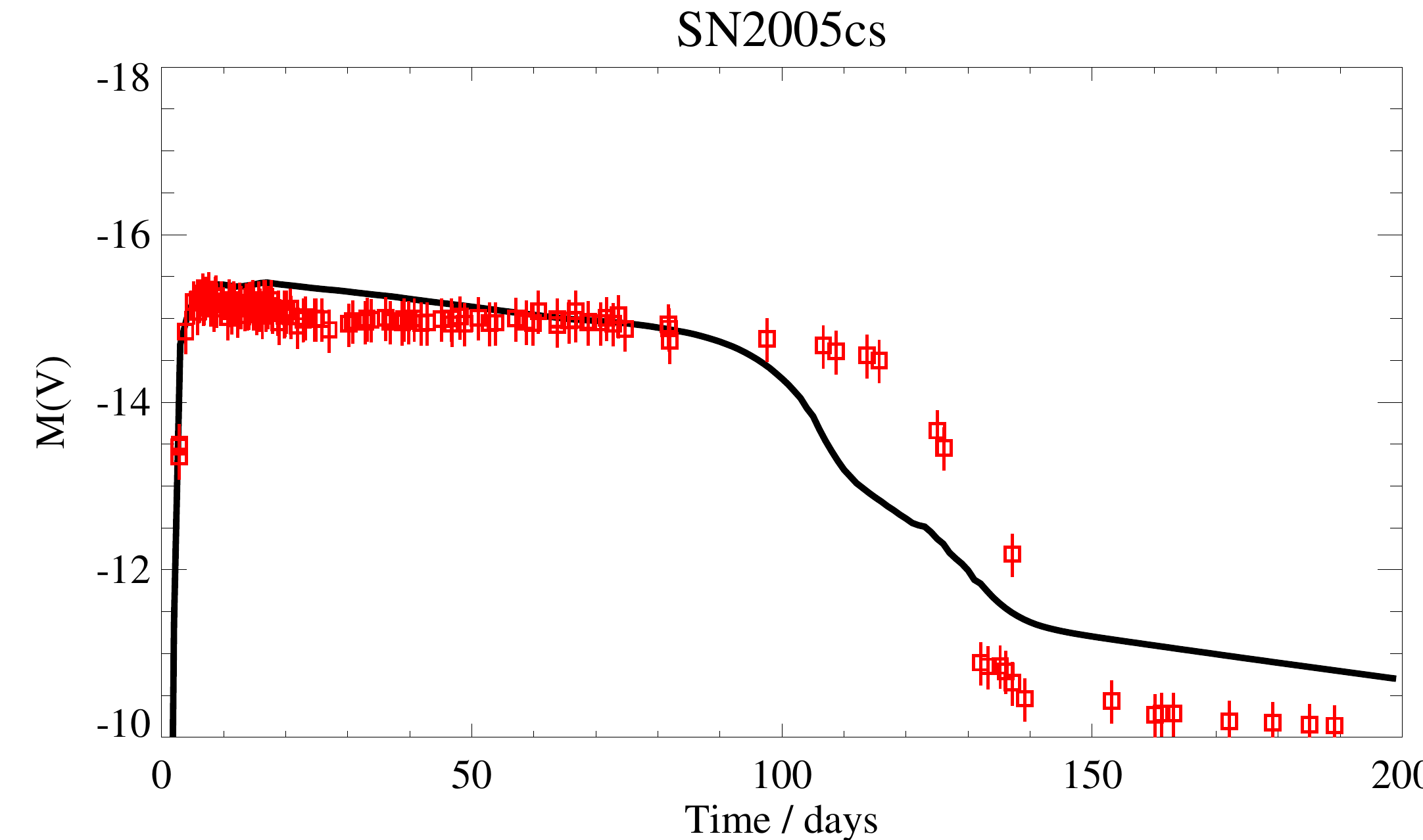}
\caption{SN2005cs Free-Fit, as in Figure \ref{fig:freefit2003gd}}
\label{fig:freefit2005cs}
\end{center}
\end{figure*}

\begin{figure*}[!h]
\begin{center}
\includegraphics[width=2\columnwidth]{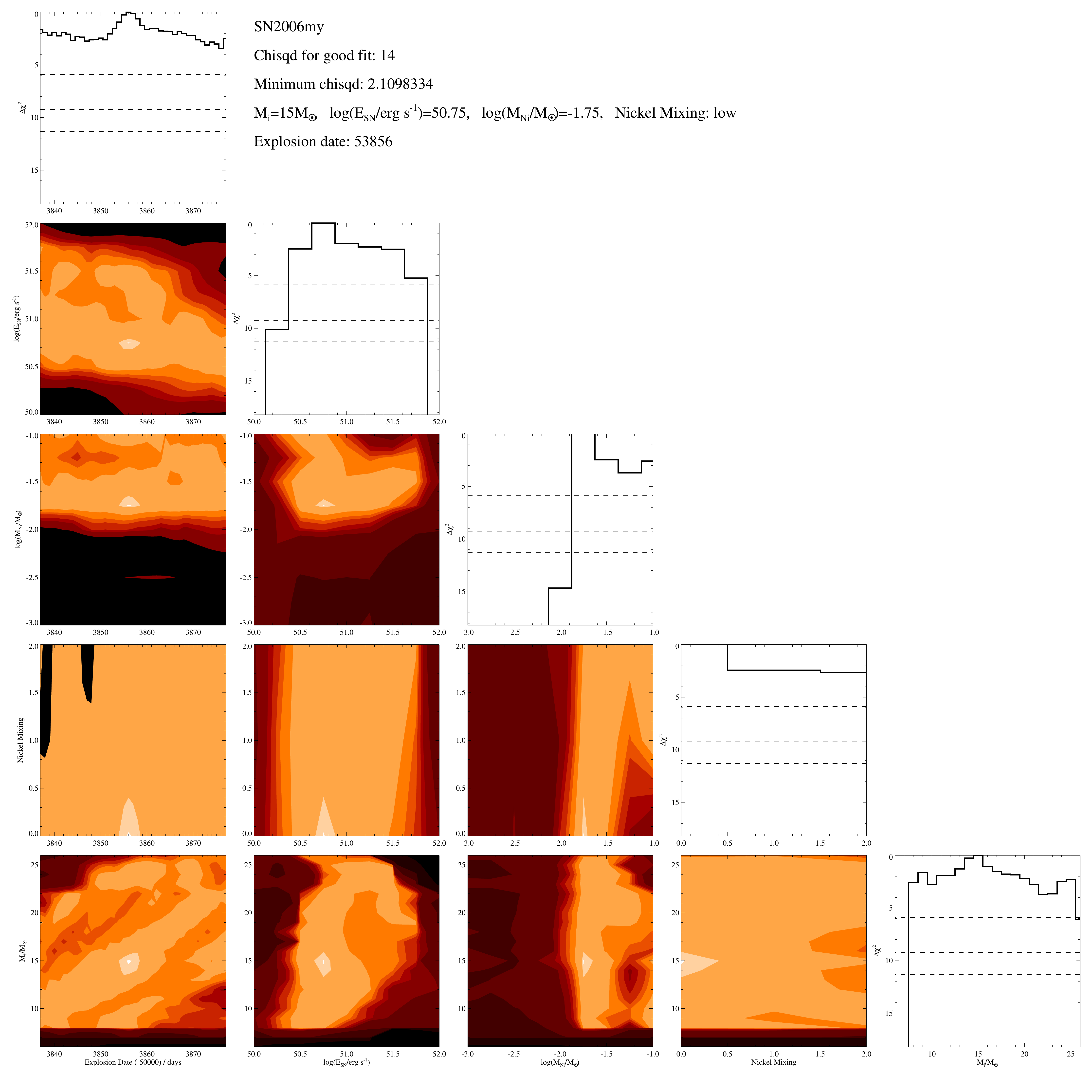}
\includegraphics[width=\columnwidth]{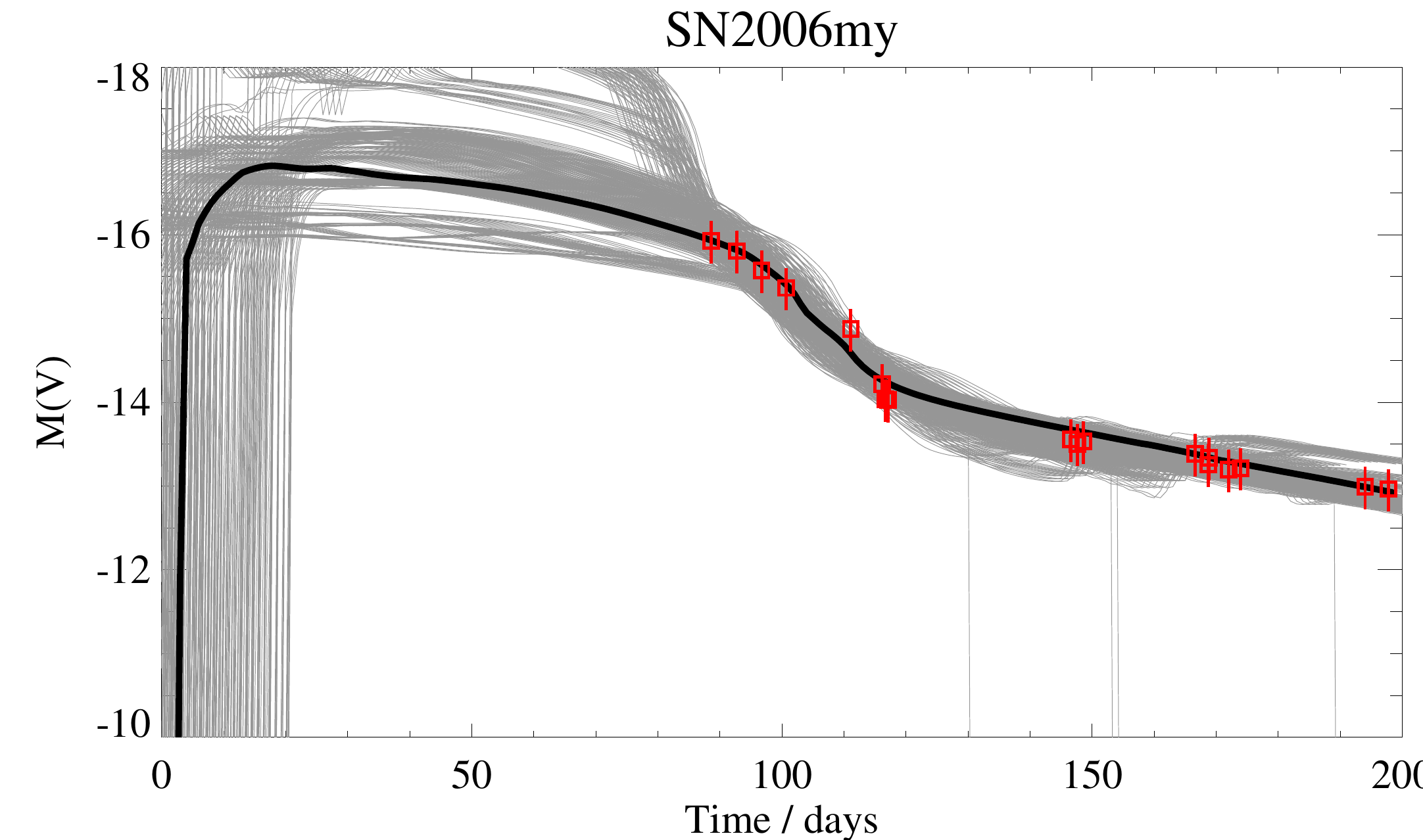}
\caption{SN2006my Free-Fit, as in Figure \ref{fig:freefit2003gd}}
\label{fig:freefit2006my}
\end{center}
\end{figure*}

\begin{figure*}[h]
\begin{center}
\includegraphics[width=2\columnwidth]{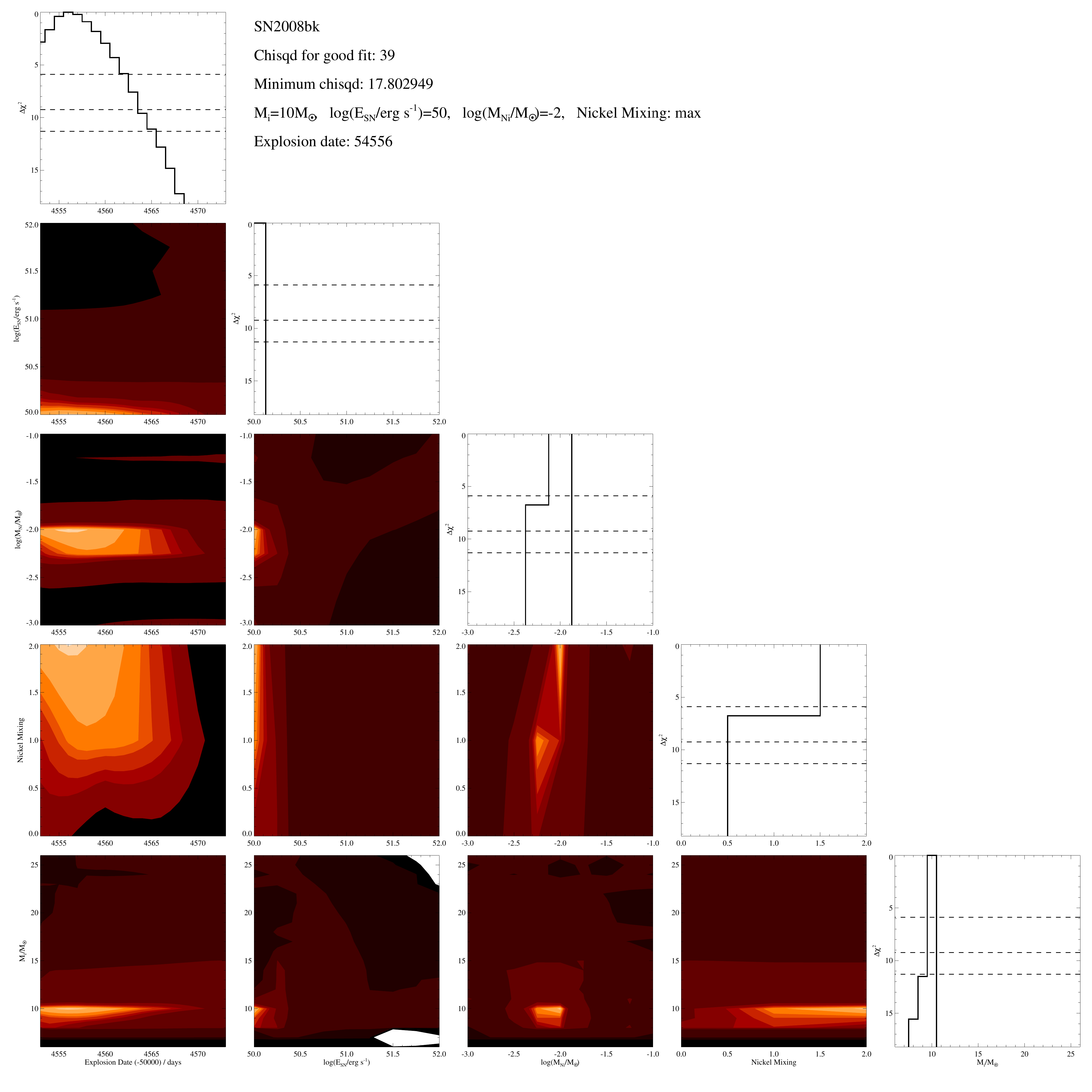}
\includegraphics[width=\columnwidth]{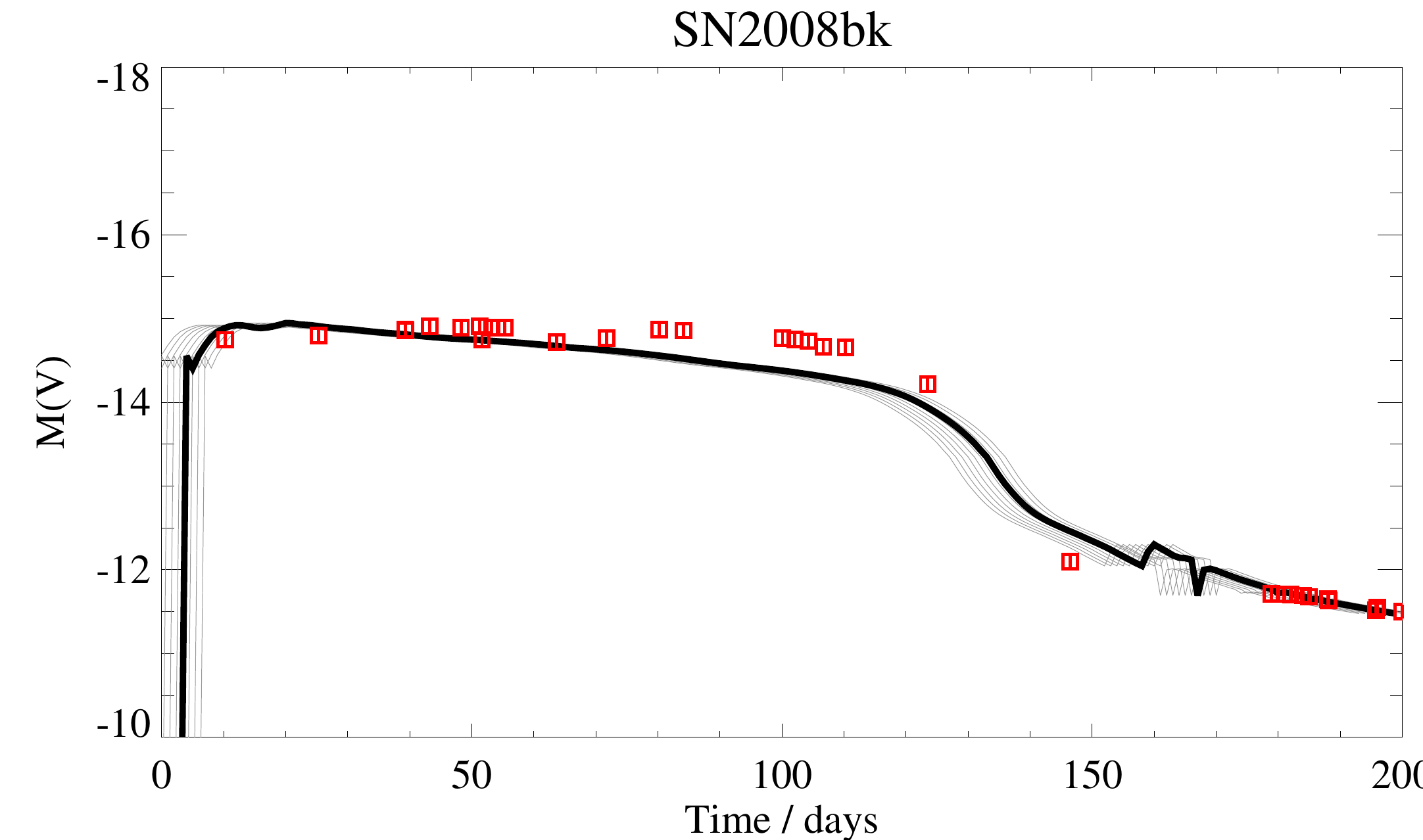}
\caption{SN2008bk Free-Fit, as in Figure \ref{fig:freefit2003gd}}
\label{fig:freefit2008bk}
\end{center}
\end{figure*}

\begin{figure*}[!h]
\begin{center}
\includegraphics[width=2\columnwidth]{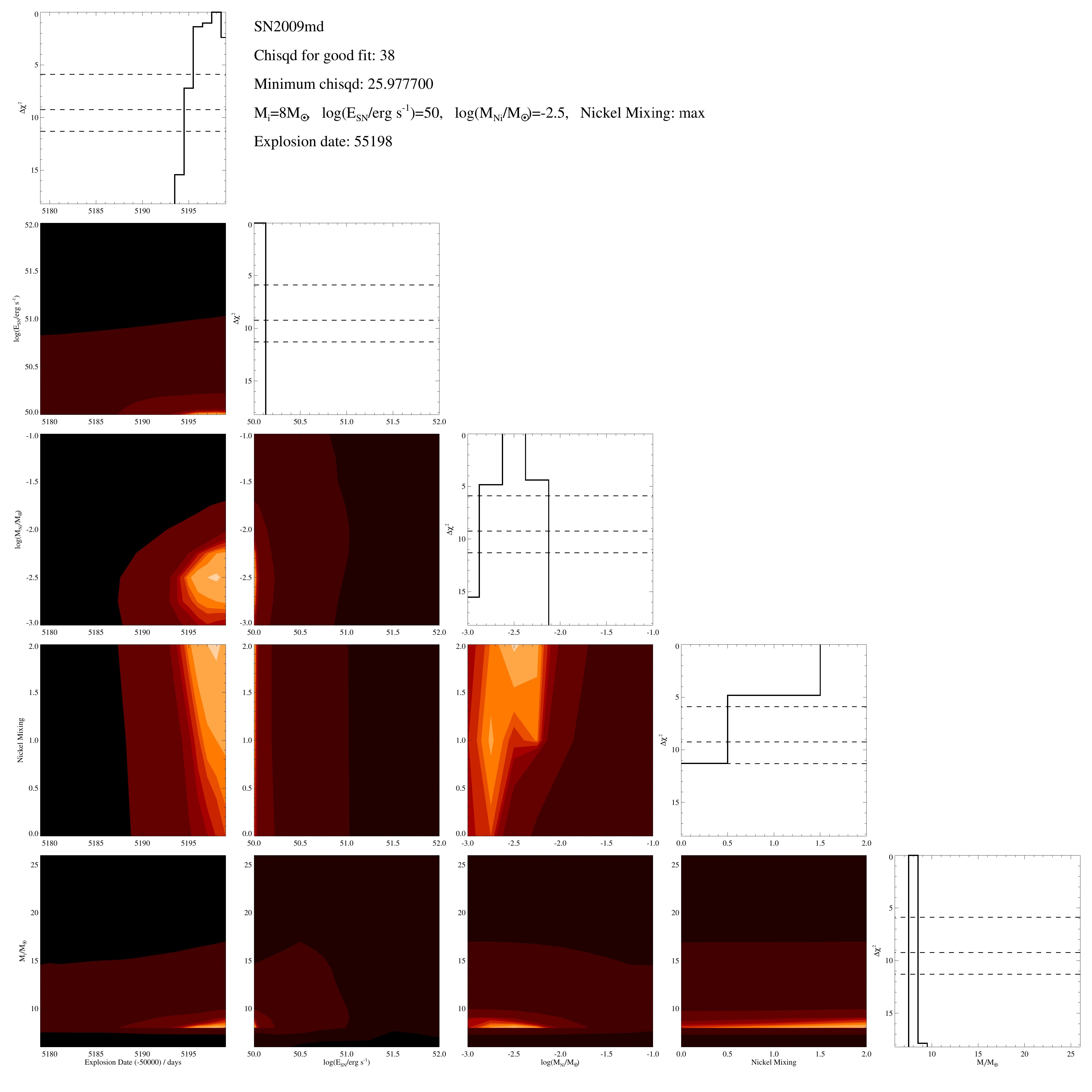}
\includegraphics[width=\columnwidth]{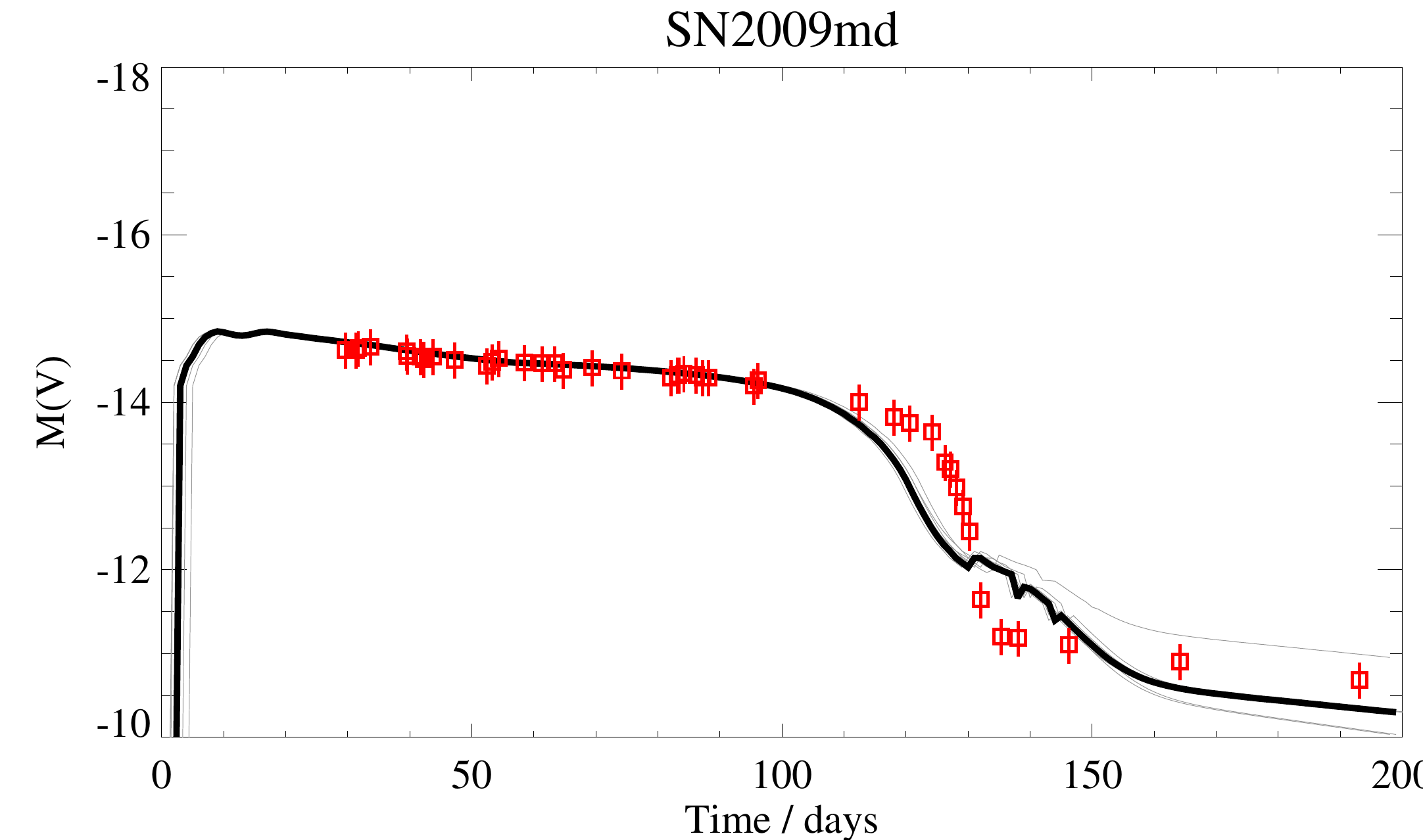}
\caption{SN2009md Free-Fit, as in Figure \ref{fig:freefit2003gd}}
\label{fig:freefit2009md}
\end{center}
\end{figure*}

\begin{figure*}[!h]
\begin{center}
\includegraphics[width=2\columnwidth]{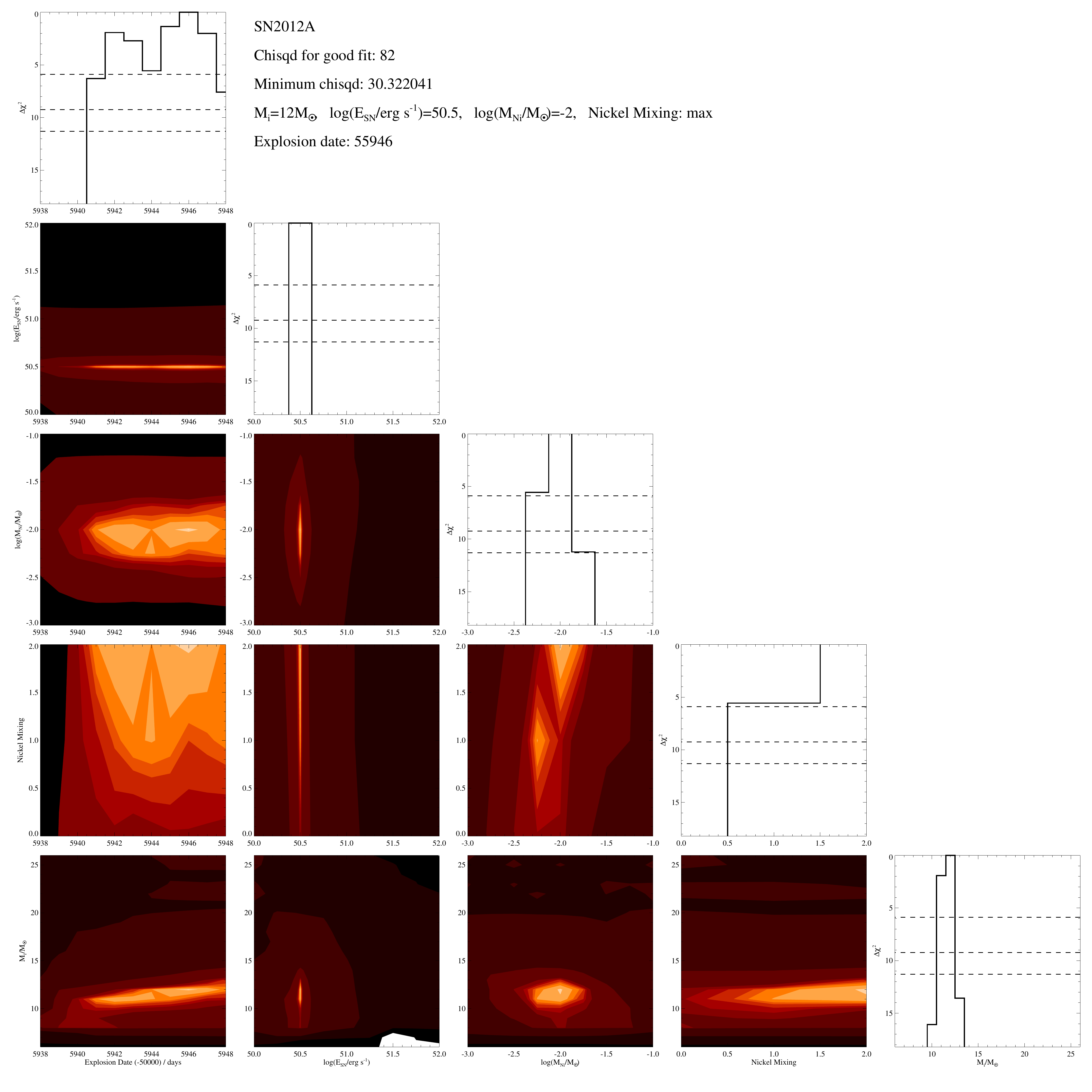}
\includegraphics[width=\columnwidth]{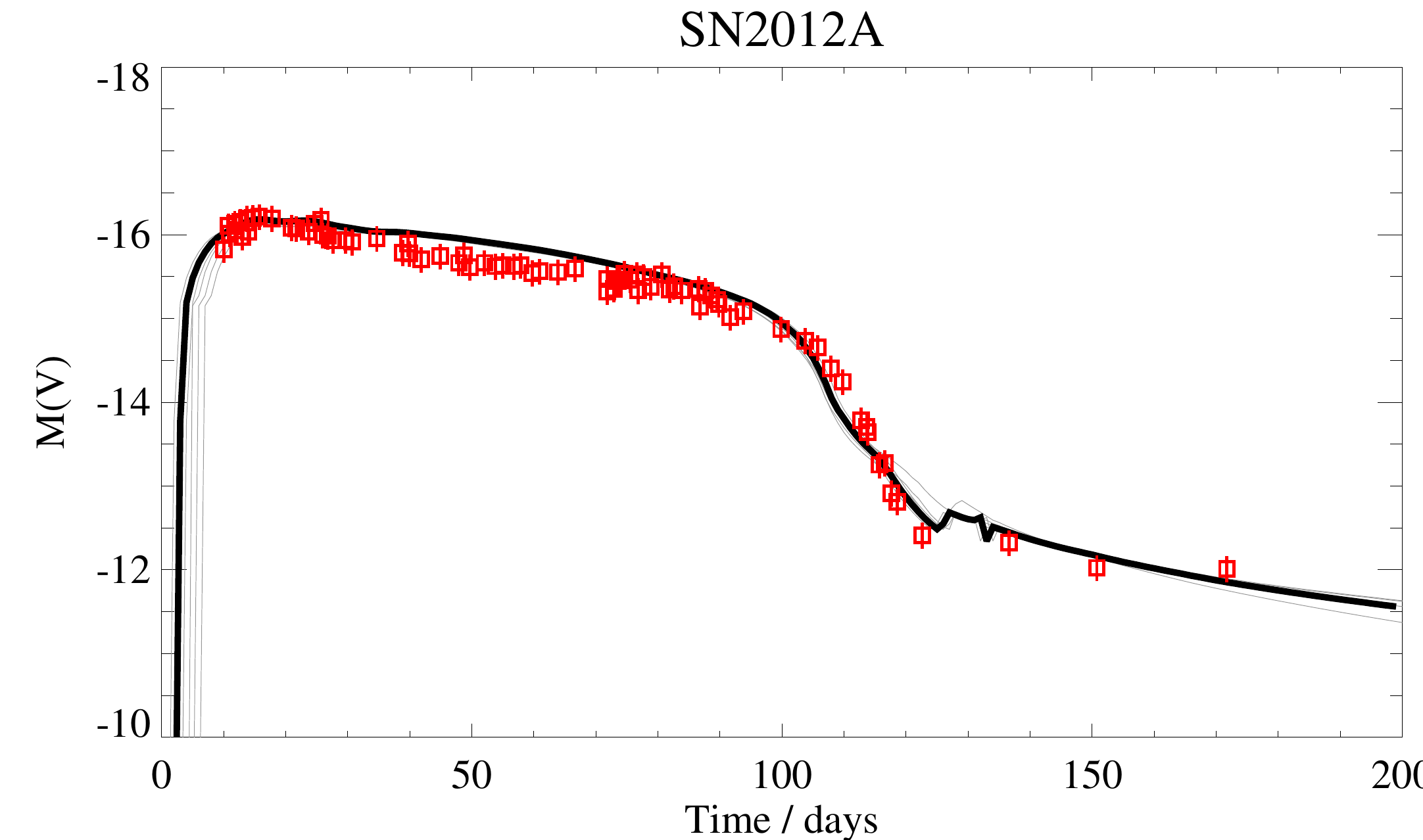}
\caption{SN2012A Free-Fit, as in Figure \ref{fig:freefit2003gd}}
\label{fig:freefit2012A}
\end{center}
\end{figure*}

\begin{figure*}[h]
\begin{center}
\includegraphics[width=2\columnwidth]{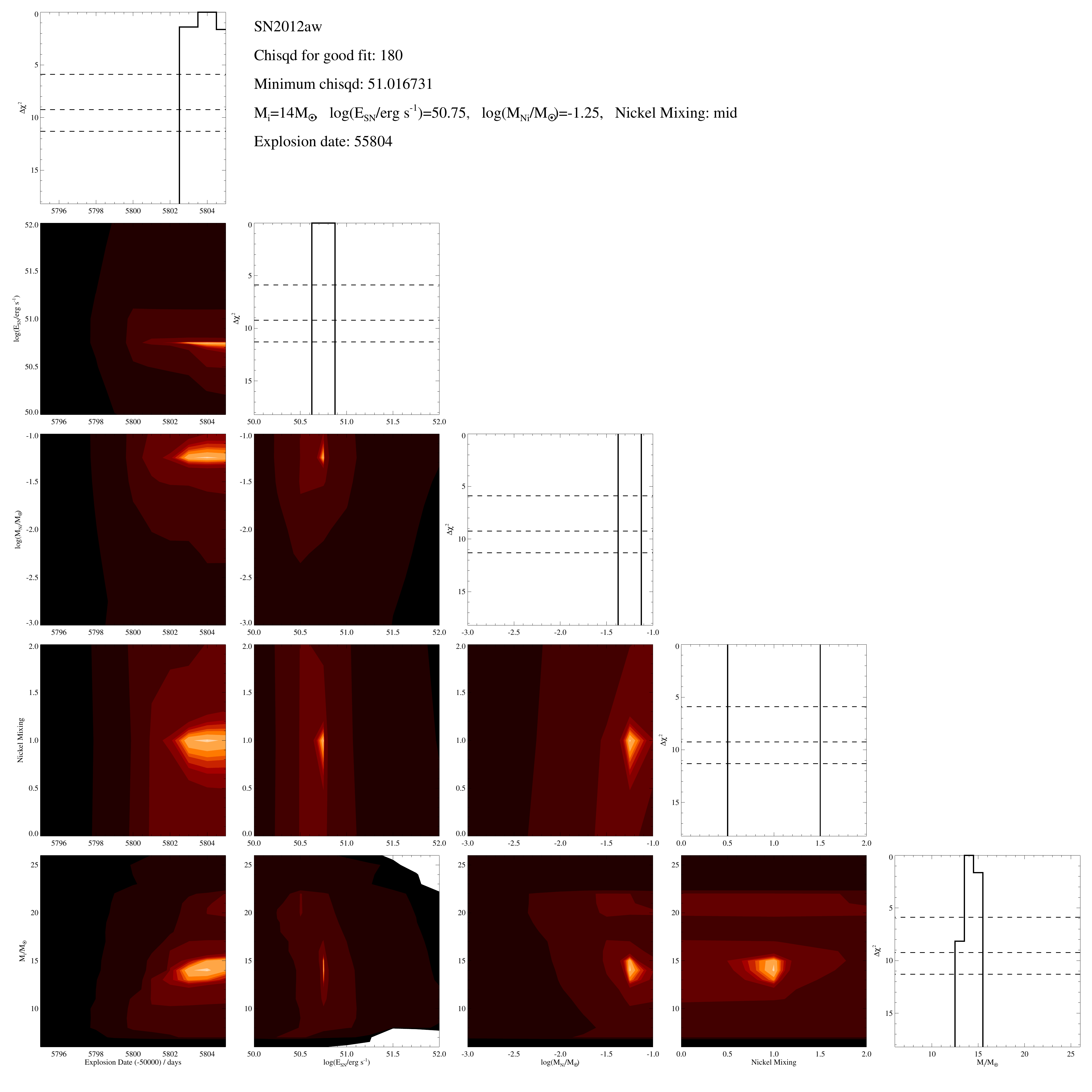}
\includegraphics[width=\columnwidth]{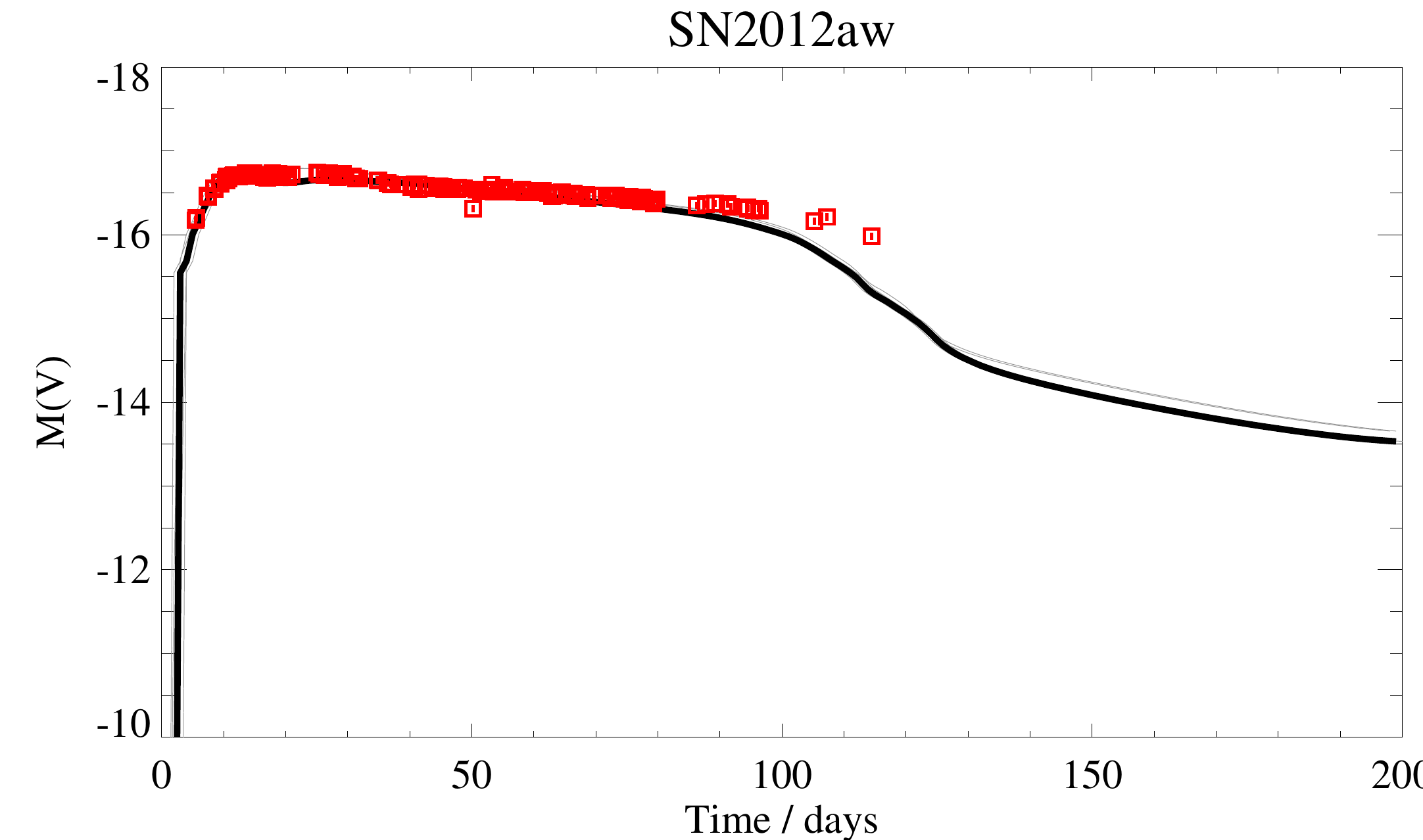}
\caption{SN2012aw Free-Fit, as in Figure \ref{fig:freefit2003gd}}
\label{fig:freefit2012aw}
\end{center}
\end{figure*}

\begin{figure*}[!h]
\begin{center}
\includegraphics[width=2\columnwidth]{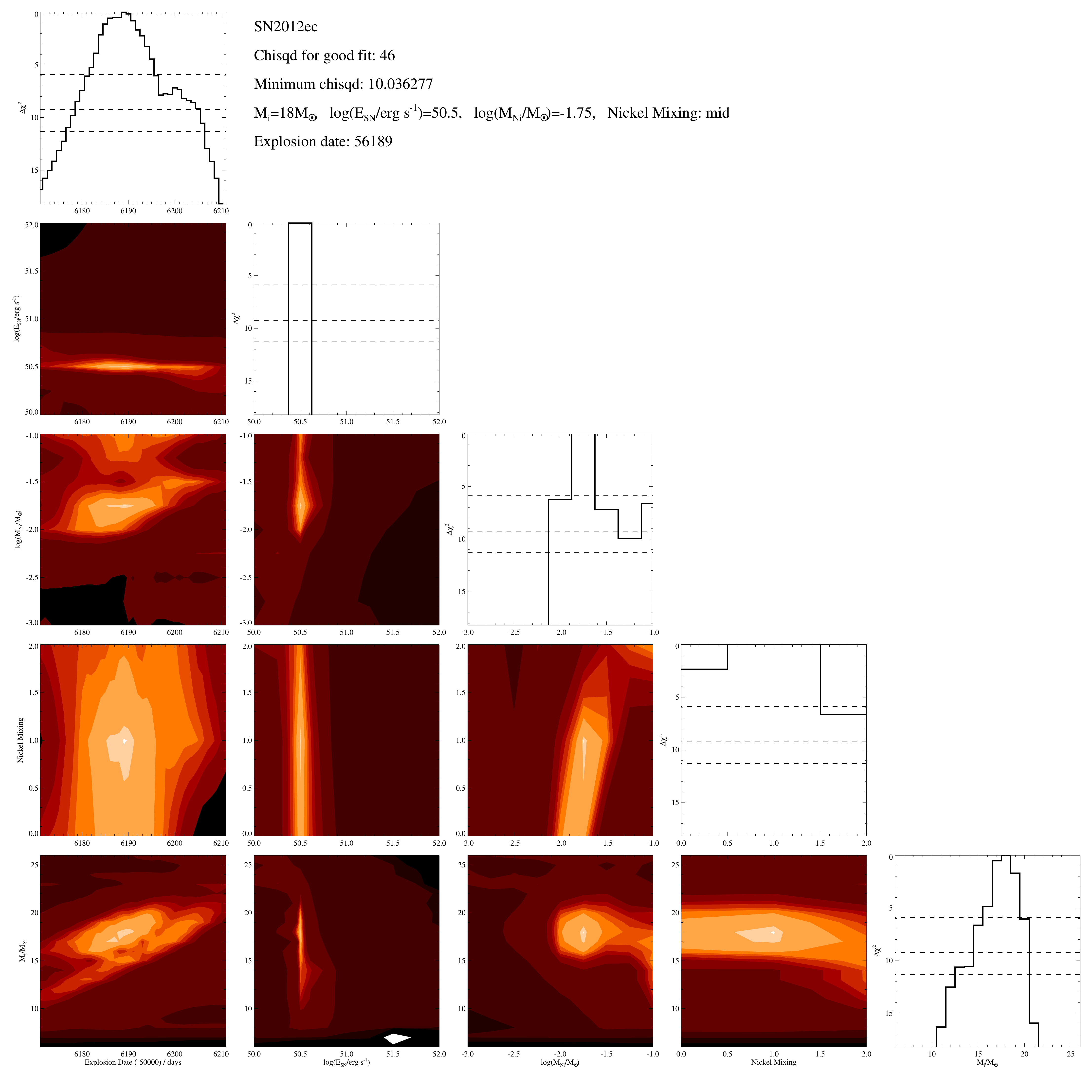}
\includegraphics[width=\columnwidth]{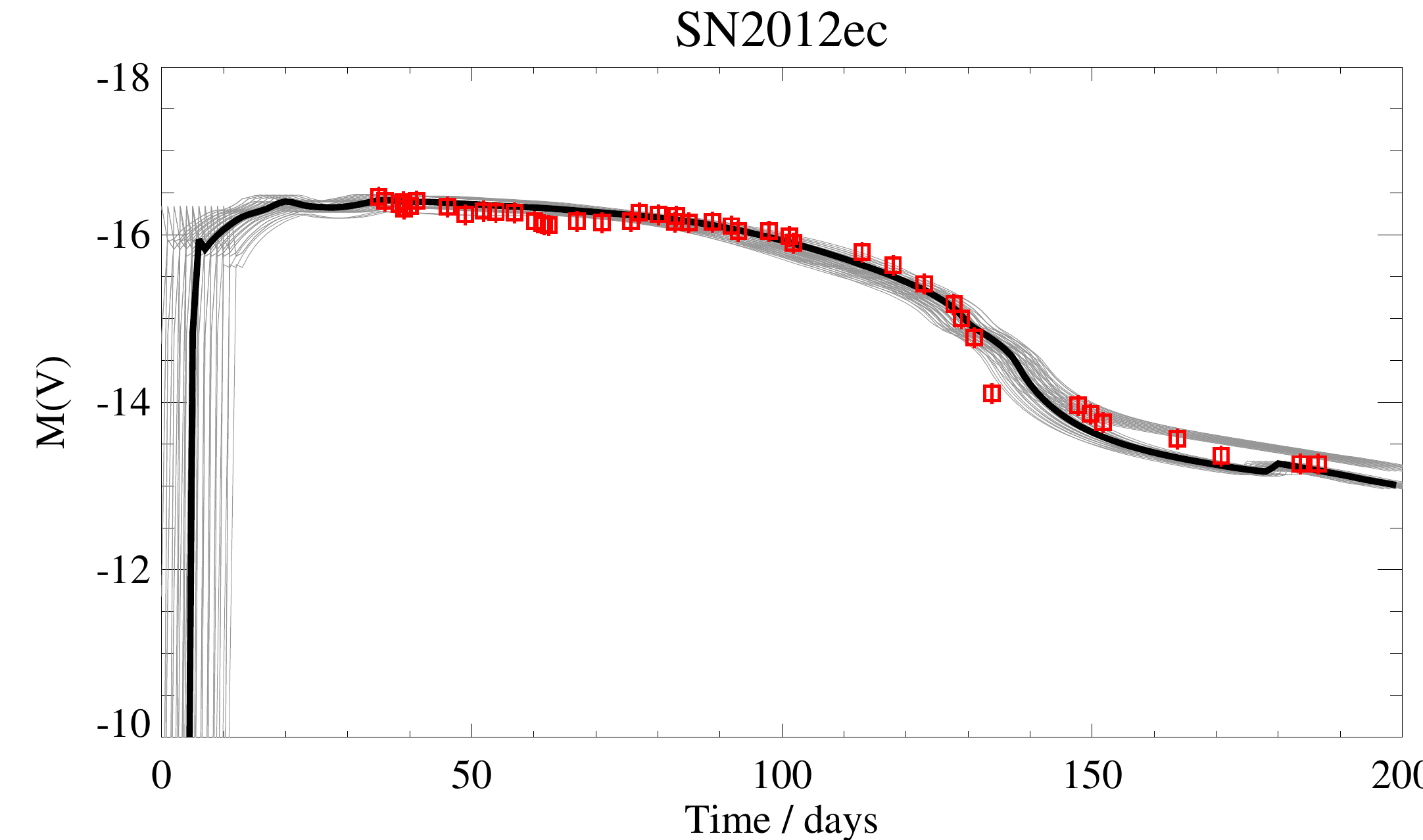}
\caption{SN2012ec Free-Fit, as in Figure \ref{fig:freefit2003gd}}
\label{fig:freefit2012ec}
\end{center}
\end{figure*}

\begin{figure*}[!h]
\begin{center}
\includegraphics[width=2\columnwidth]{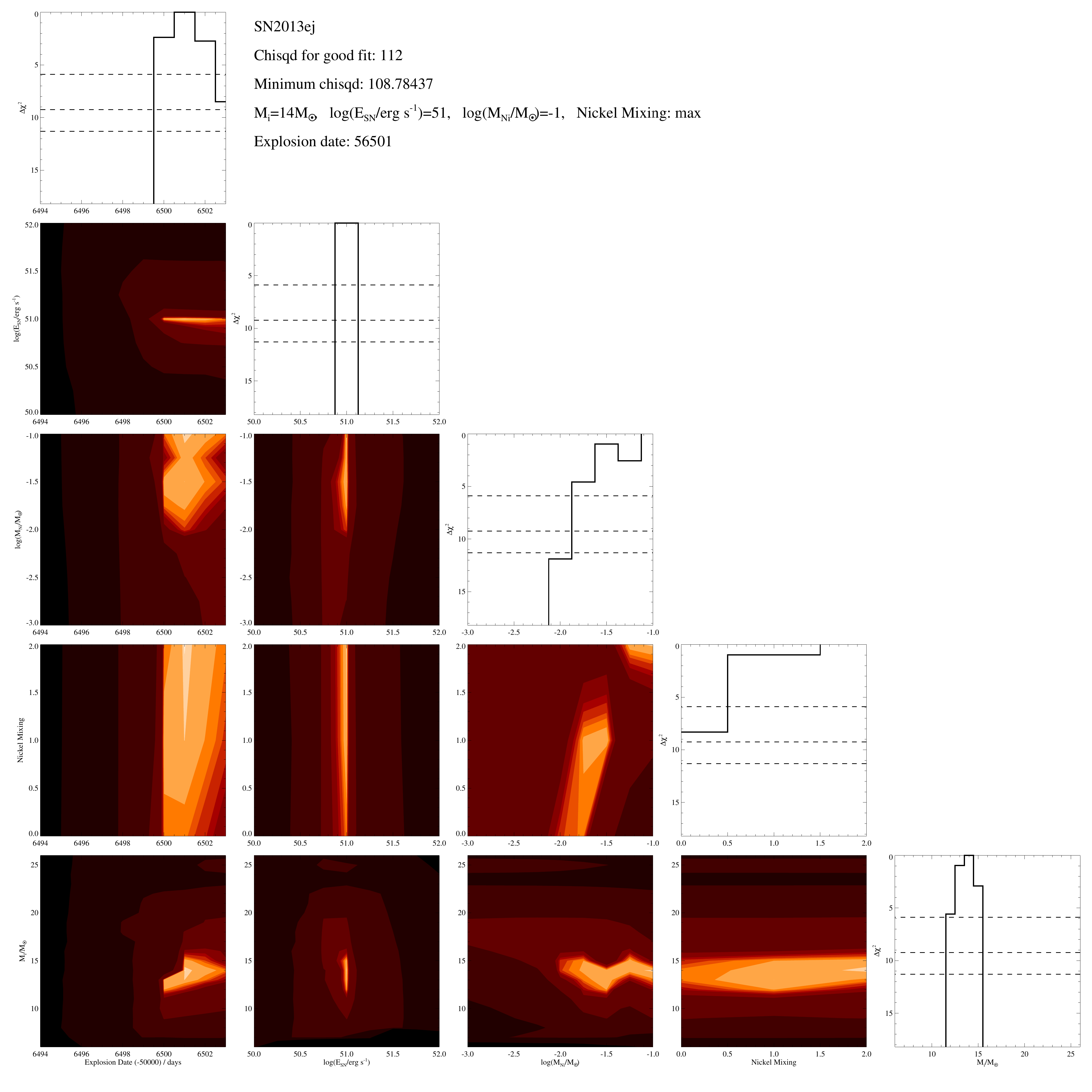}
\includegraphics[width=\columnwidth]{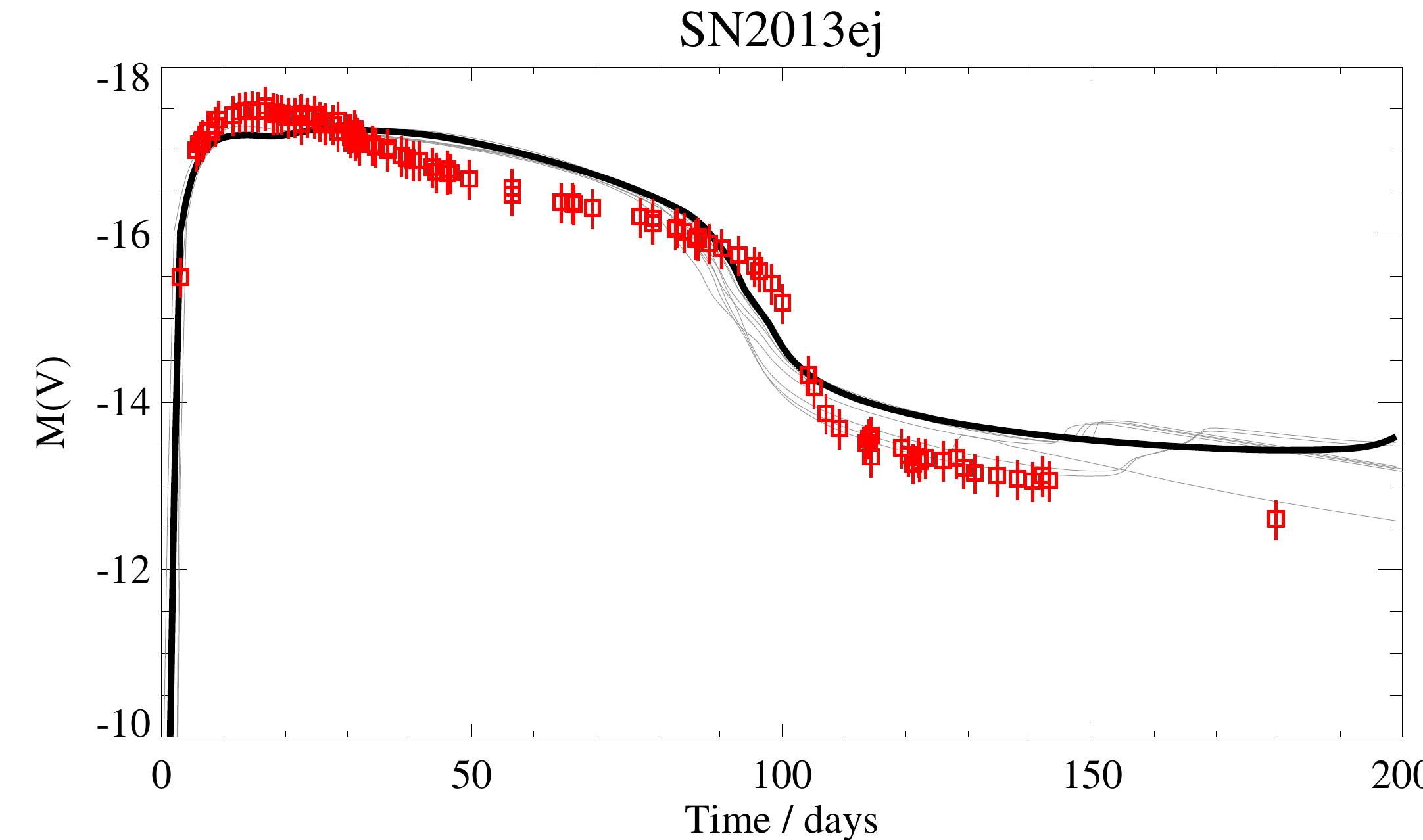}
\caption{SN2013ej Free-Fit, as in Figure \ref{fig:freefit2003gd}}
\label{fig:freefit2013ej}
\end{center}
\end{figure*}
\raggedbottom

\clearpage

\section{Results from a progenitor constrained fit of the light curves}
Here we present the best fitting V-band magnitude lightcurves for the supernovae when a constrained fit across the 68\% uncertainty range of modelled initial masses allowed by the masses derived by Smartt et al. (2009). We also include corner plots showing how the $\chi^2$ varies with the 5 parameters we fit. We also include plots showing how $\chi^2$ only depends on one parameter at a time as shown by the solid black line while the horizontal dashed lines show the $\Delta \chi^2$ for the 1$\sigma$, 2$\sigma$ and 3$\sigma$. In the bottom figure of the lightcurves we plot the best fitting model (black line) along with the lightcurves at are within the 1$\sigma$ uncertainty in grey while the observations are shown in red. 
\begin{figure*}[ht]
\begin{center}
\includegraphics[width=2\columnwidth]{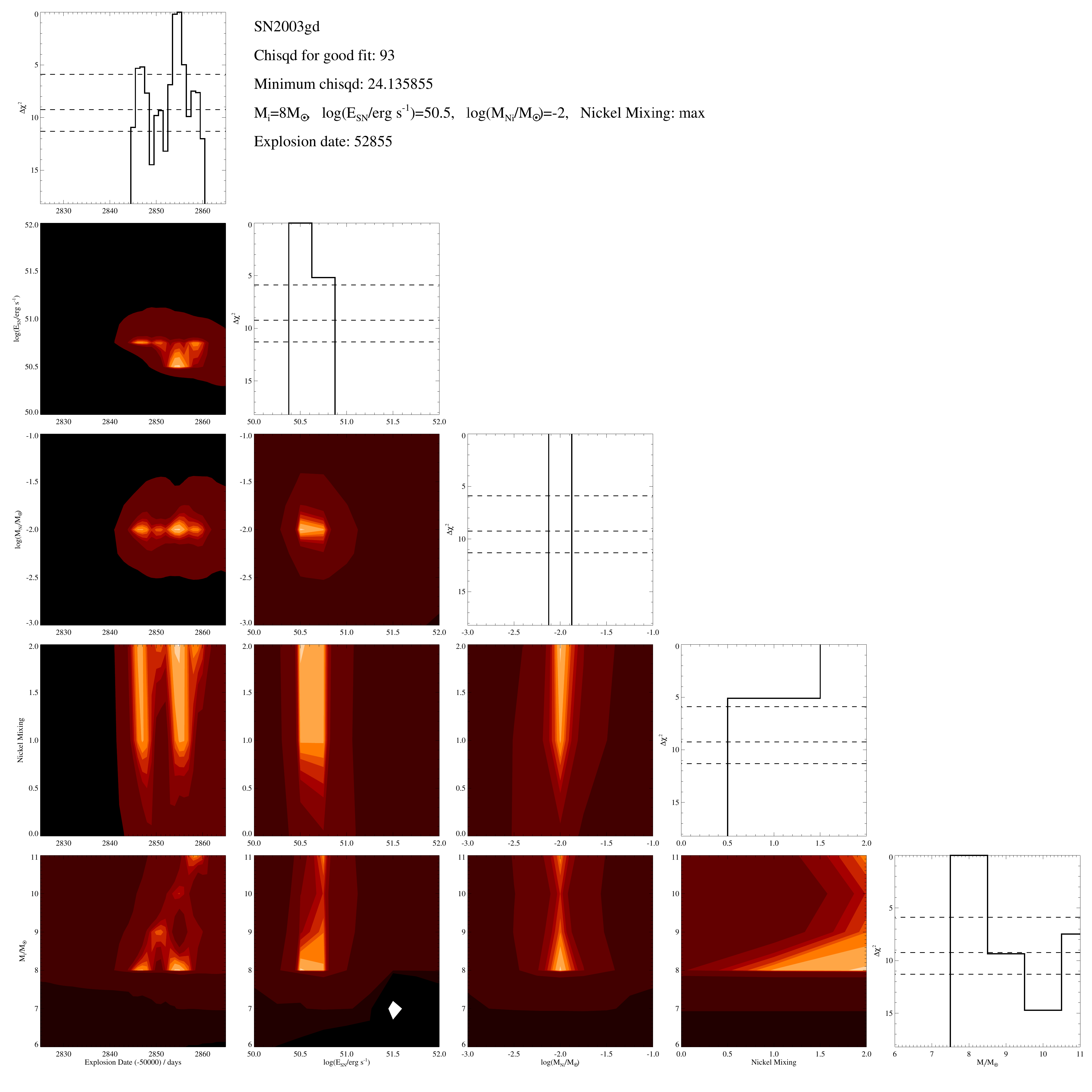}
\includegraphics[width=\columnwidth]{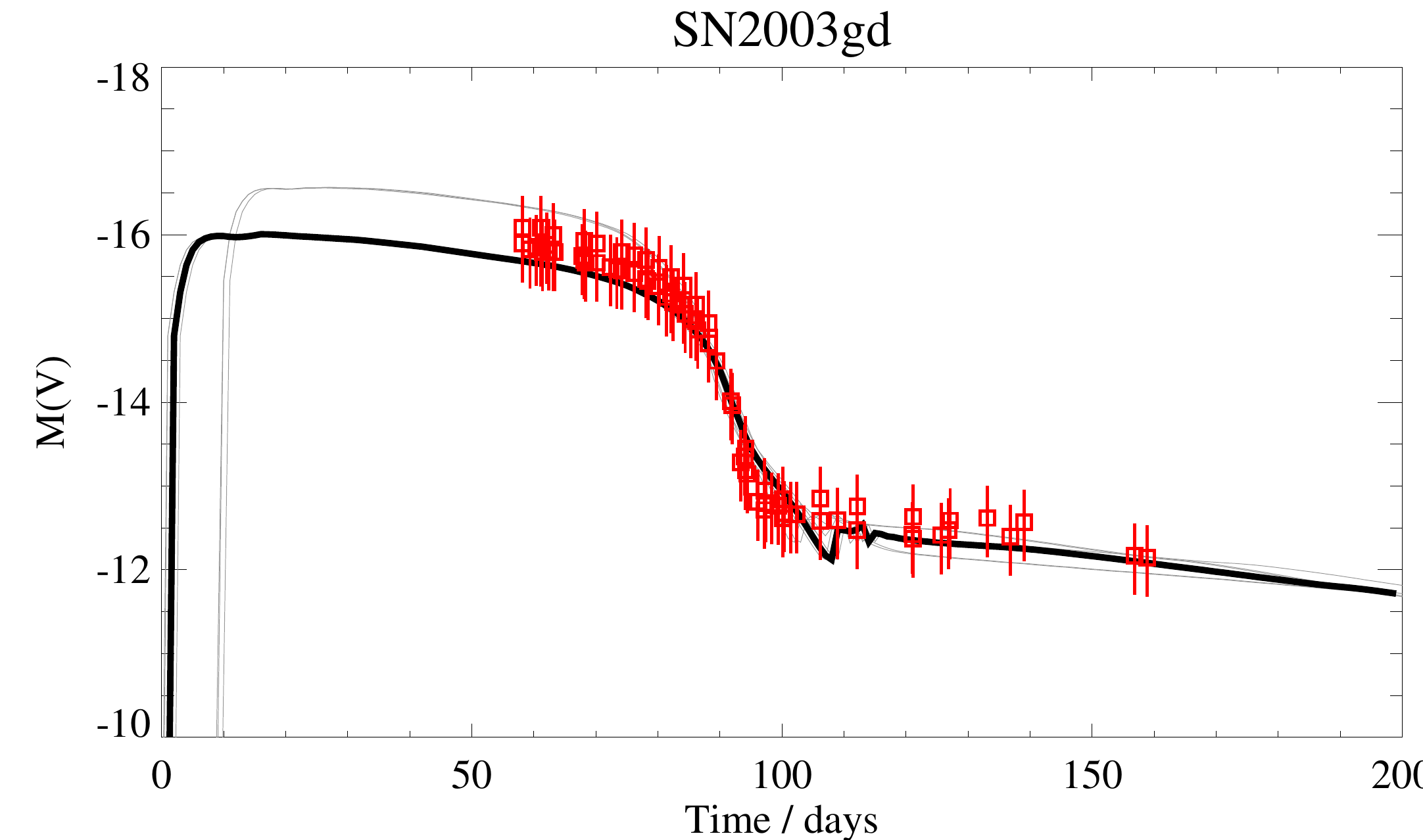}
\caption{SN2003gd Constrained, corner plots showing how $\chi^2$ varies over the 5 parameters as well a plot comparing the observed lightcurves to the matching theoretical models.}
\label{fig:constrained2003gd}
\end{center}
\end{figure*}

\begin{figure*}[!ht]
\begin{center}
\includegraphics[width=2\columnwidth]{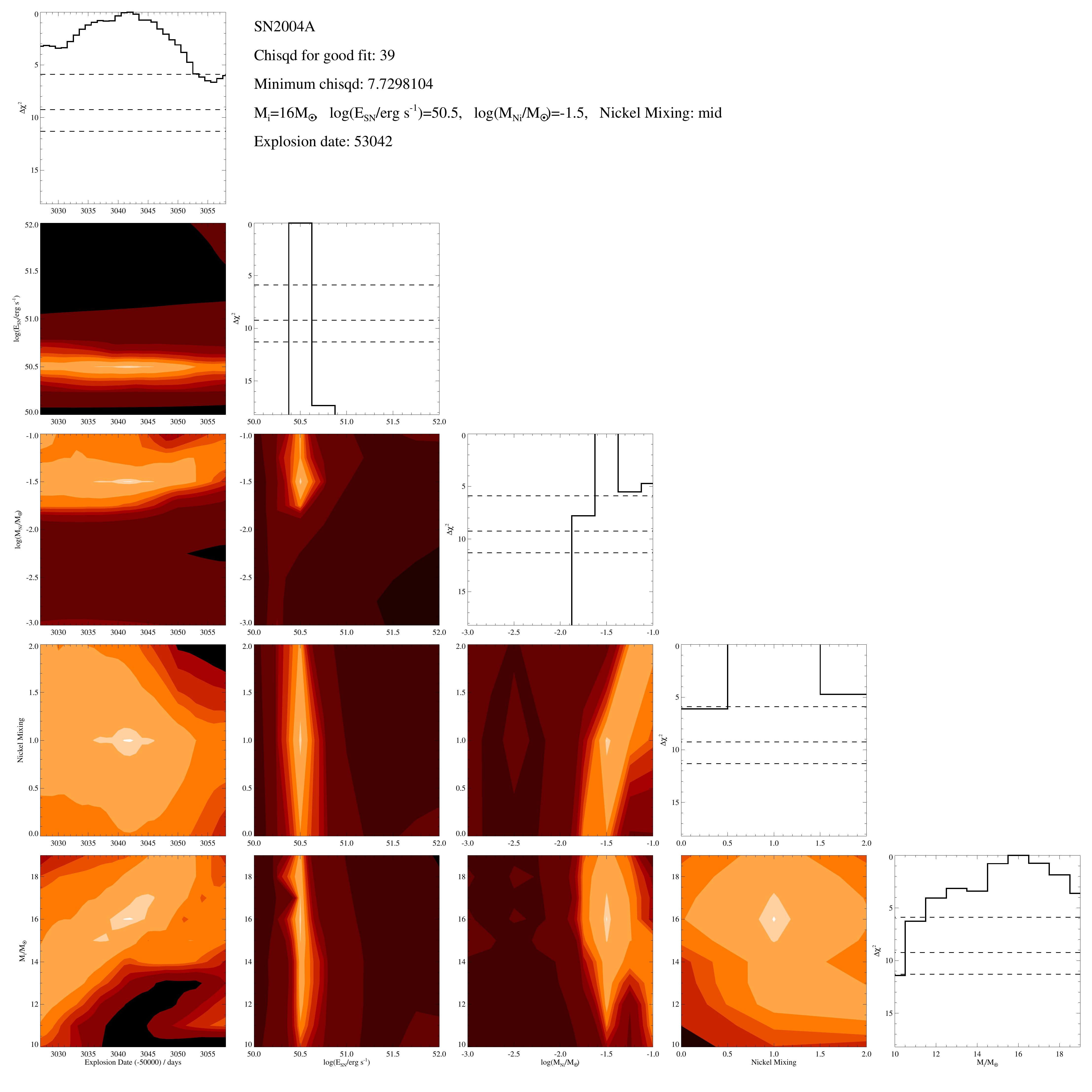}
\includegraphics[width=\columnwidth]{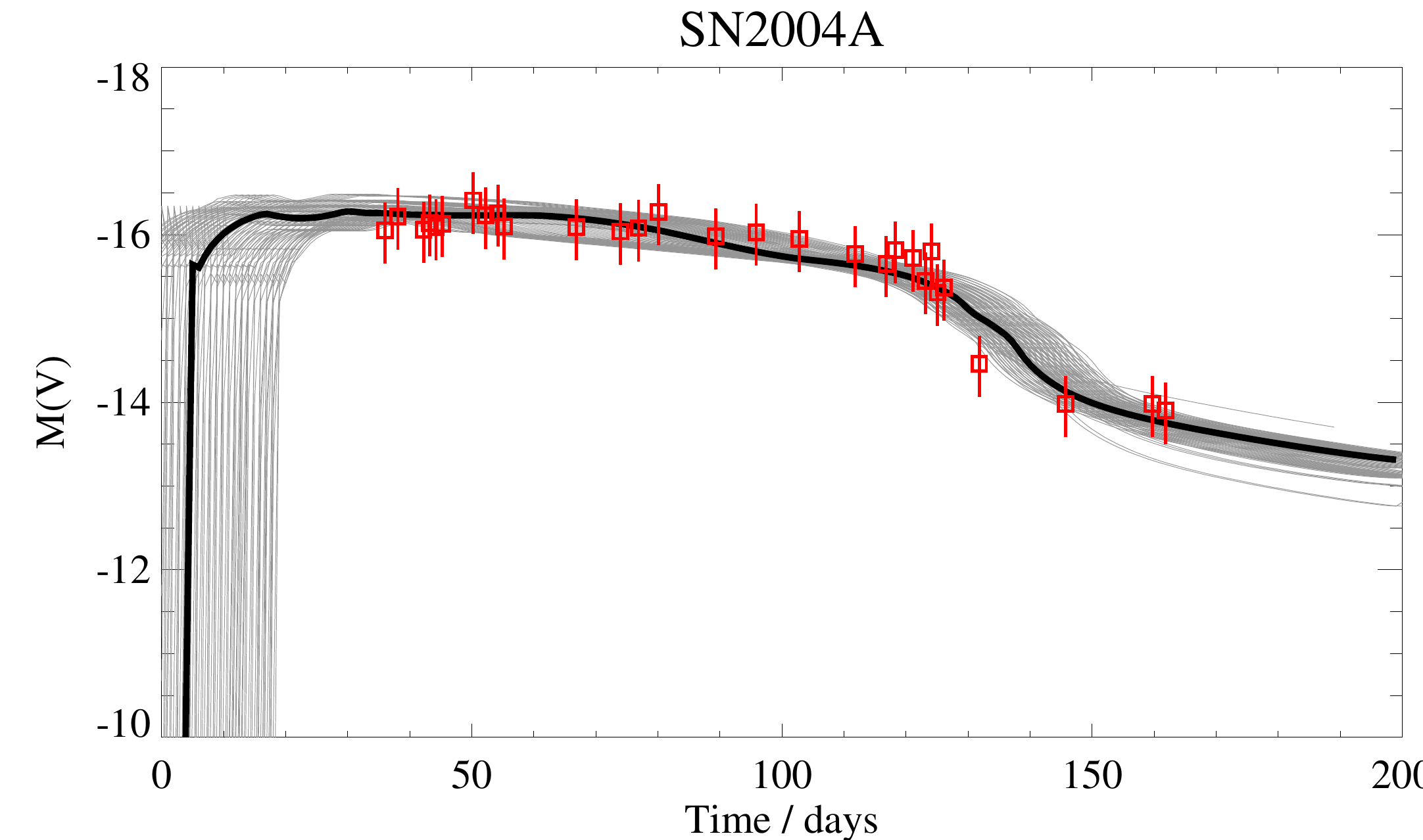}
\caption{SN2004A Constrained, as in Figure \ref{fig:constrained2003gd}}
\label{fig:constrained2004A}
\end{center}
\end{figure*}

\begin{figure*}[h]
\begin{center}
\includegraphics[width=2\columnwidth]{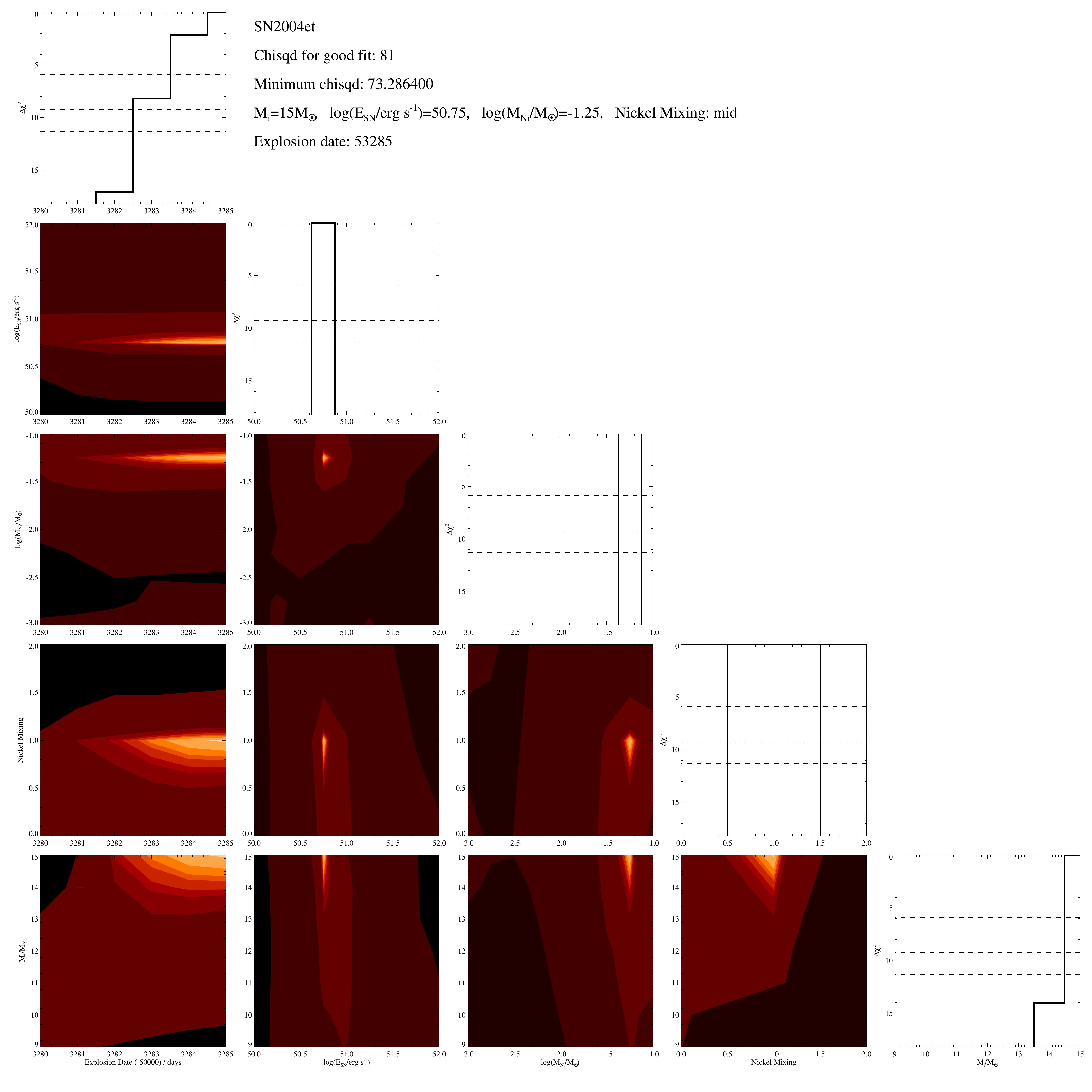}
\includegraphics[width=\columnwidth]{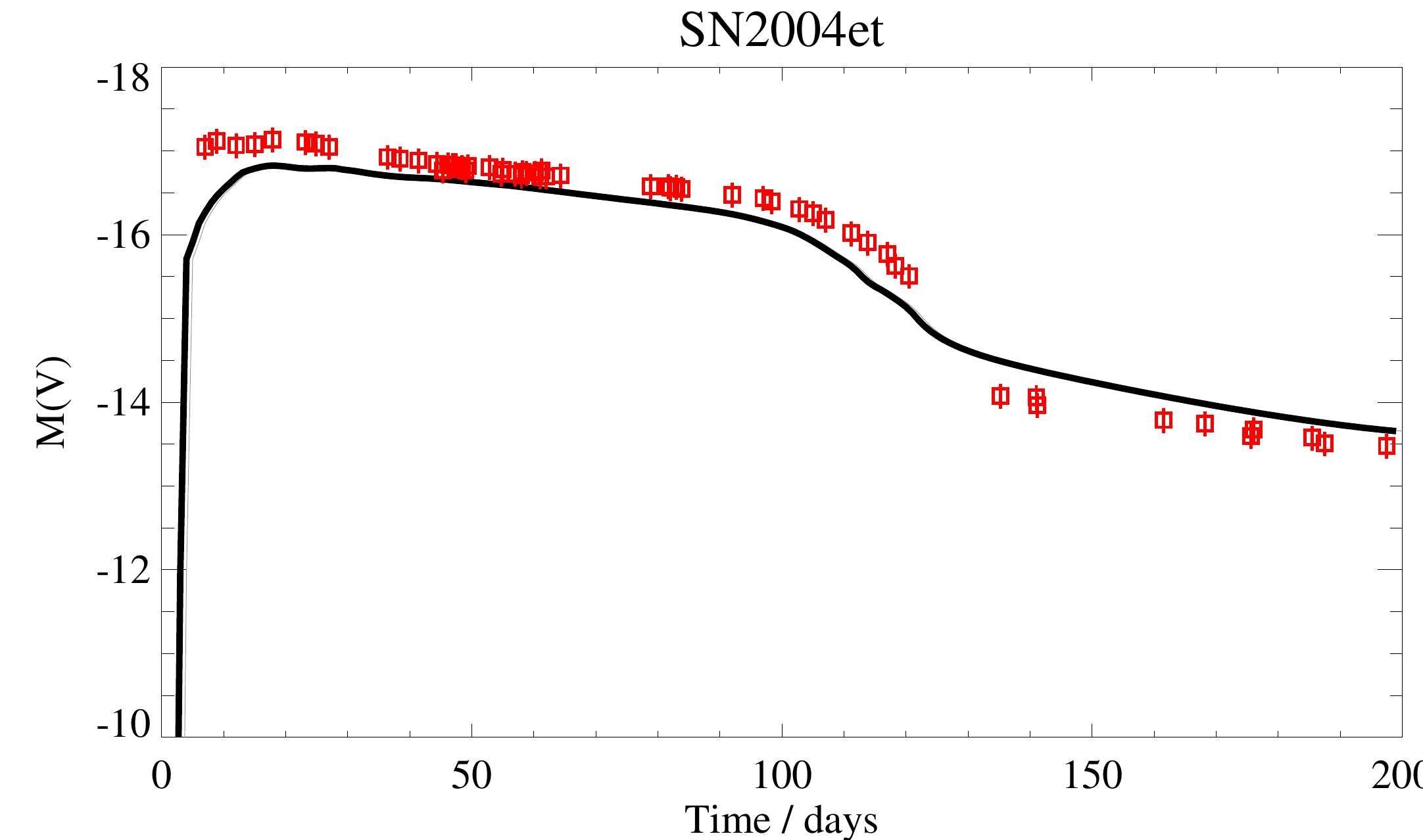}
\caption{SN2004et Constrained, as in Figure \ref{fig:constrained2003gd}}
\label{fig:constrained2004et}
\end{center}
\end{figure*}

\begin{figure*}[!h]
\begin{center}
\includegraphics[width=2\columnwidth]{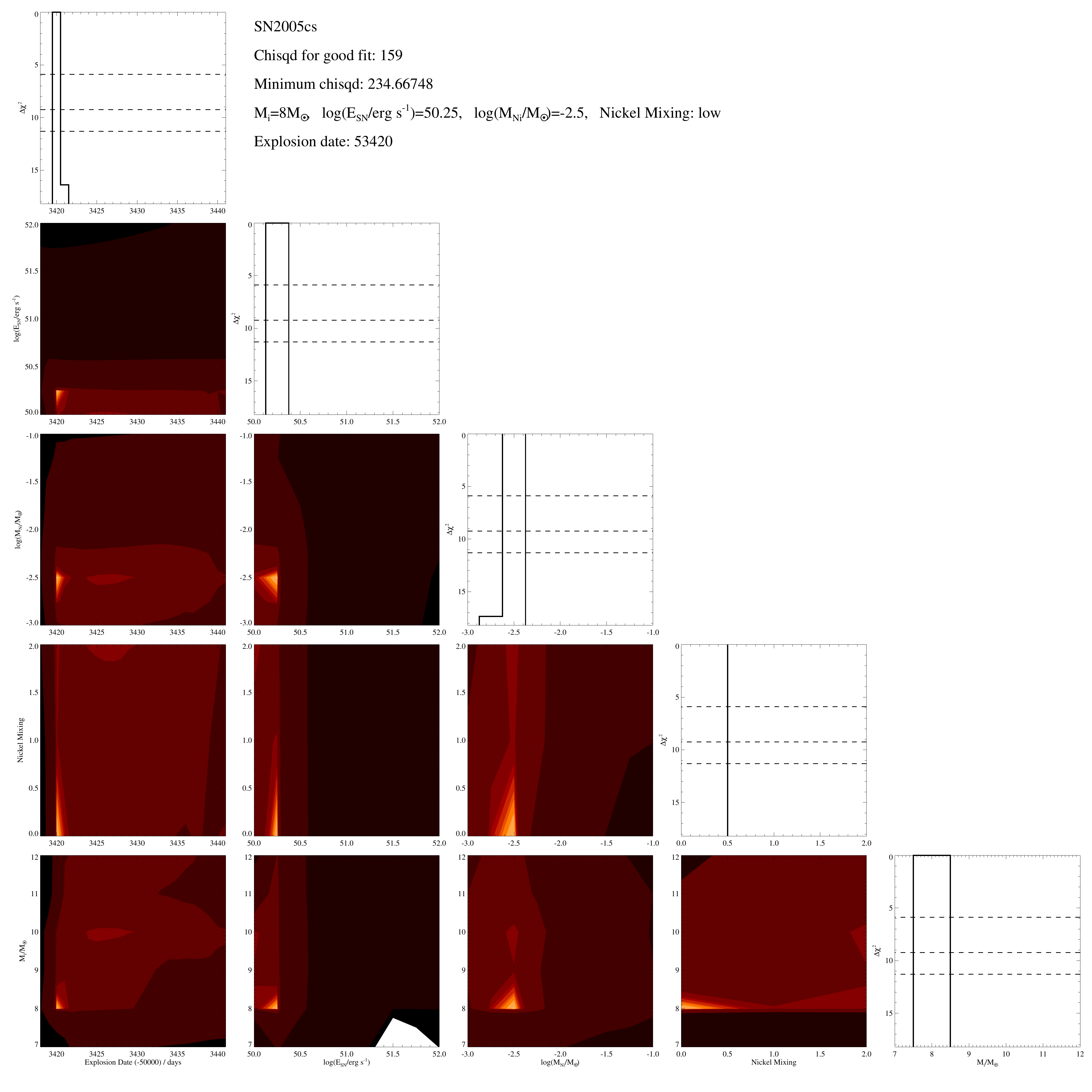}
\includegraphics[width=\columnwidth]{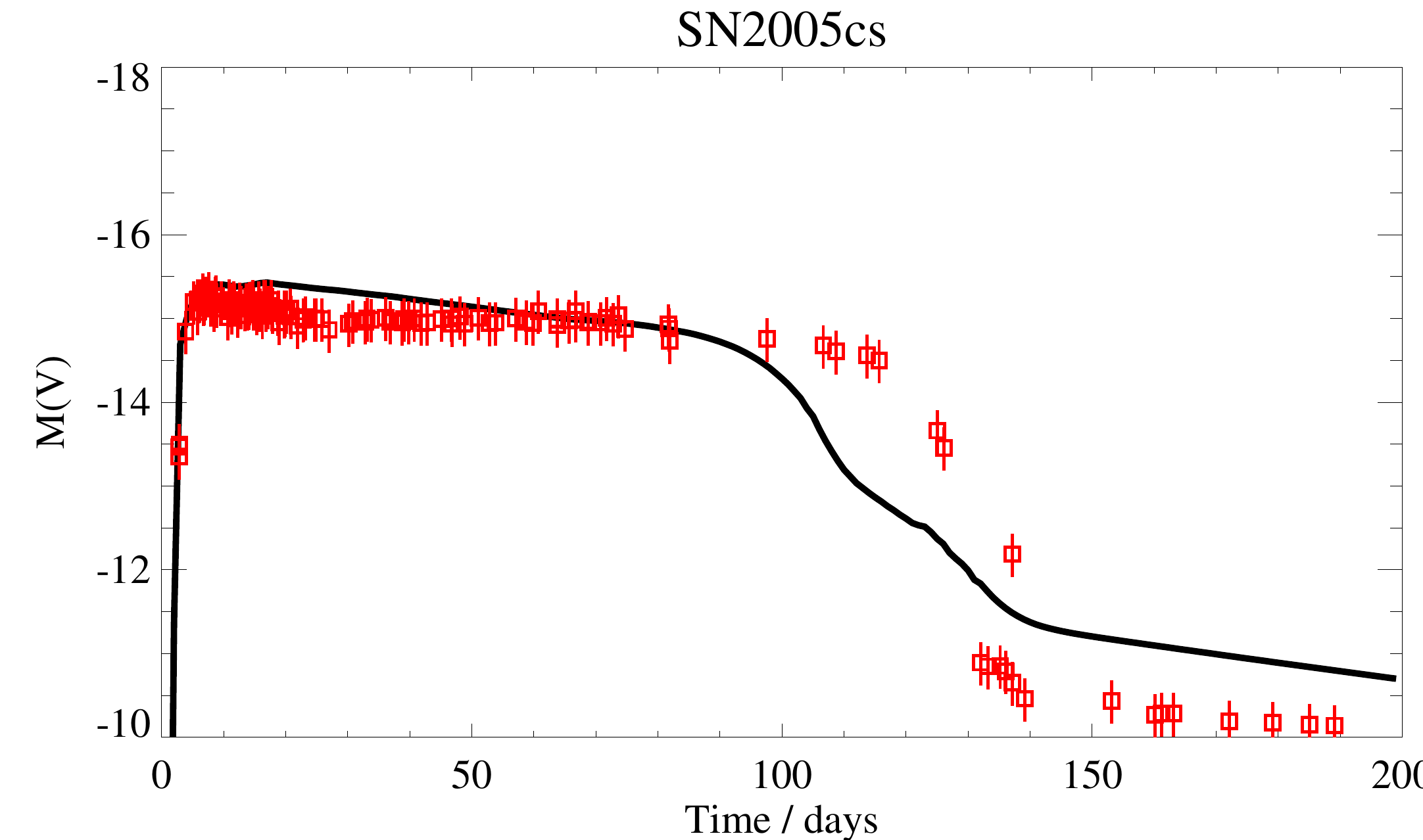}
\caption{SN2005cs Constrained, as in Figure \ref{fig:constrained2003gd}}
\label{fig:constrained2005cs}
\end{center}
\end{figure*}

\begin{figure*}[!h]
\begin{center}
\includegraphics[width=2\columnwidth]{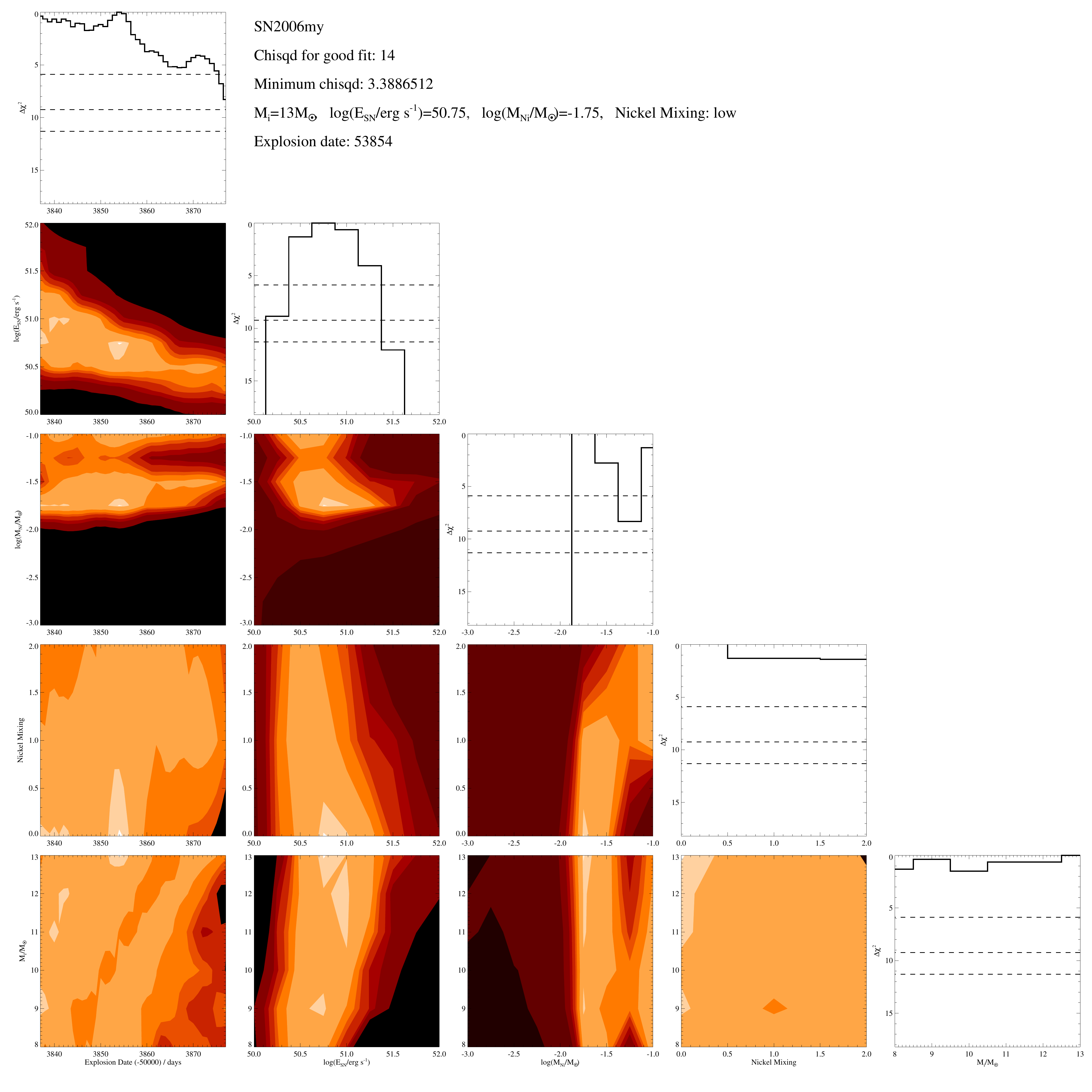}
\includegraphics[width=\columnwidth]{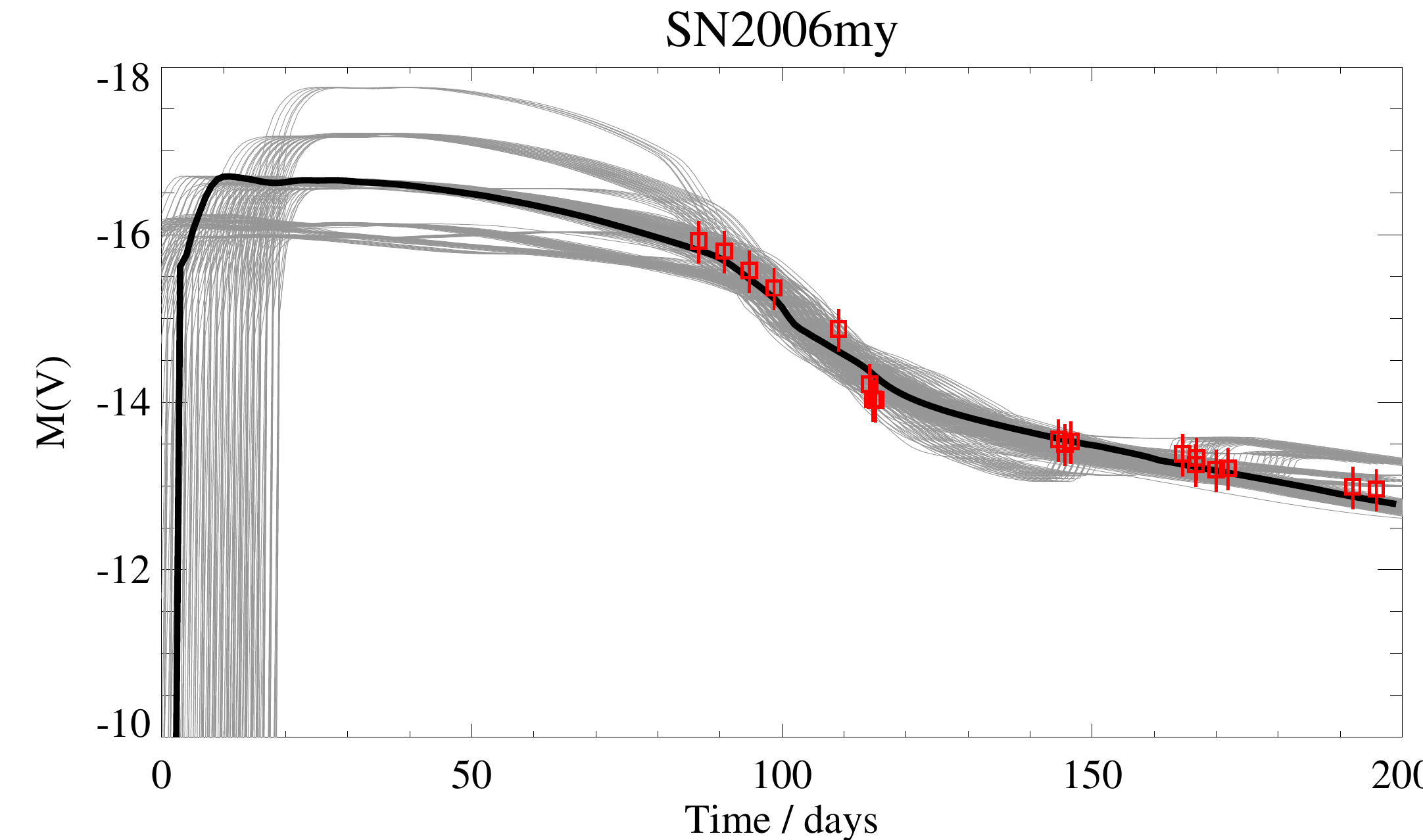}
\caption{SN2006my Constrained, as in Figure \ref{fig:constrained2003gd}}
\label{fig:constrained2006my}
\end{center}
\end{figure*}

\begin{figure*}[h]
\begin{center}
\includegraphics[width=2\columnwidth]{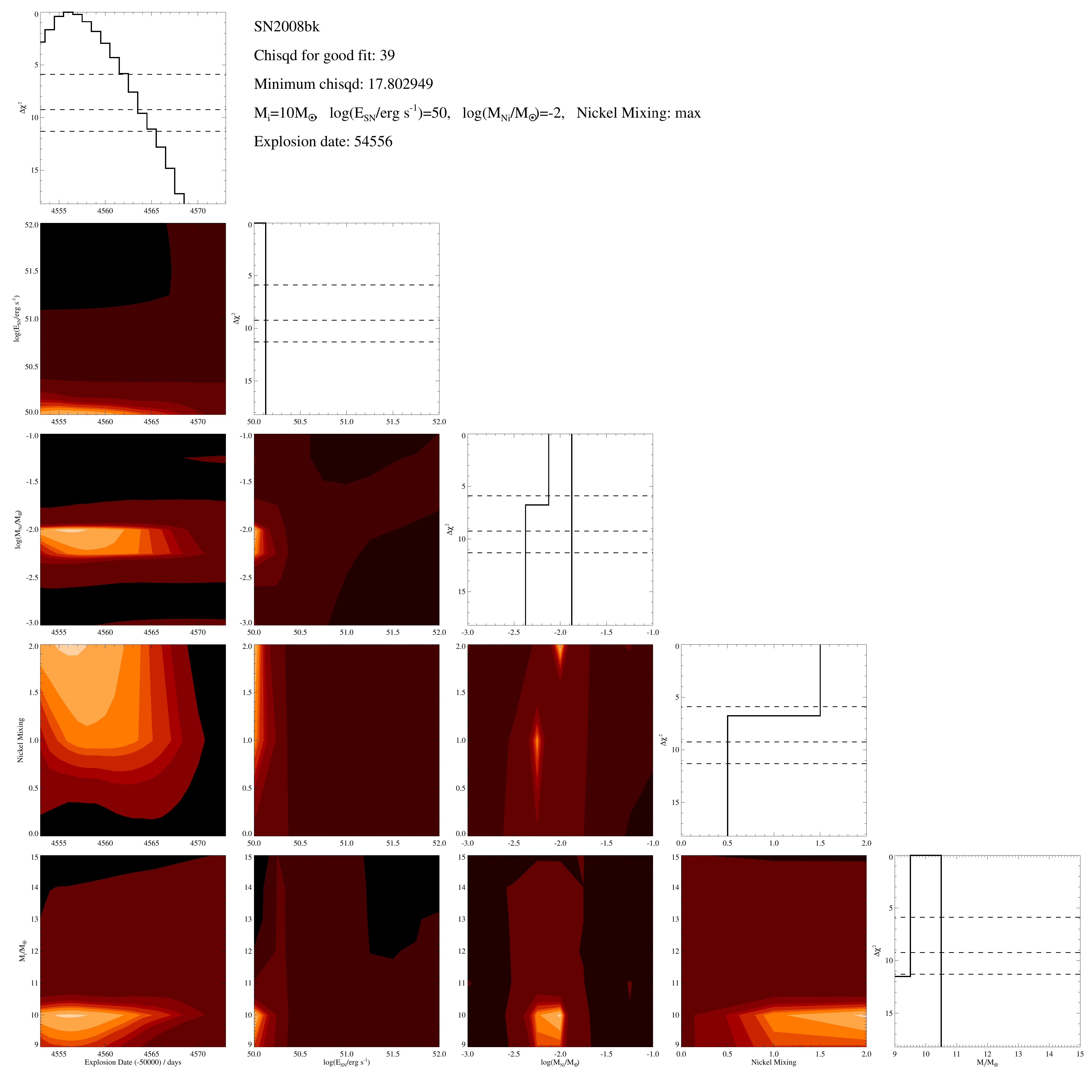}
\includegraphics[width=\columnwidth]{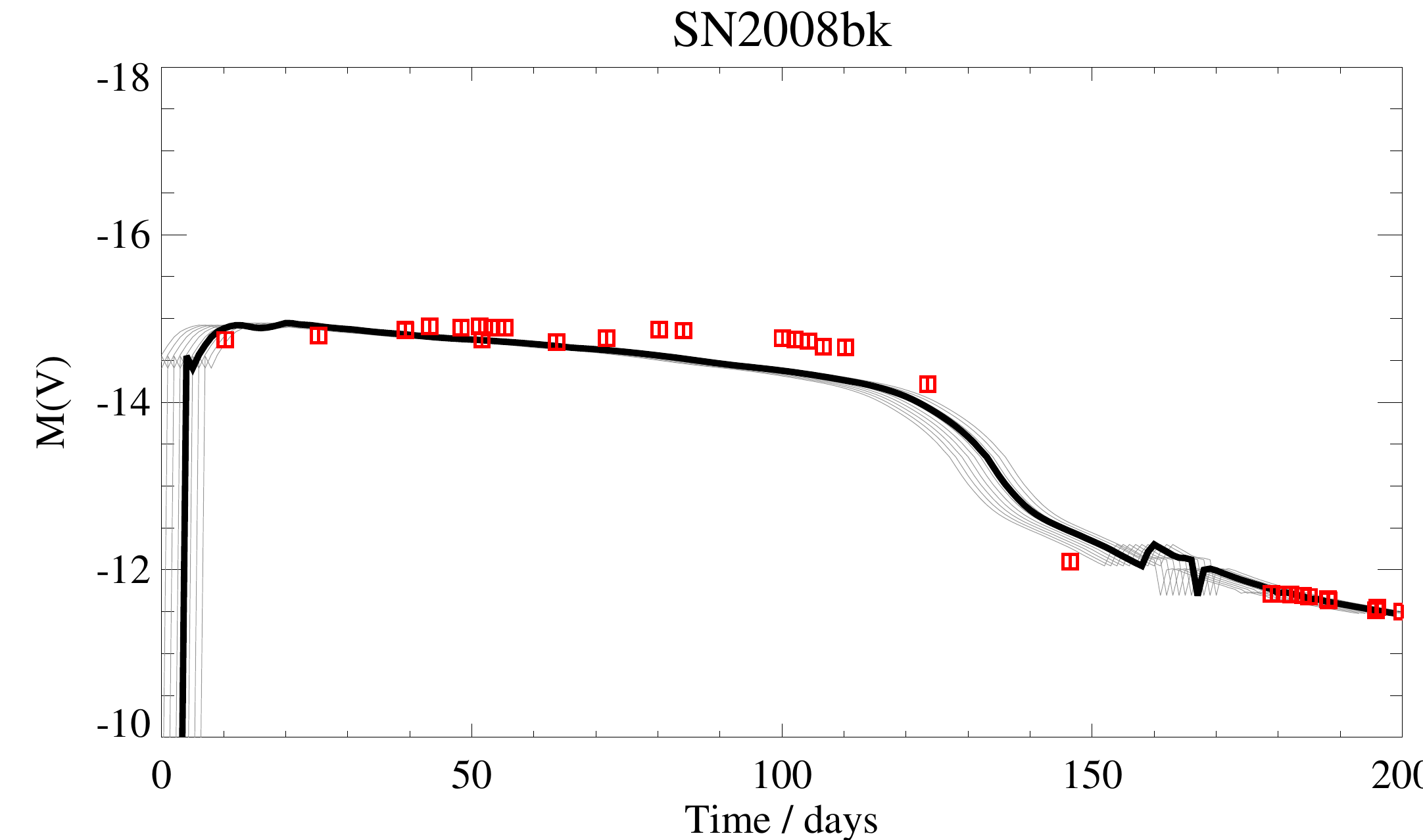}
\caption{SN2008bk Constrained, as in Figure \ref{fig:constrained2003gd}}
\label{fig:constrained2008bk}
\end{center}
\end{figure*}

\begin{figure*}[!h]
\begin{center}
\includegraphics[width=2\columnwidth]{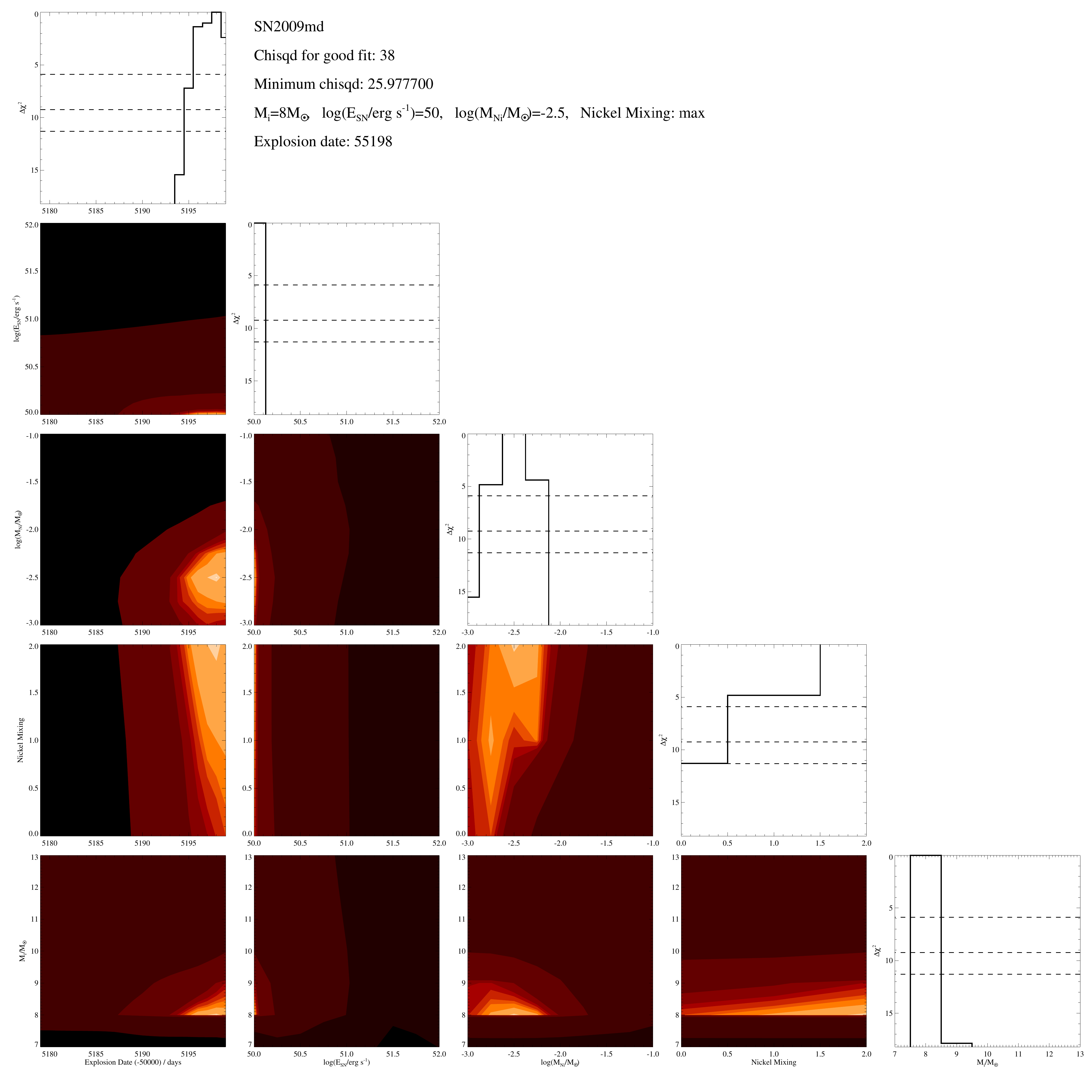}
\includegraphics[width=\columnwidth]{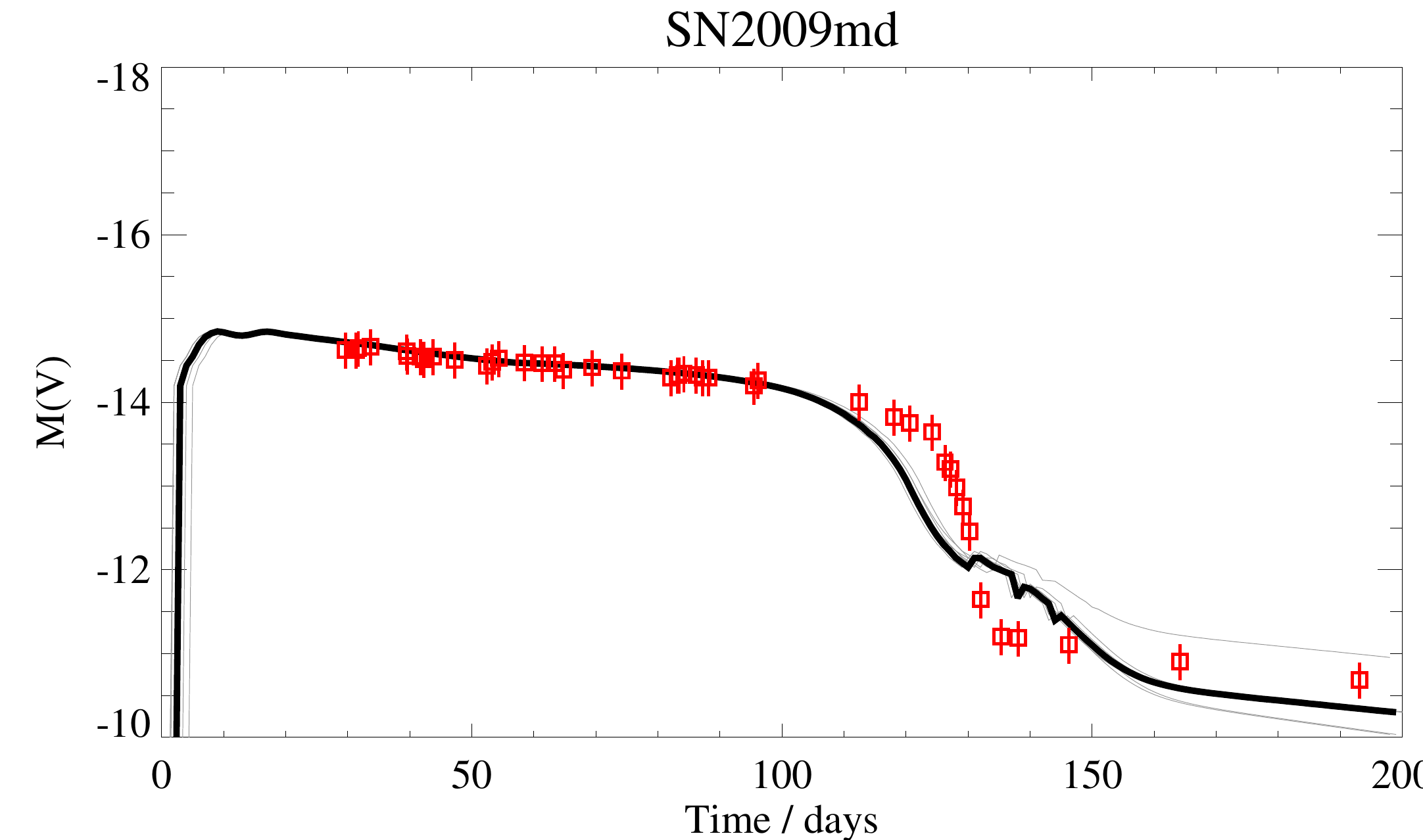}
\caption{SN2009md Constrained, as in Figure \ref{fig:constrained2003gd}}
\label{fig:constrained2009md}
\end{center}
\end{figure*}

\begin{figure*}[!h]
\begin{center}
\includegraphics[width=2\columnwidth]{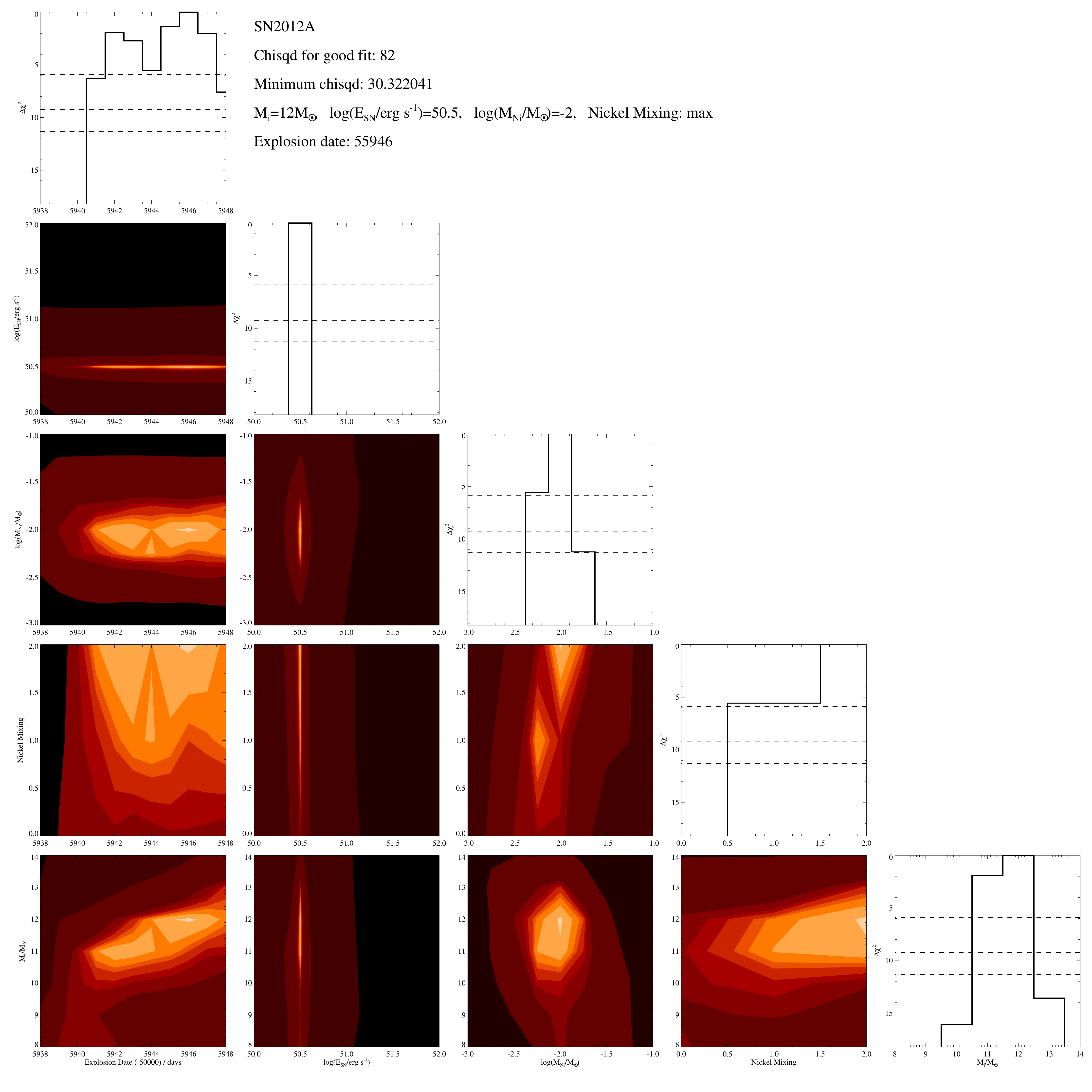}
\includegraphics[width=\columnwidth]{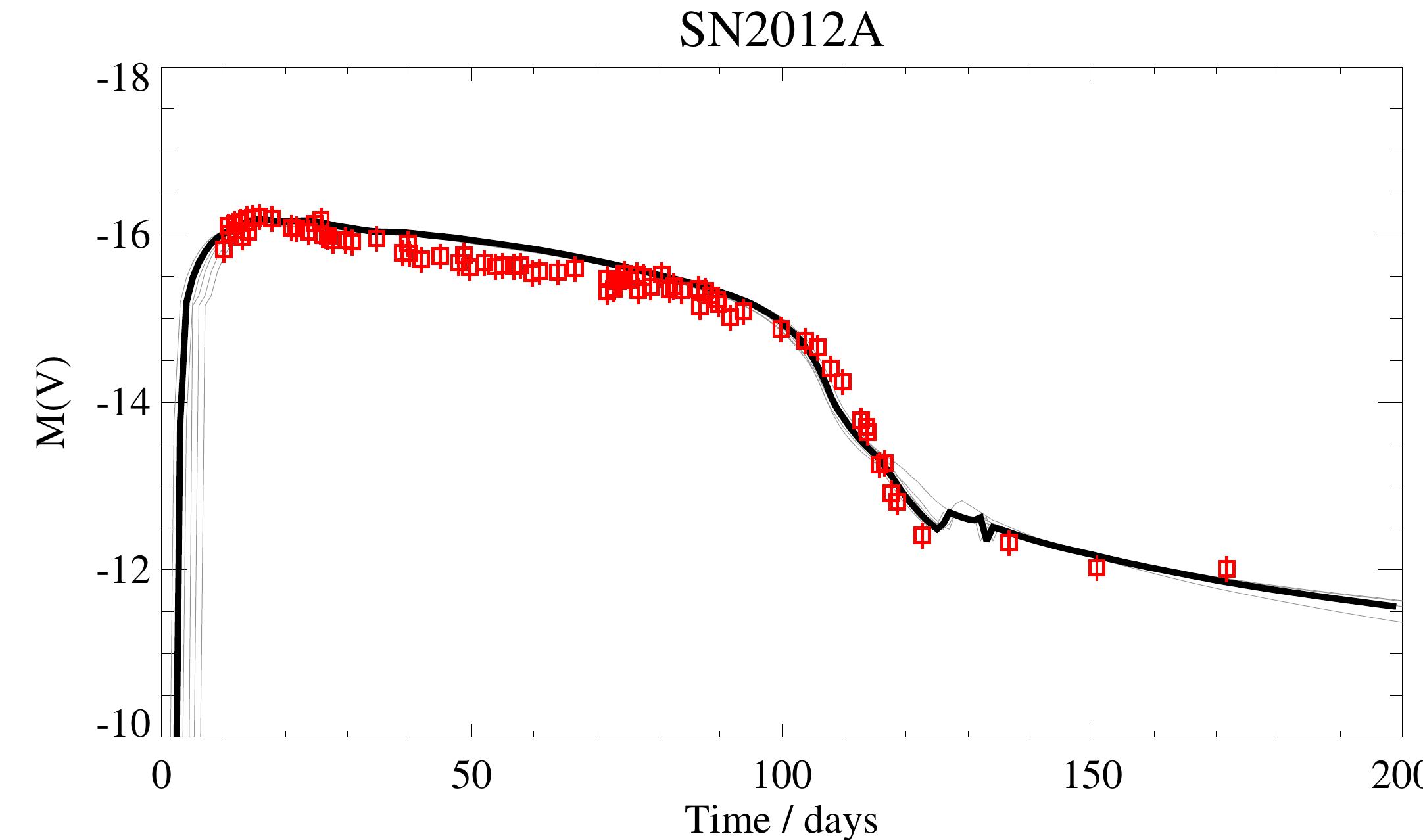}
\caption{SN2012A Constrained, as in Figure \ref{fig:constrained2003gd}}
\label{fig:constrained2012A}
\end{center}
\end{figure*}

\begin{figure*}[h]
\begin{center}
\includegraphics[width=2\columnwidth]{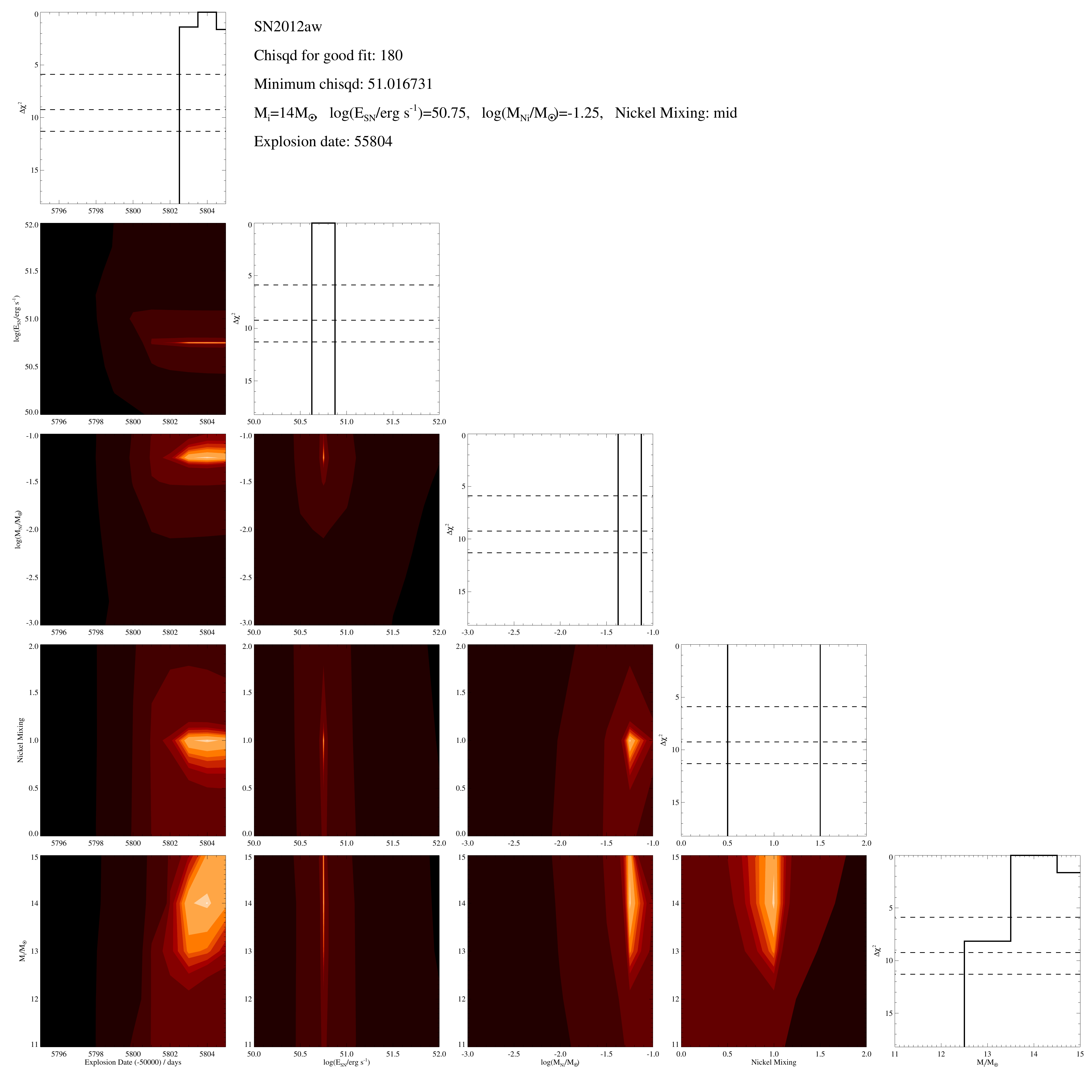}
\includegraphics[width=\columnwidth]{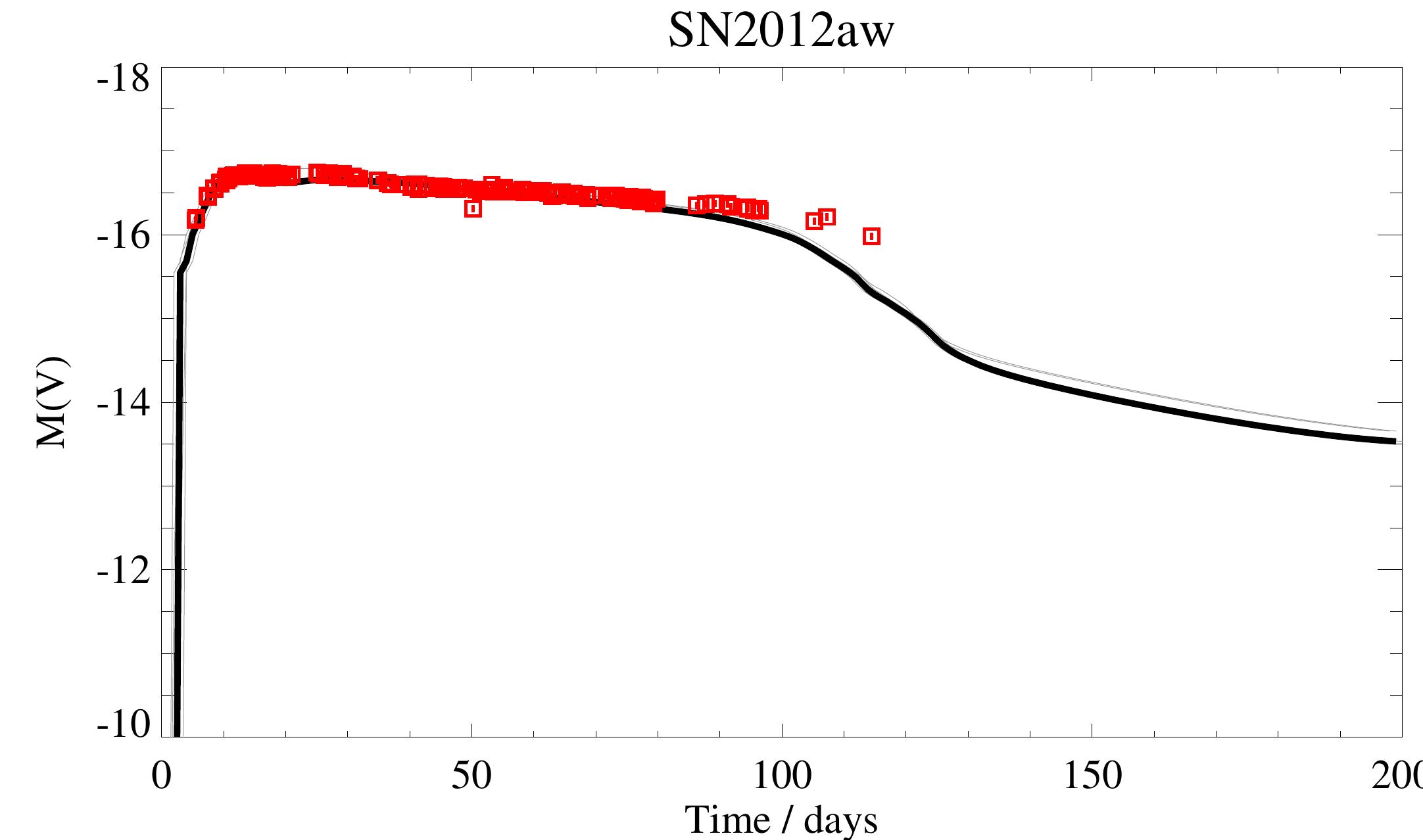}
\caption{SN2012aw Constrained, as in Figure \ref{fig:constrained2003gd}}
\label{fig:constrained2012aw}
\end{center}
\end{figure*}

\begin{figure*}[!h]
\begin{center}
\includegraphics[width=2\columnwidth]{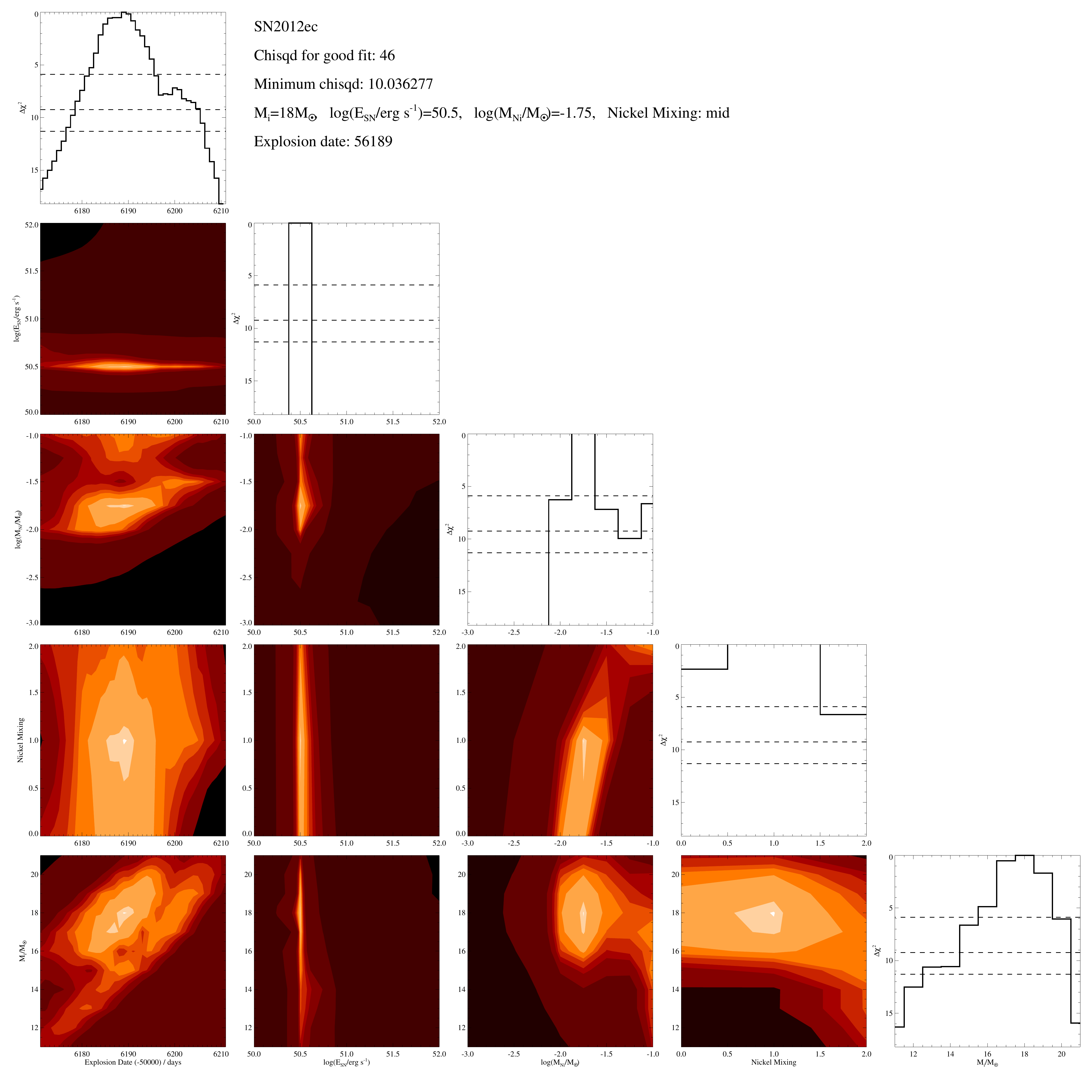}
\includegraphics[width=\columnwidth]{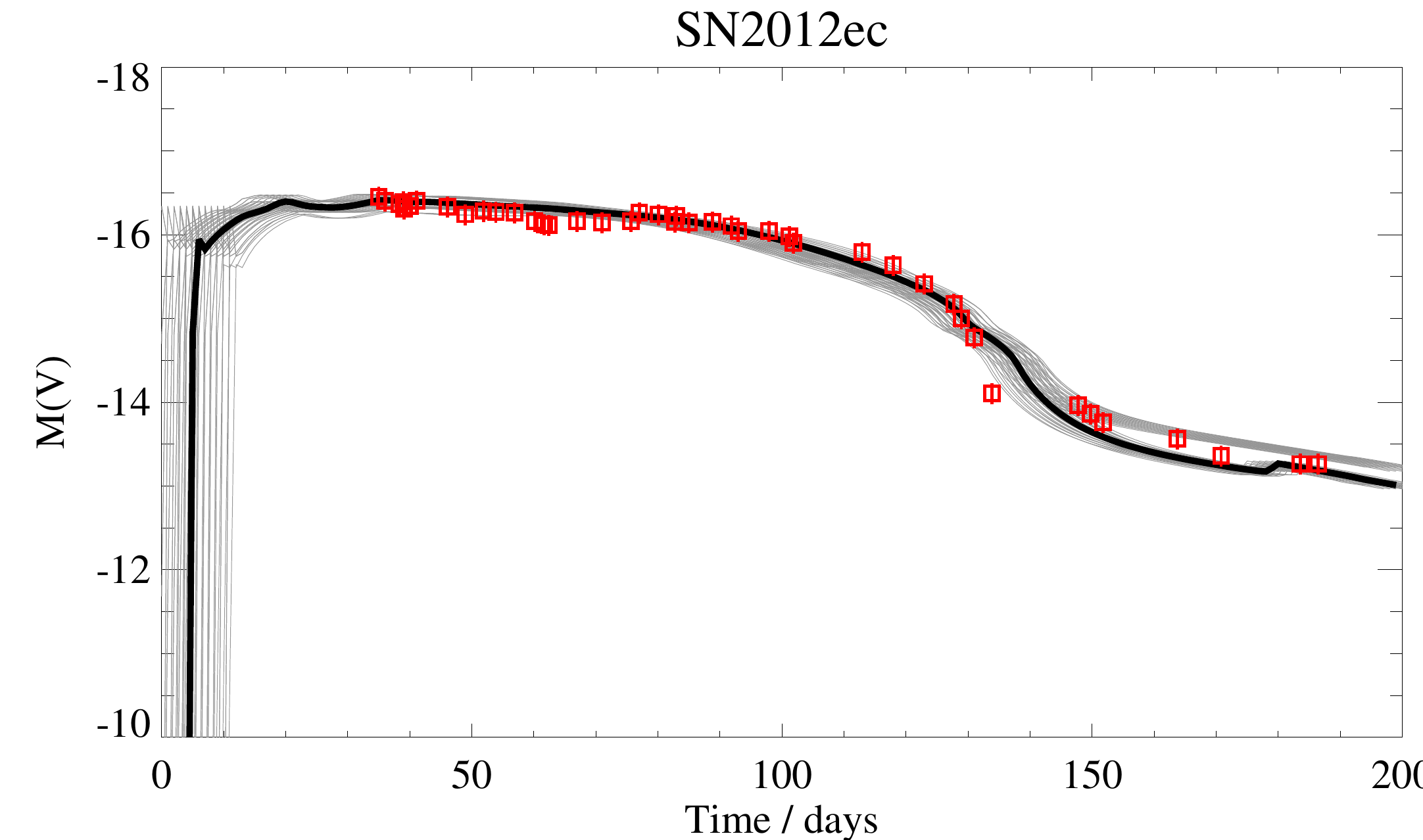}
\caption{SN2012ec Constrained, as in Figure \ref{fig:constrained2003gd}}
\label{fig:constrained2012ec}
\end{center}
\end{figure*}

\begin{figure*}[!h]
\begin{center}
\includegraphics[width=2\columnwidth]{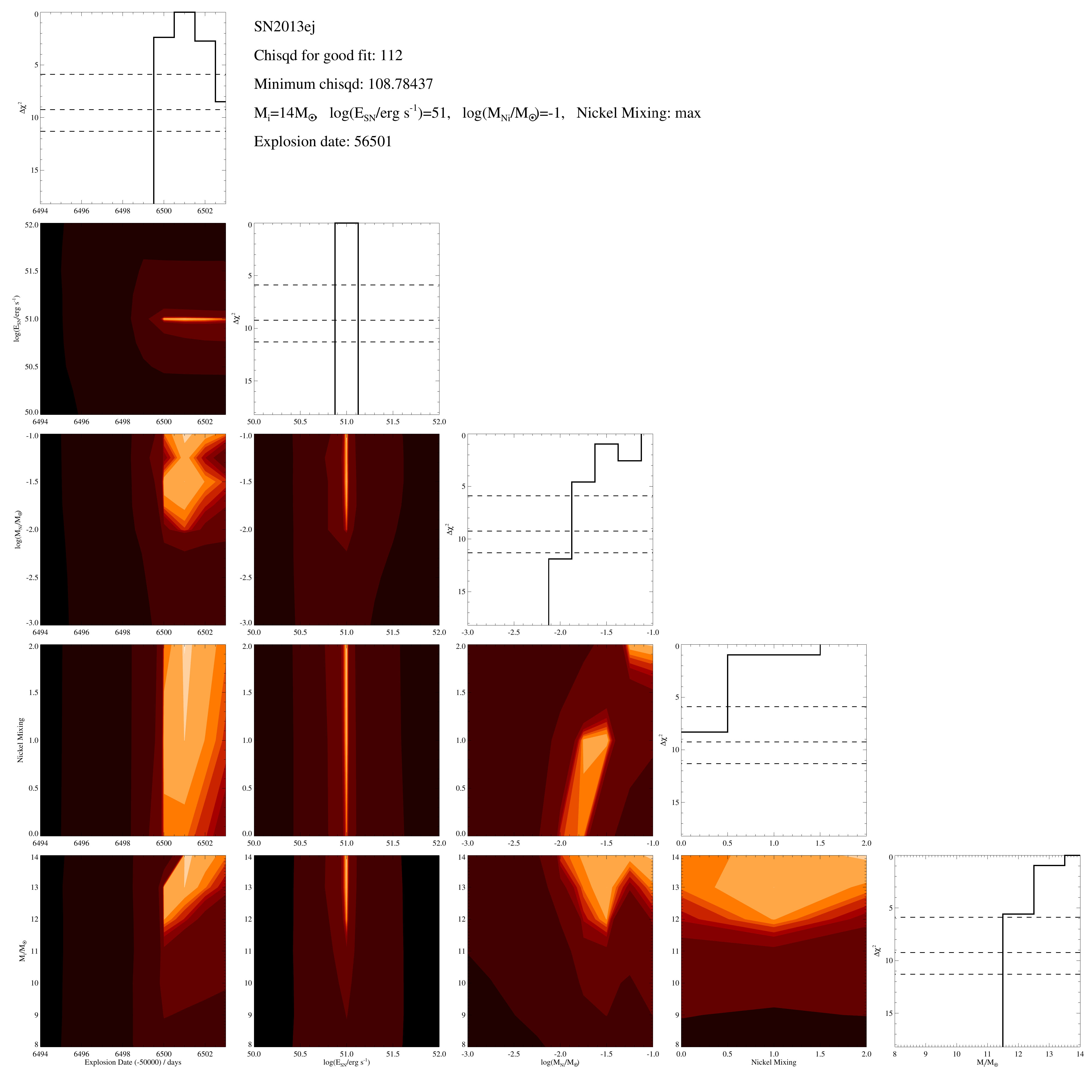}
\includegraphics[width=\columnwidth]{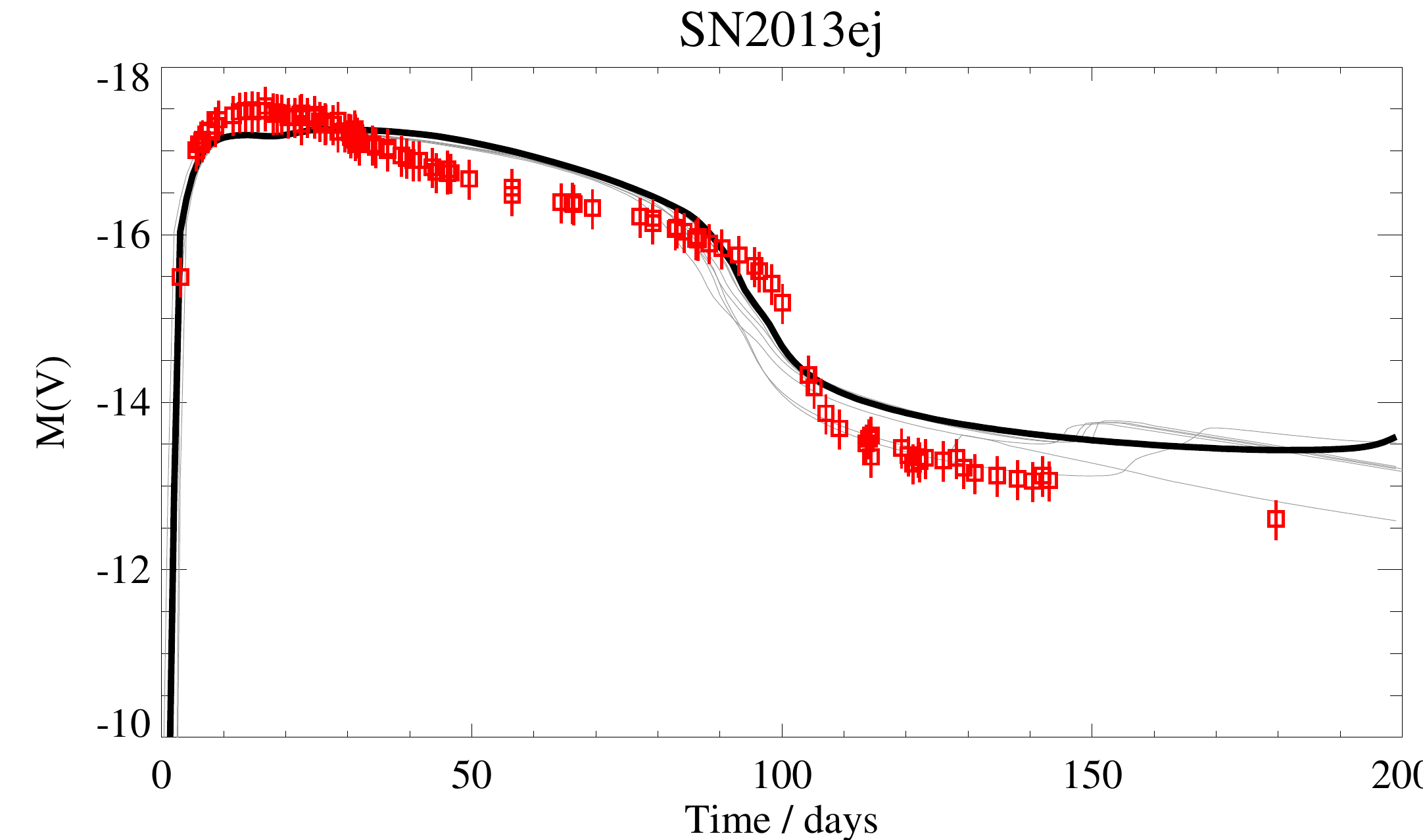}
\caption{SN2013ej Constrained, as in Figure \ref{fig:constrained2003gd}}
\label{fig:constrained2013ej}
\end{center}
\end{figure*}
\raggedbottom

\clearpage

\section{Test of changes at end of stellar models}

Our input grid of stellar evolution and structure models, based on the BPASS modification of the STARS code, was selected for its utility in modelling binary interactions and its compatibility with a large population synthesis modelling project, and thus will allow future generalisations of this study. However an important drawback in the BPASS models is that they only follow stellar evolution to the end of core carbon burning. It is entirely possible that important structural changes may occur to massive stars in the very rapid and complex late stages of core nuclear burning, immediately before supernova. This is impossible to explore within BPASS, but is accessible to other stellar evolution codes.

\begin{figure}[!h]
\begin{center}
\includegraphics[width=0.99\columnwidth]{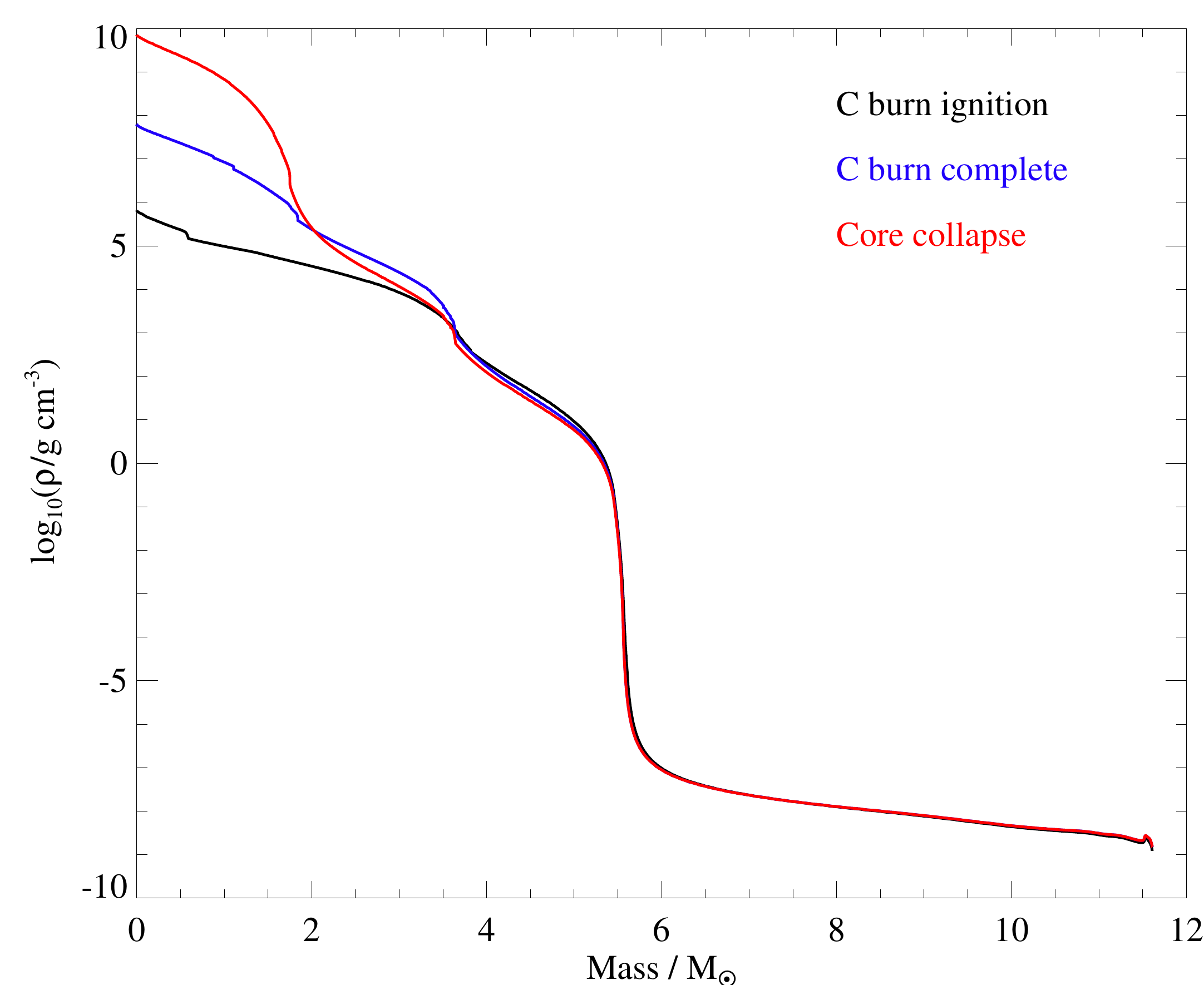}
\includegraphics[width=0.99\columnwidth]{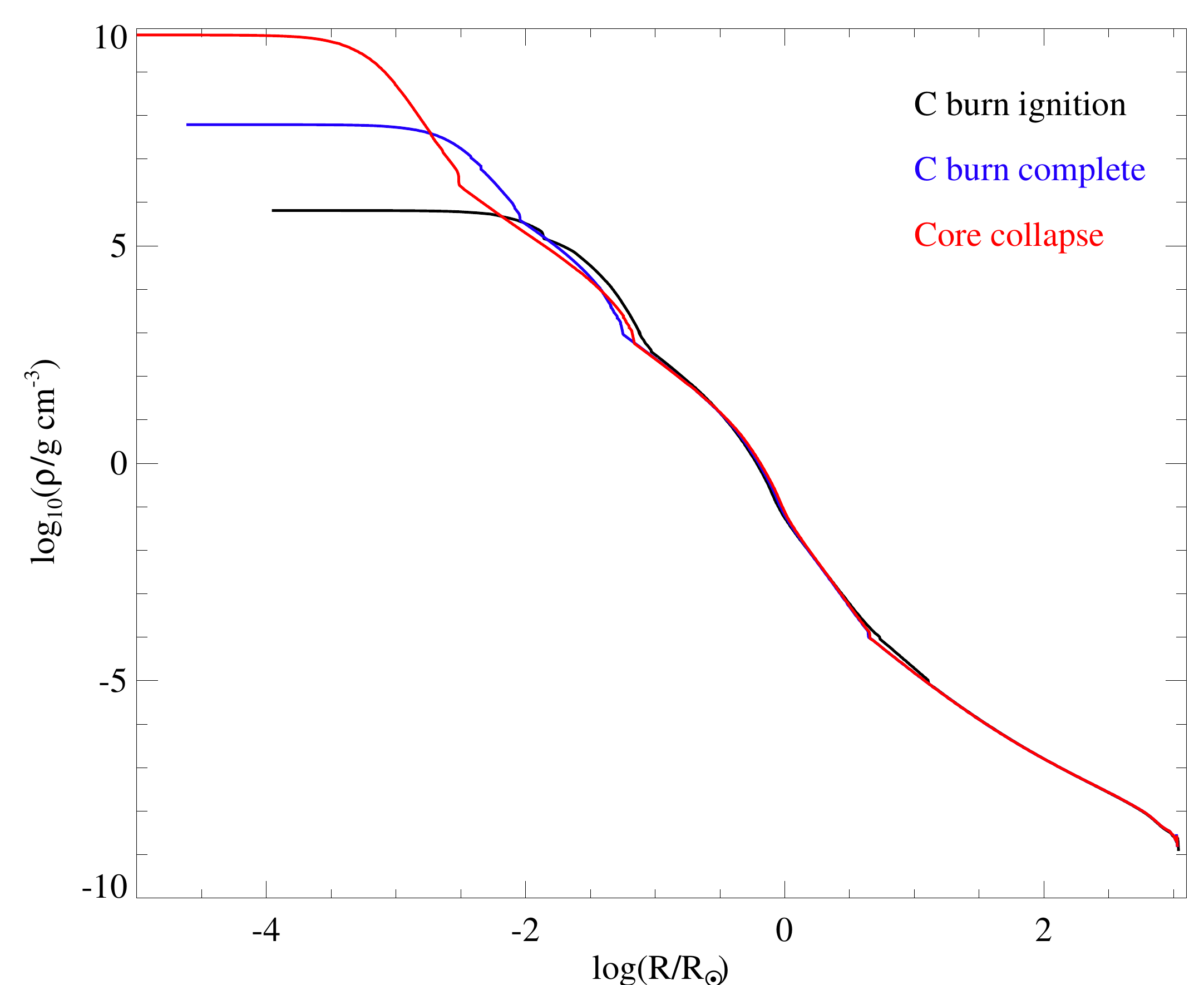}
\includegraphics[width=0.99\columnwidth]{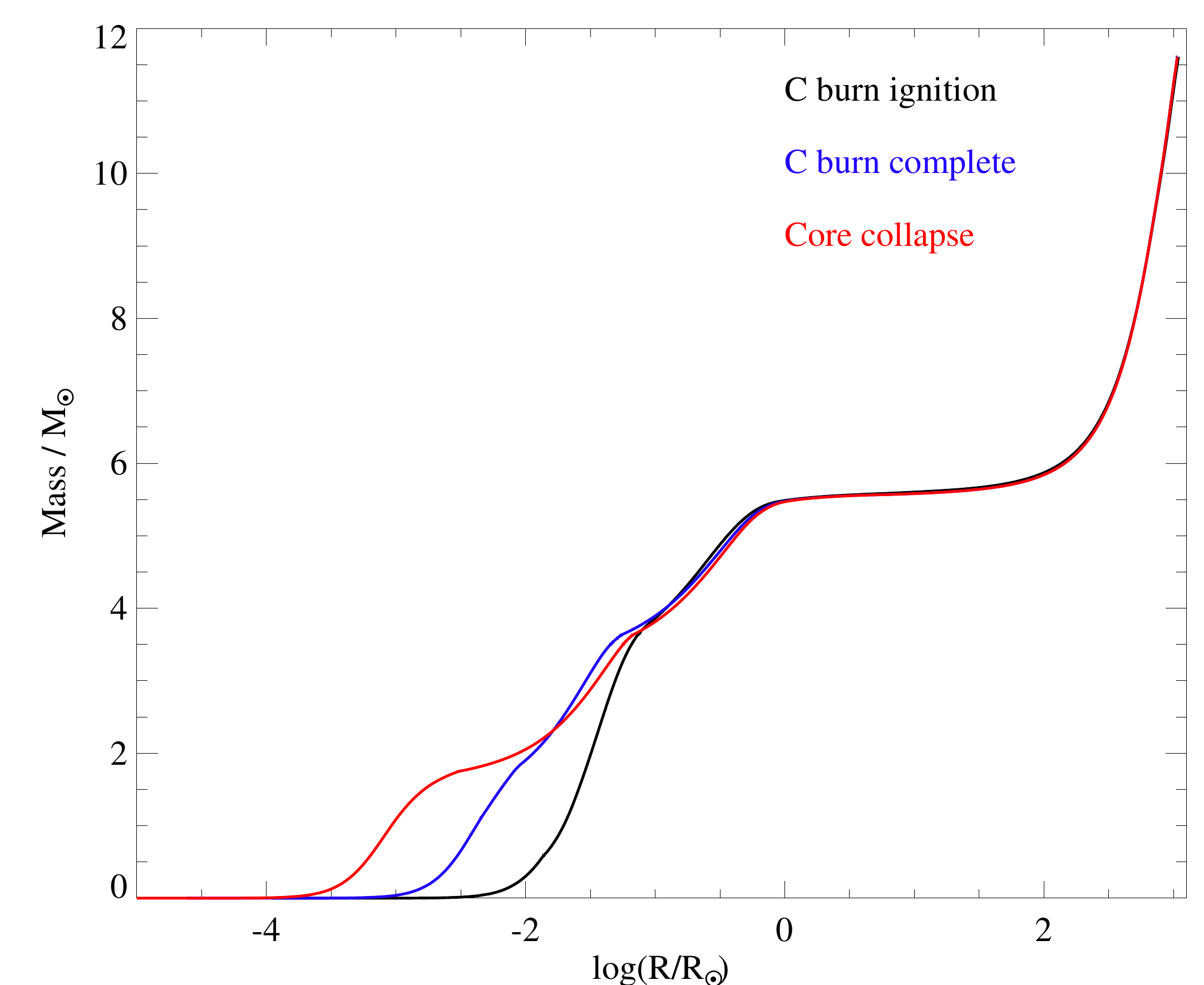}
\caption{An initially 15.6M$_{\odot}$ stellar model evolved with MESA\,v10398. The models are taken at the beginning of carbon burning, the end of core carbon burning and the final model output before core-collapse. The BPASS models we use are taken after the point of carbon burning completing in the stellar core. }
\label{fig:mesa_comp}
\end{center}
\end{figure}

To evaluate the principal effects of late core-burning stages, we have considered the evolution of a 15.6\,M$_\odot$ single-star model (i.e. a typical SN\,IIP progenitor) with the Modules for Experiments in Stellar Astrophysics  \citep[MESA-r10398,][]{Paxton2011, Paxton2013, Paxton2015, Paxton2018} stellar evolution code, which follows the stars through to core collapse. In figure \ref{fig:mesa_comp} we illustrate the difference in stellar structure observed as a function of initial mass for the models captured at three stages: at core carbon ignition, the end of core carbon burning and at core collapse.

As the figure illustrates, structural differences are observed in the very inner regions of the stellar core (i.e. within $R<0.01\,R_\ast$). In our formalism, these regions are extremely likely to be subsumed within the stellar remnant and so would contribute little if anything to the evolution of the supernova lightcurve. Nonetheless, we note this as a limitation of our modelling which may be addressed in future work.

\clearpage

\section{Effect of circumstellar medium on supernova lightcurves}

\begin{figure*}[!h]
\begin{center}

\includegraphics[width=0.65\columnwidth]{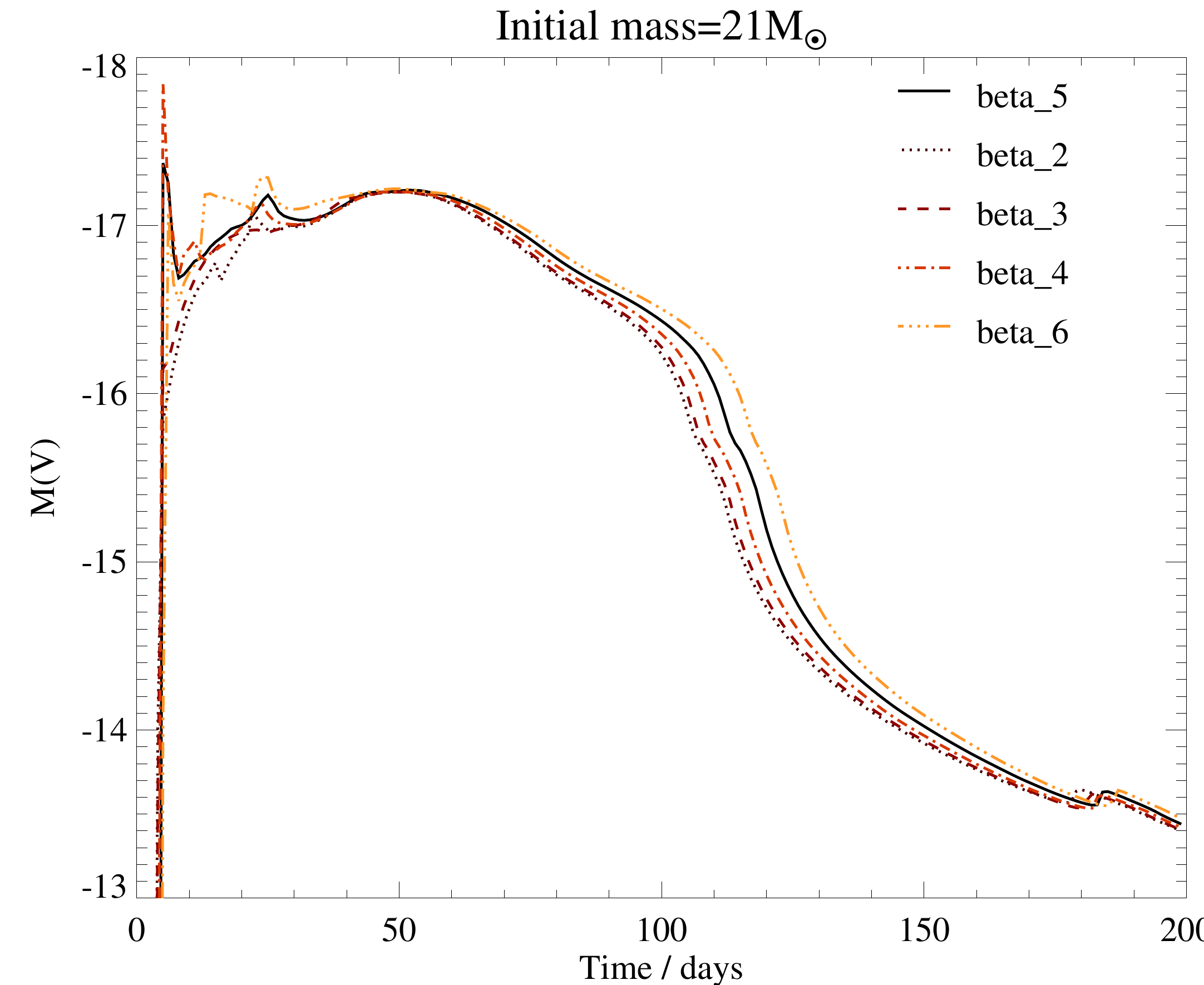}
\includegraphics[width=0.65\columnwidth]{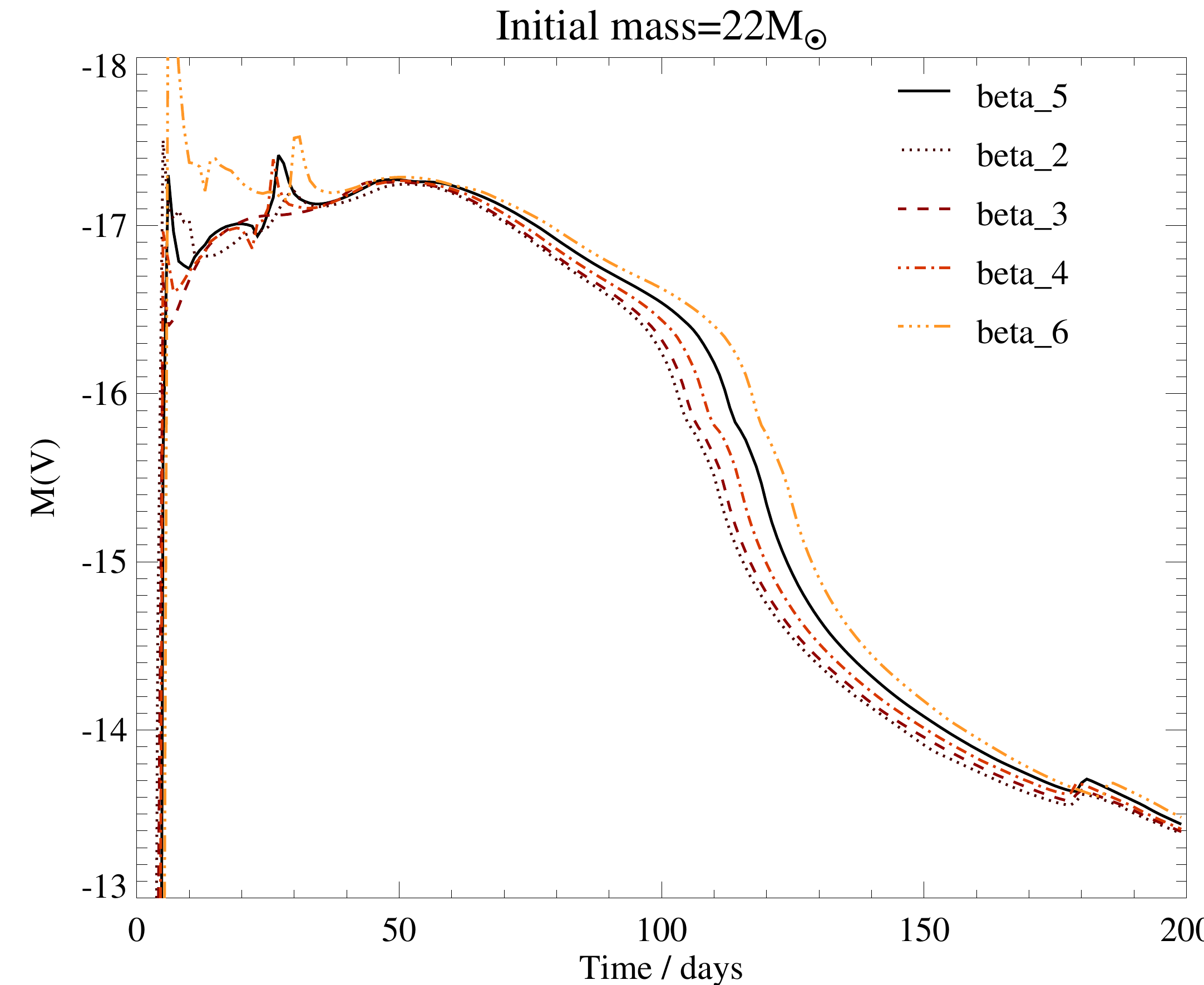}
\includegraphics[width=0.65\columnwidth]{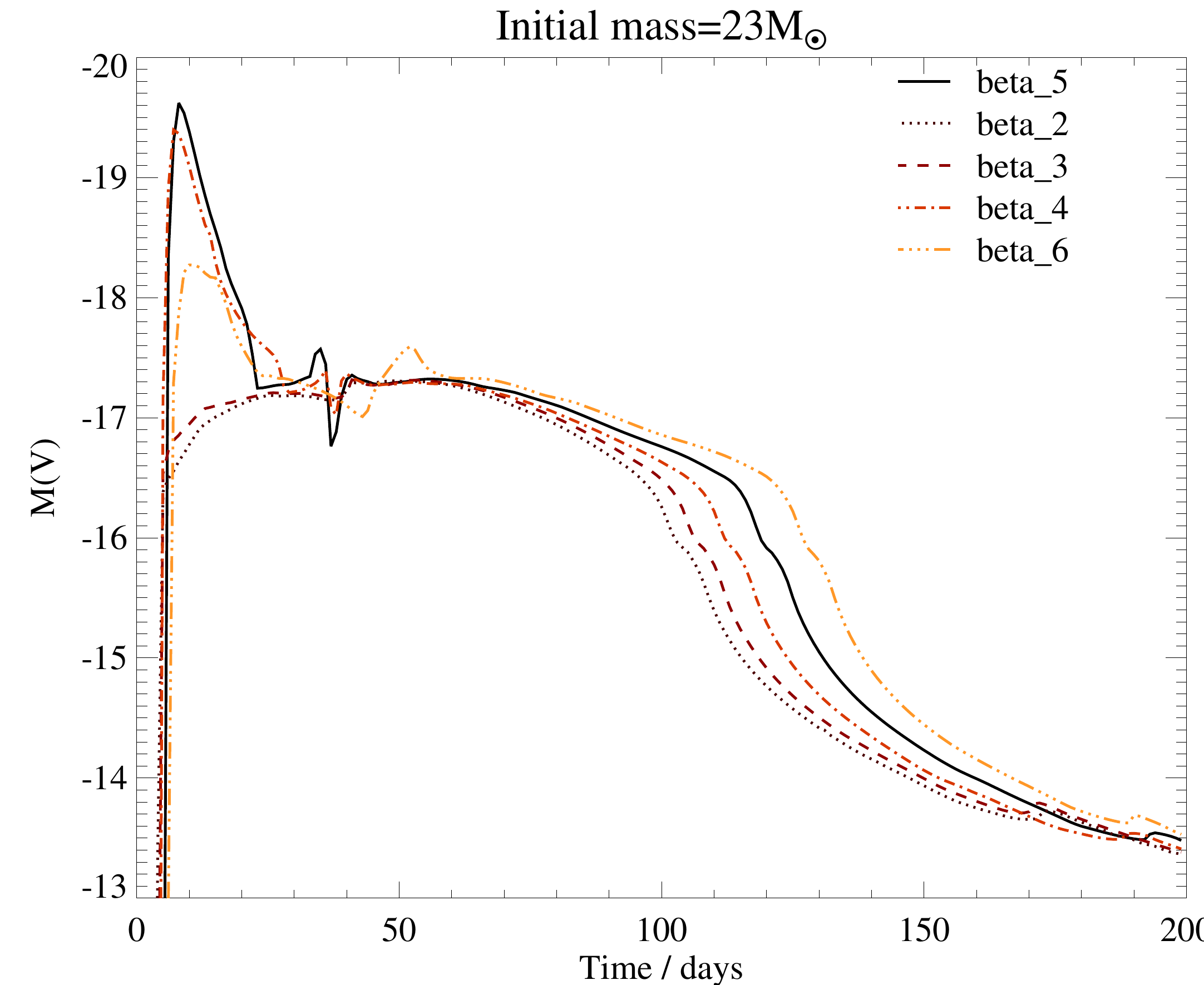}\\
\includegraphics[width=0.65\columnwidth]{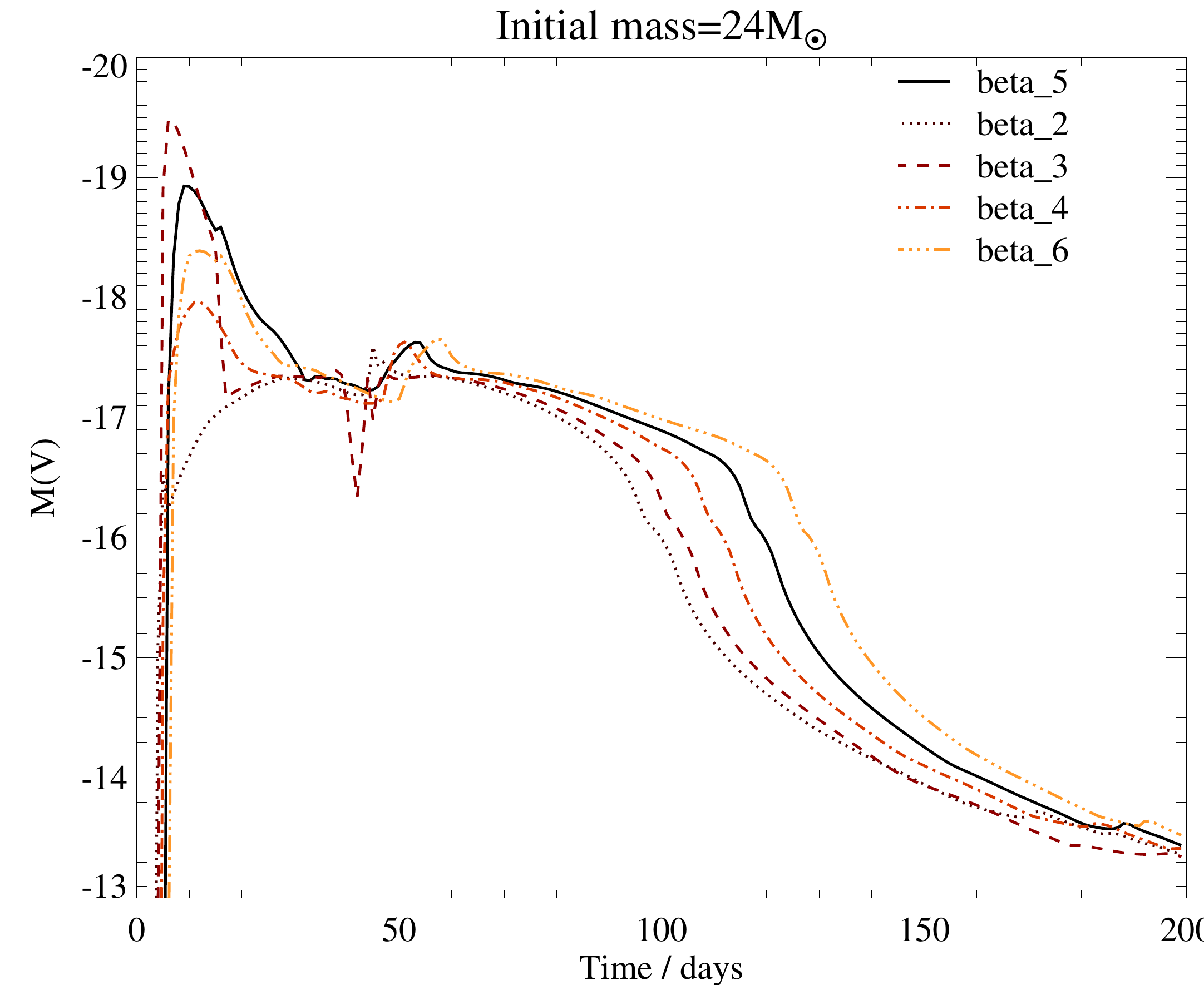}
\includegraphics[width=0.65\columnwidth]{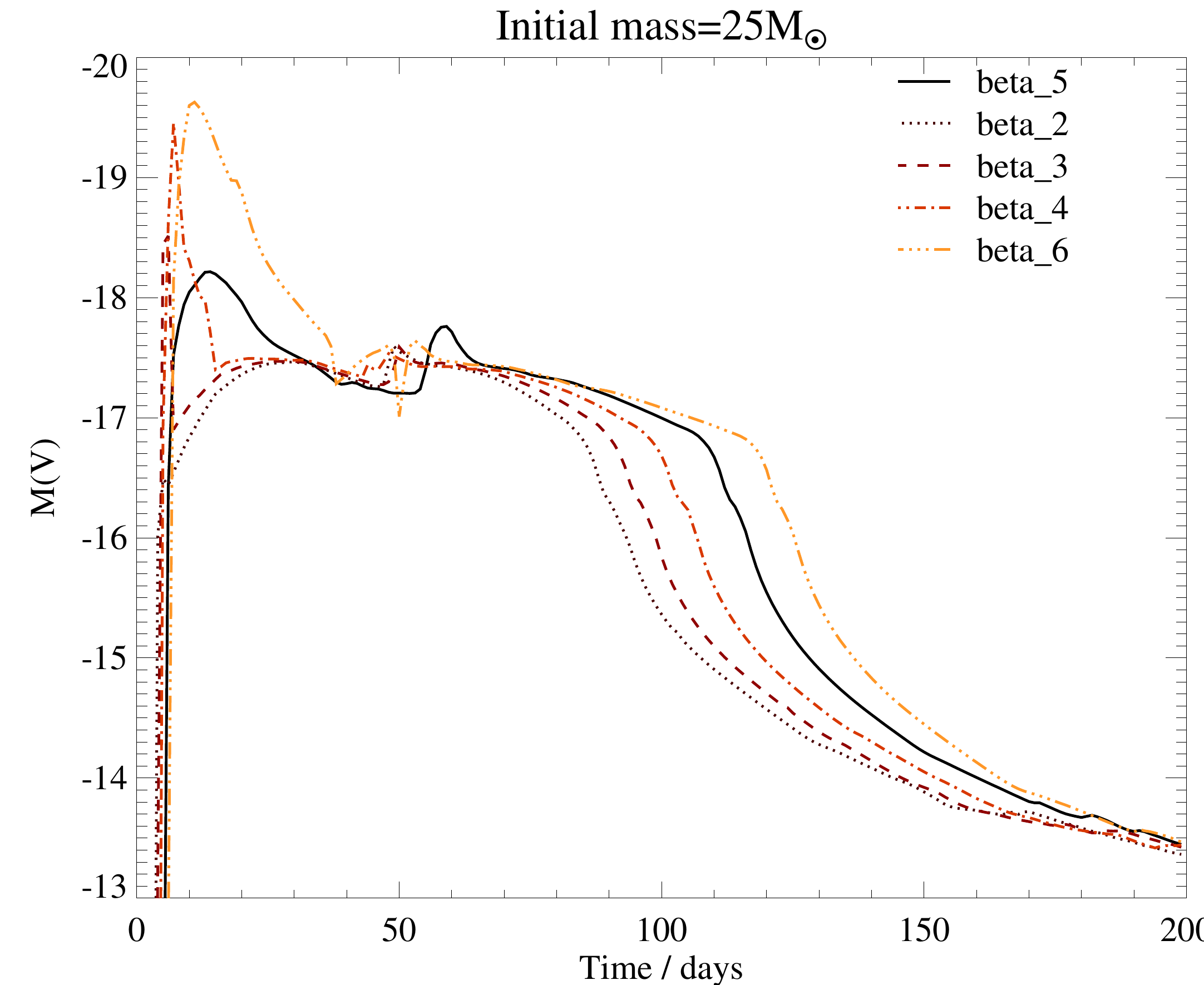}
\includegraphics[width=0.65\columnwidth]{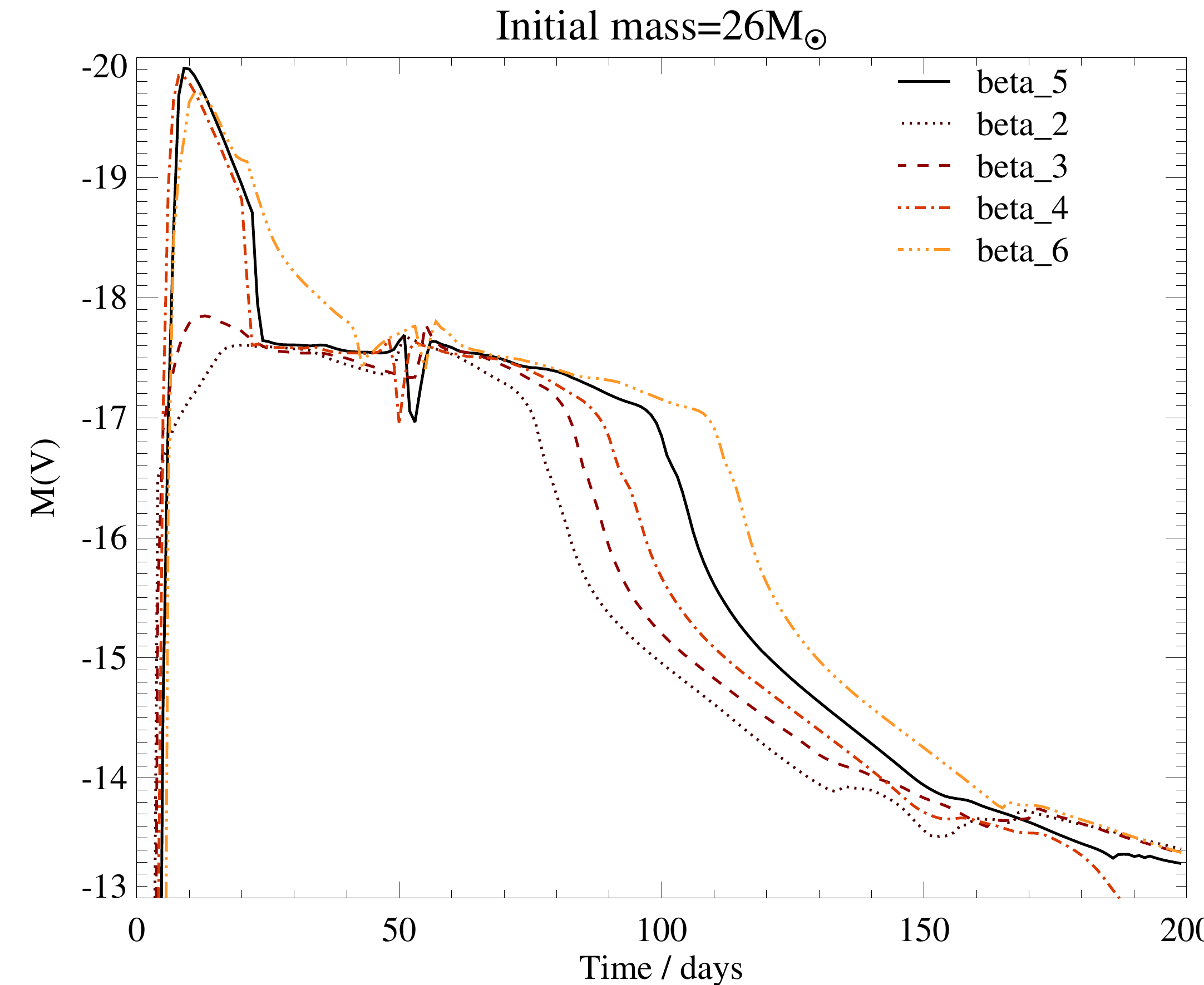}\\

\caption{Model lightcurves in the V-band for stars with initial masses of 21M$_{\odot}$ and above. Each panel now only shows one stellar initial mass, while the different lightcurves have varying values of $\beta$ the wind acceleration parameter.}
\label{fig:csmtest3}
\end{center}
\end{figure*}

\begin{figure*}[!h]
\begin{center}

\includegraphics[width=0.65\columnwidth]{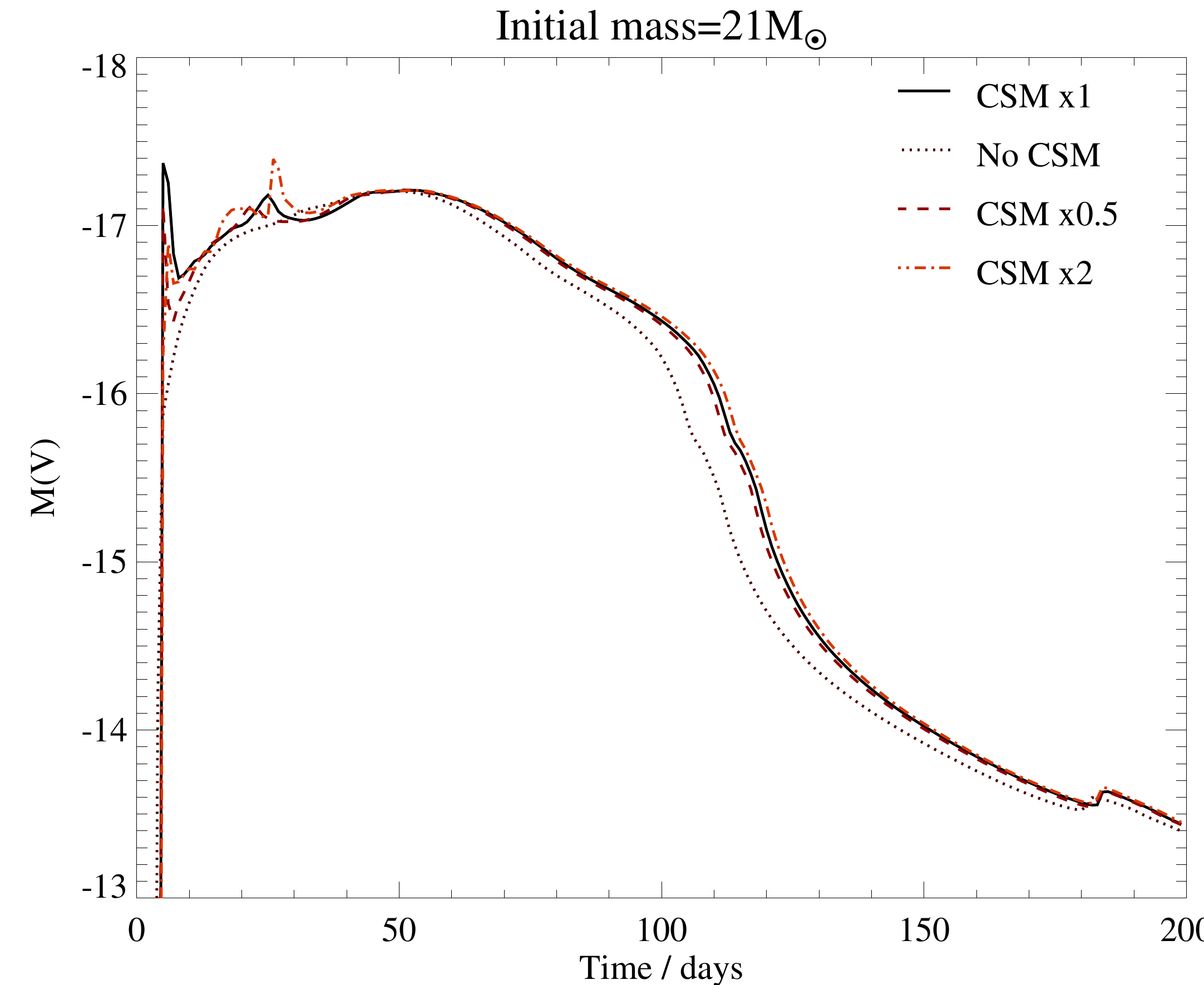}
\includegraphics[width=0.65\columnwidth]{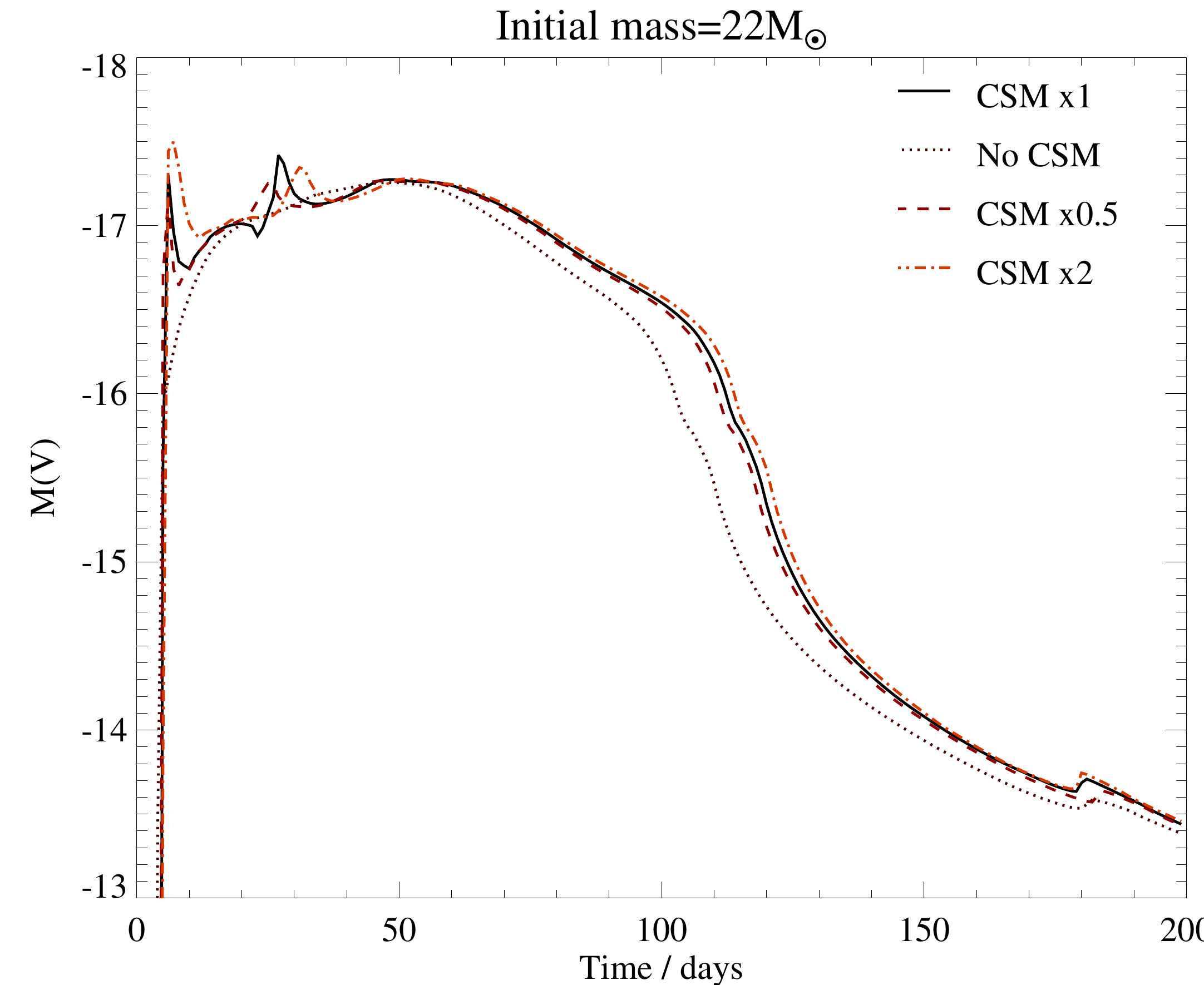}
\includegraphics[width=0.65\columnwidth]{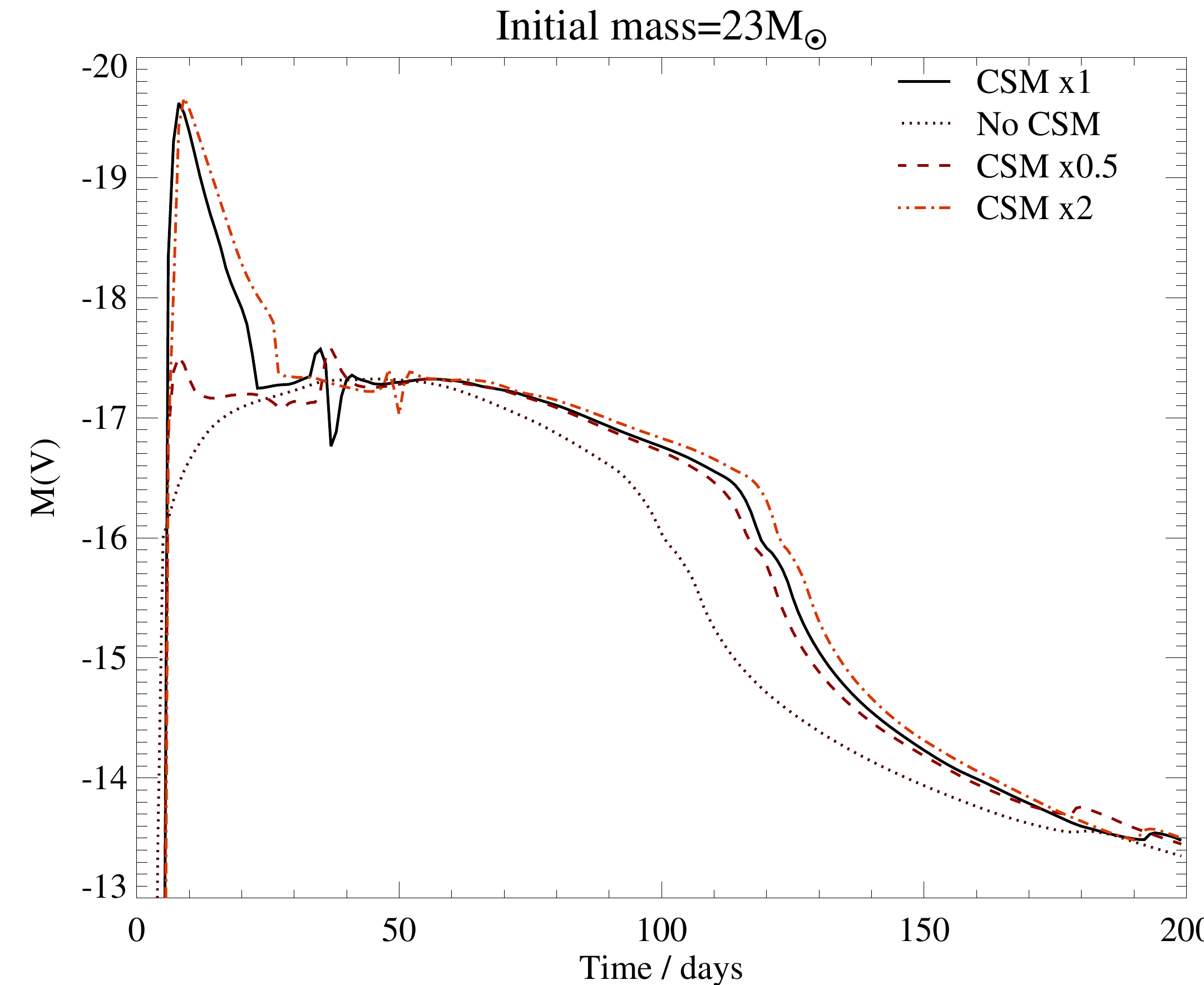}\\
\includegraphics[width=0.65\columnwidth]{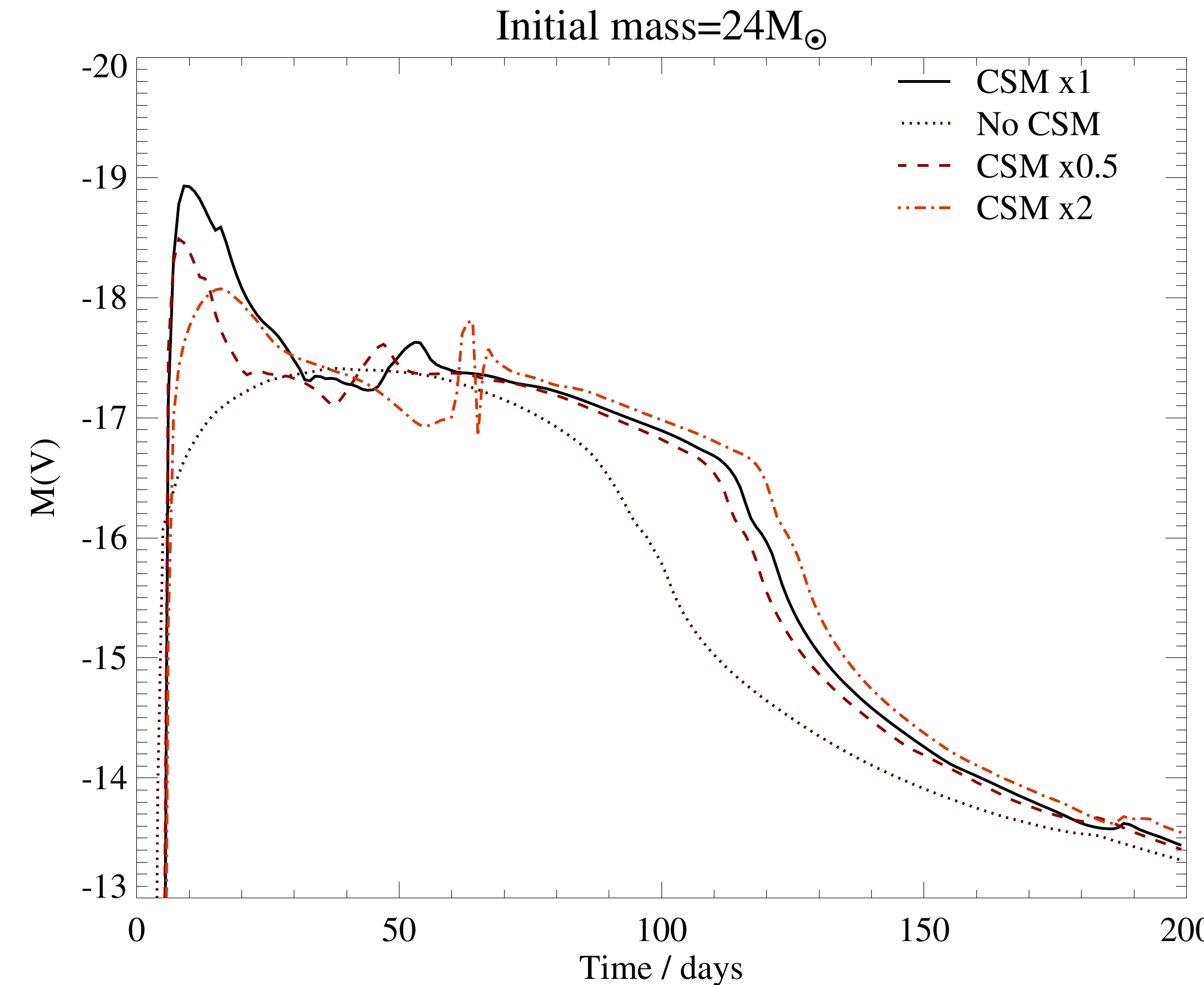}
\includegraphics[width=0.65\columnwidth]{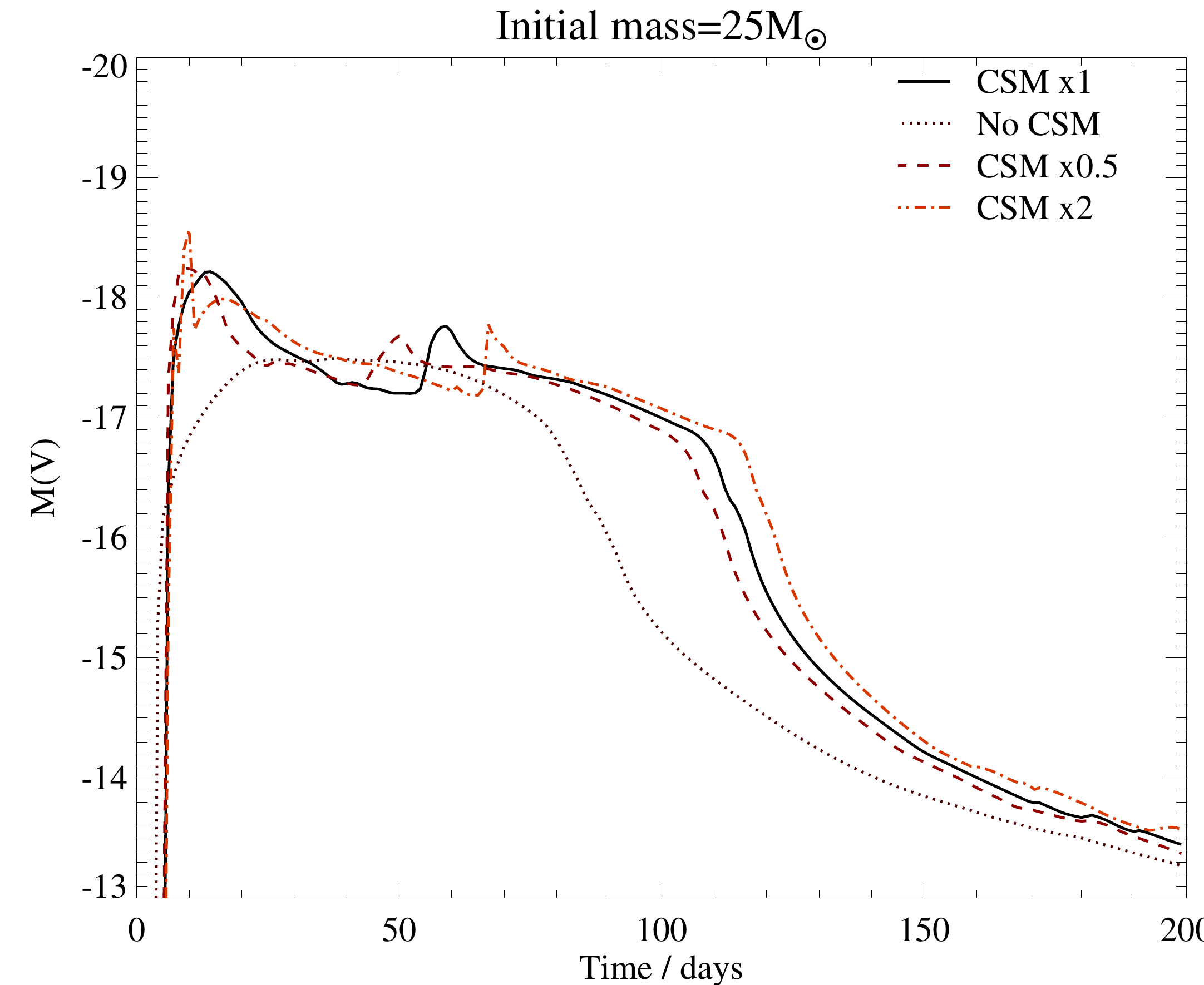}
\includegraphics[width=0.65\columnwidth]{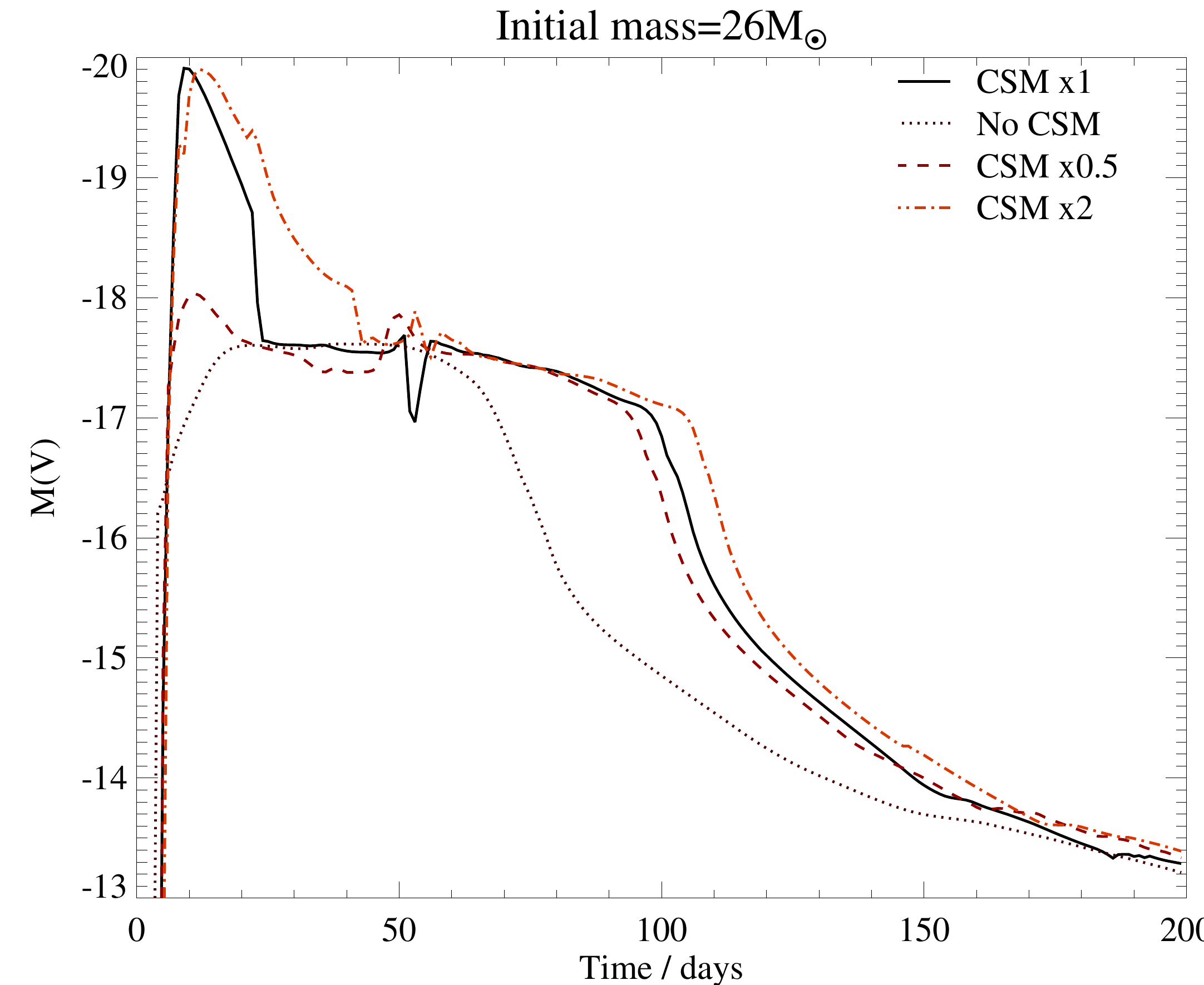}\\

\caption{As in Figure \ref{fig:csmtest3}, but here we vary the density of the circumstellar medium around the star. }
\label{fig:csmtest2}
\end{center}
\end{figure*}

\begin{figure*}[!h]
\begin{center}
\includegraphics[width=0.65\columnwidth]{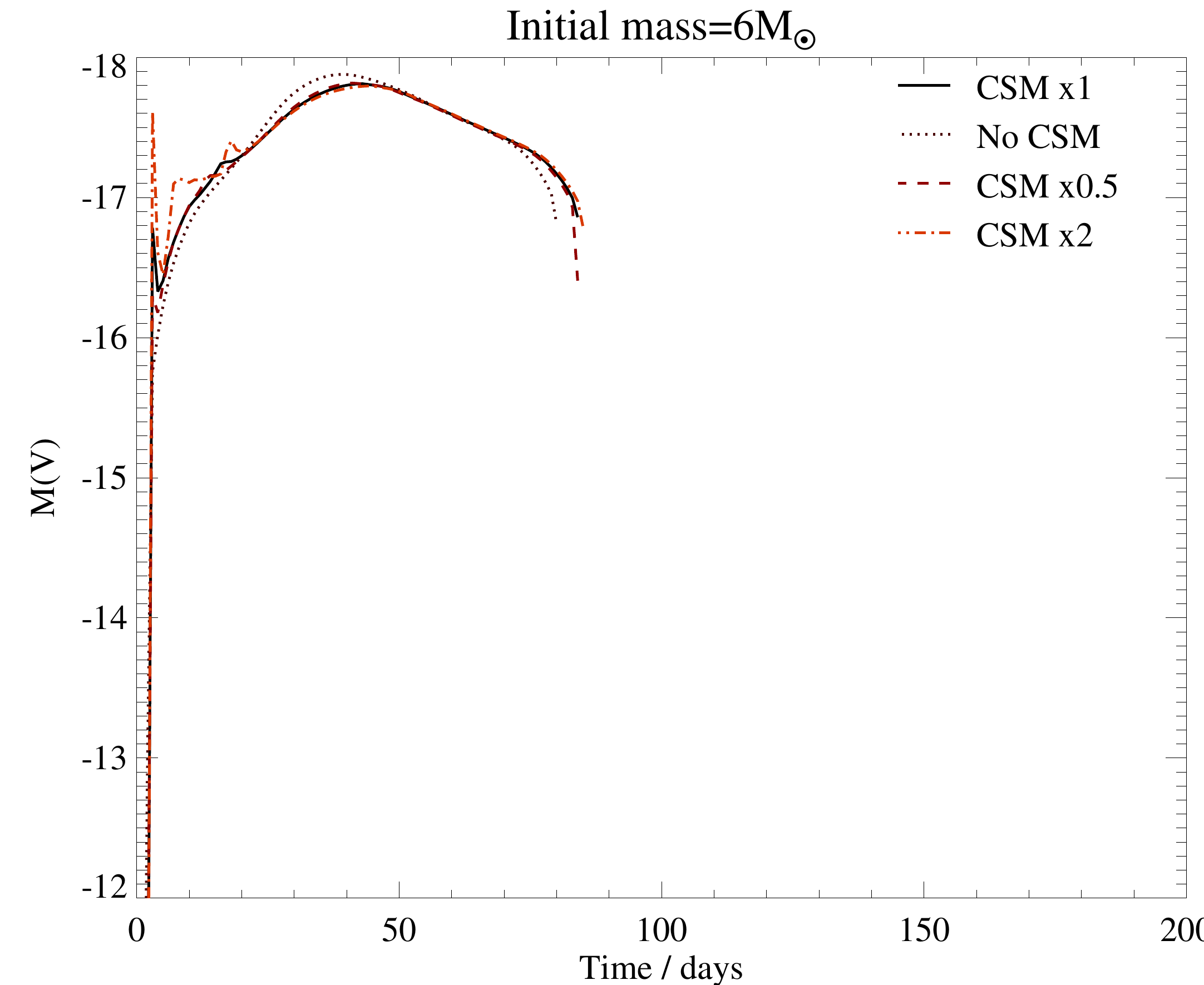}
\includegraphics[width=0.65\columnwidth]{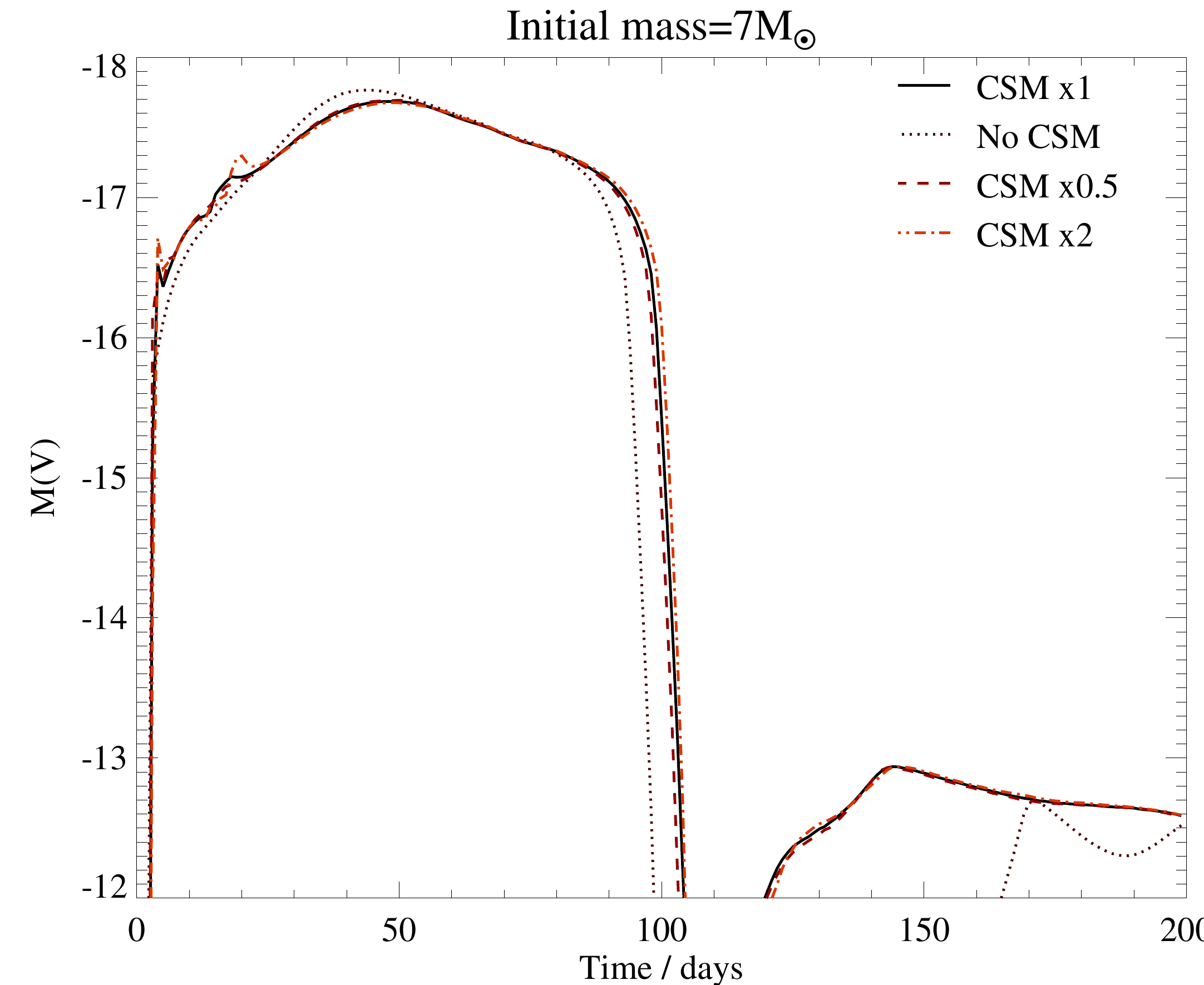}
\includegraphics[width=0.65\columnwidth]{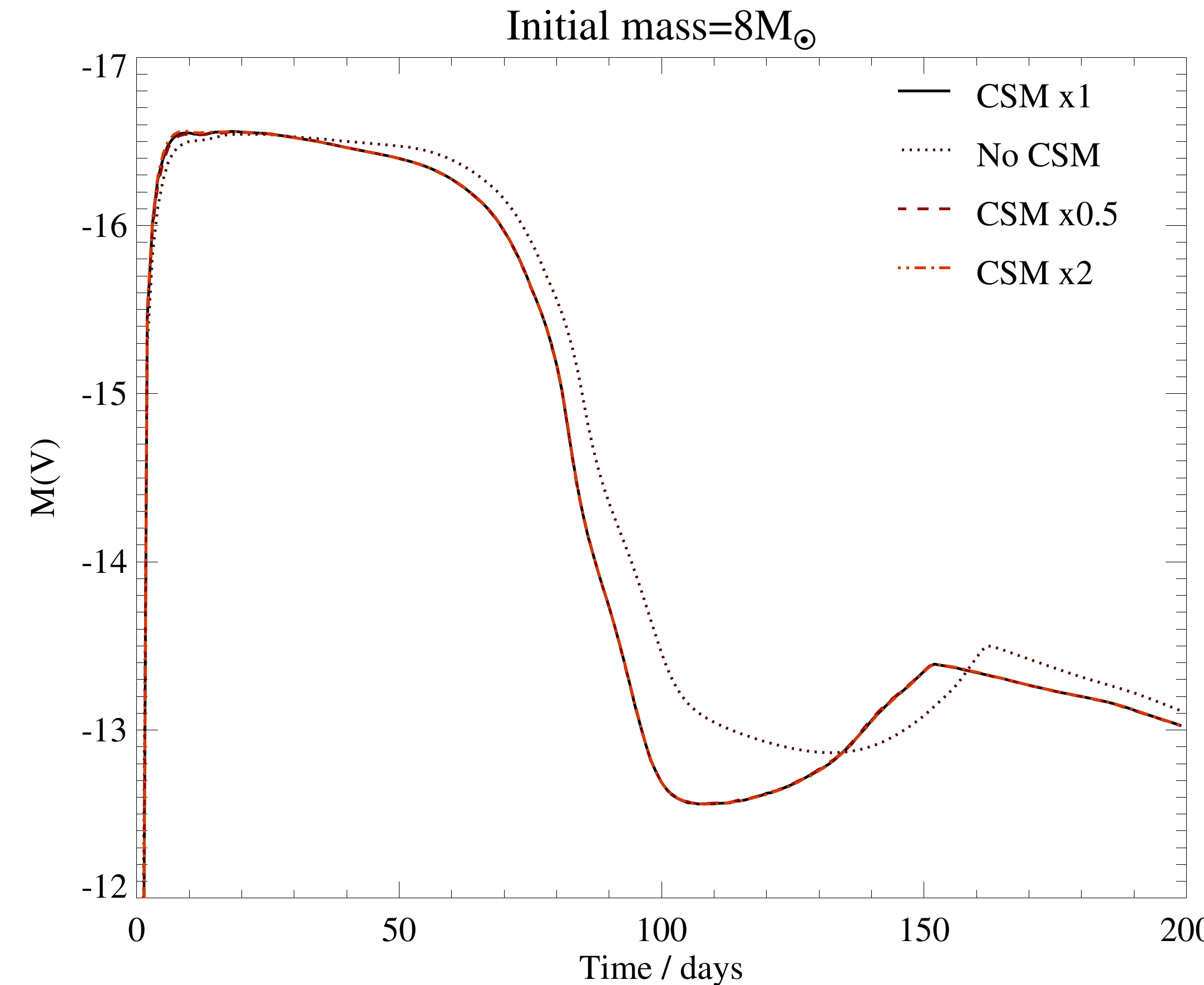}\\
\includegraphics[width=0.65\columnwidth]{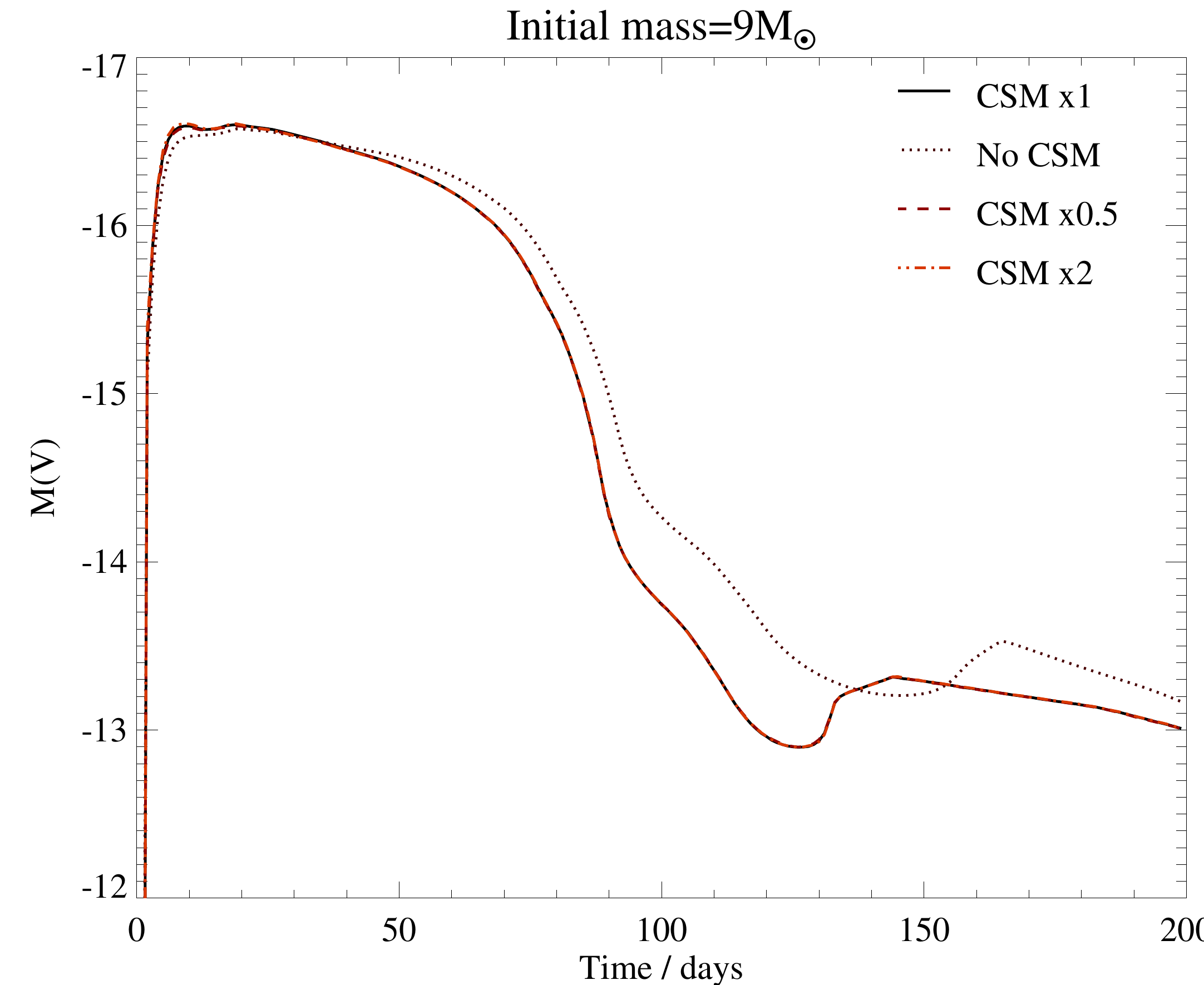}
\includegraphics[width=0.65\columnwidth]{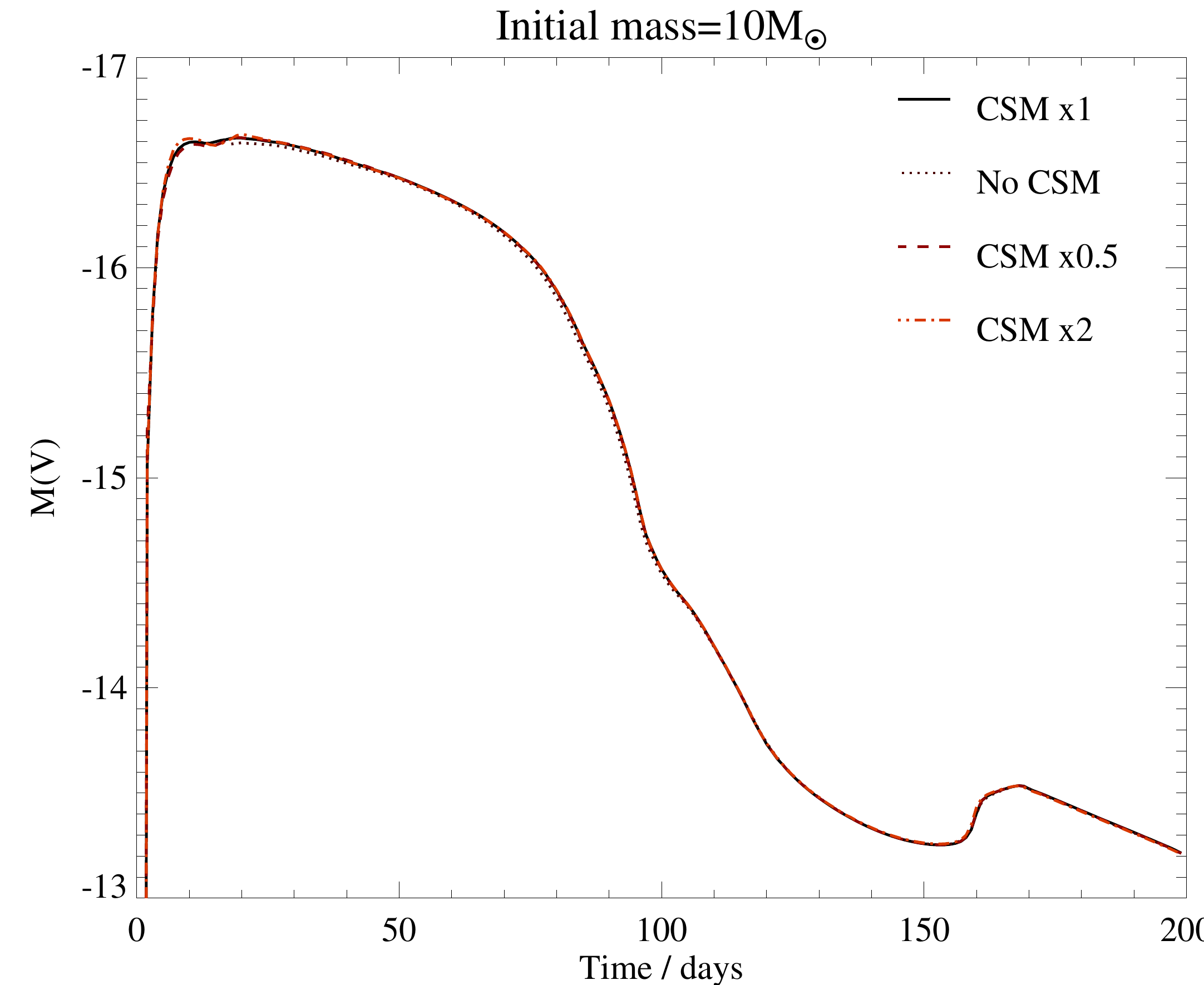}
\includegraphics[width=0.65\columnwidth]{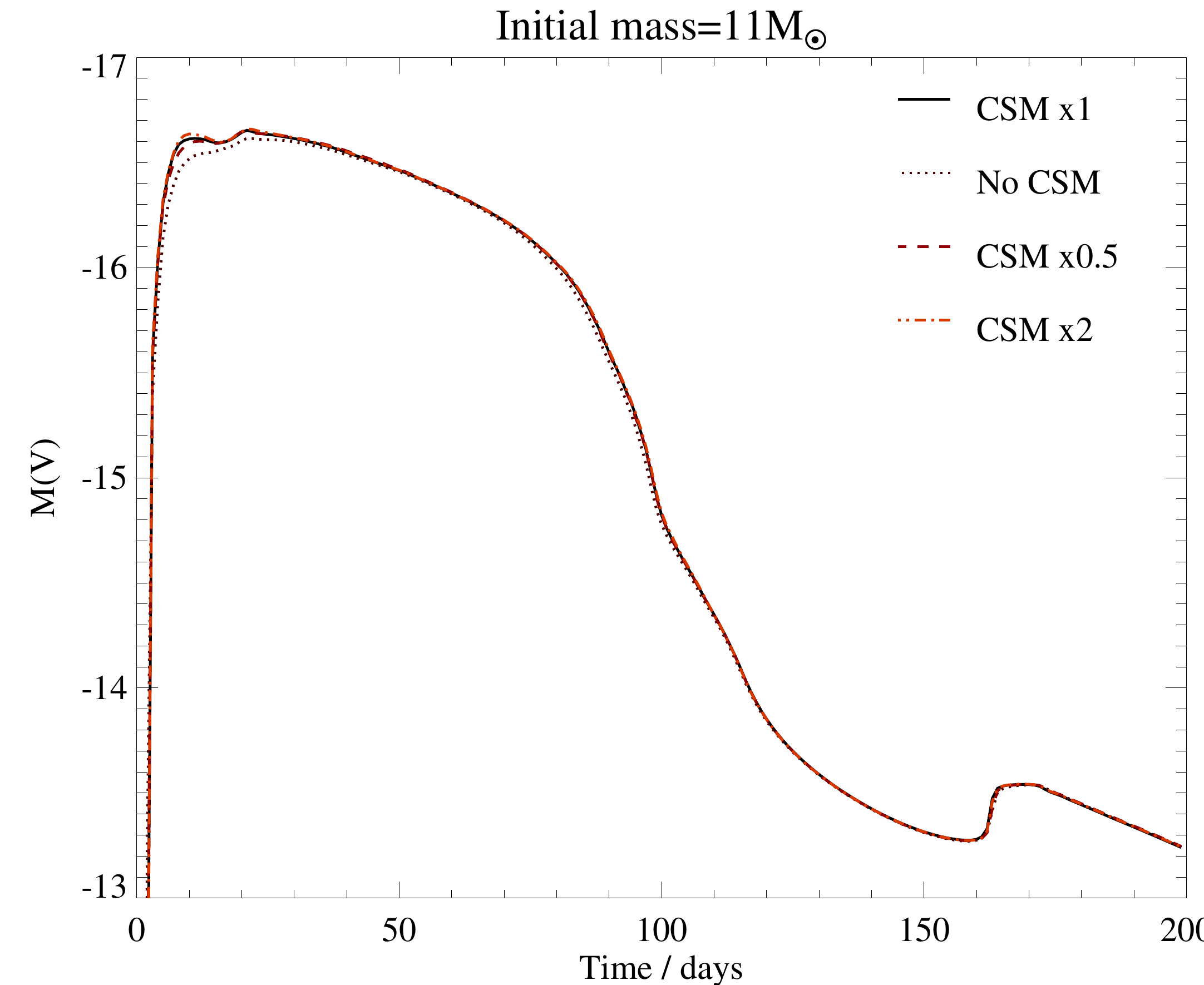}\\
\includegraphics[width=0.65\columnwidth]{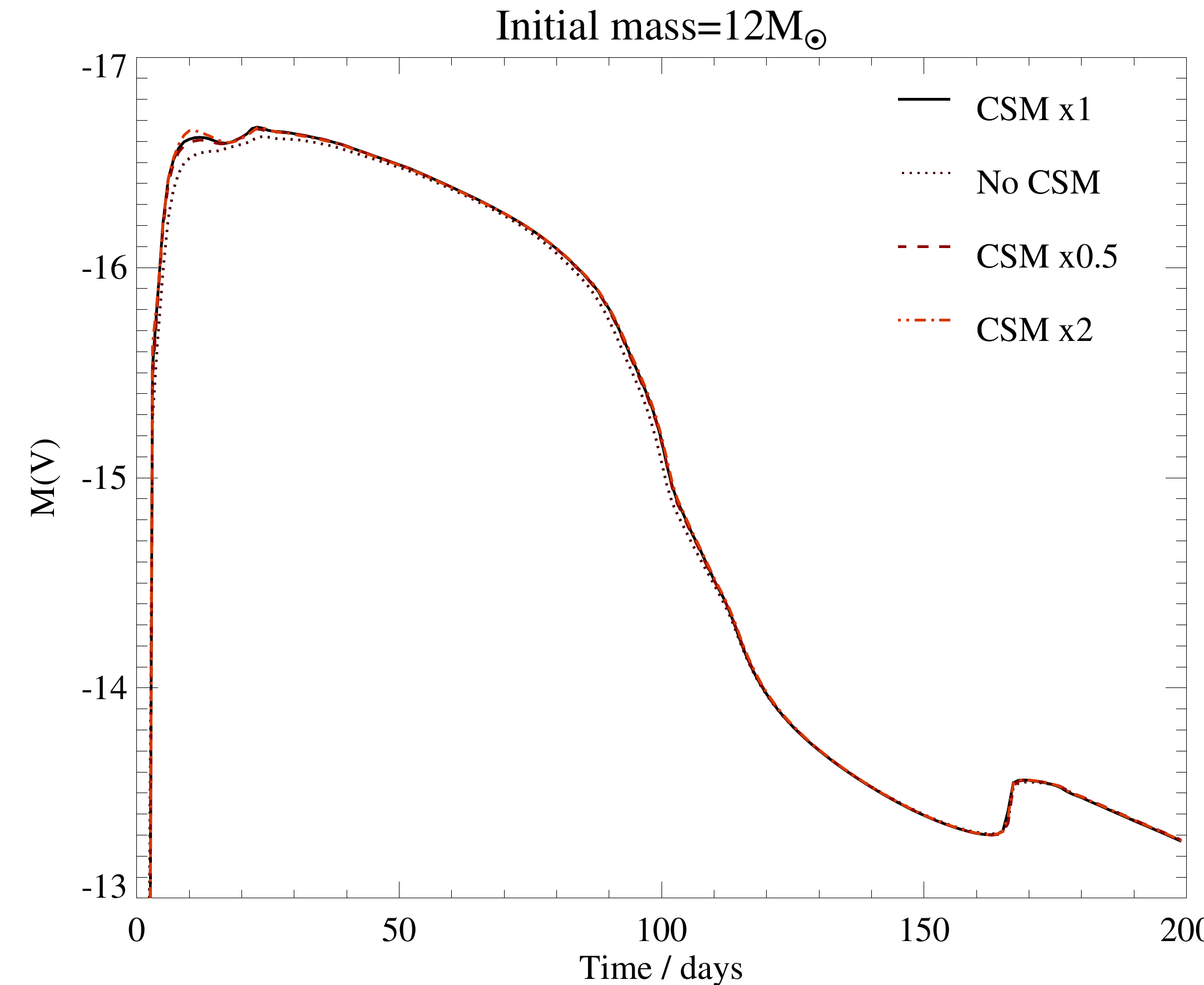}
\includegraphics[width=0.65\columnwidth]{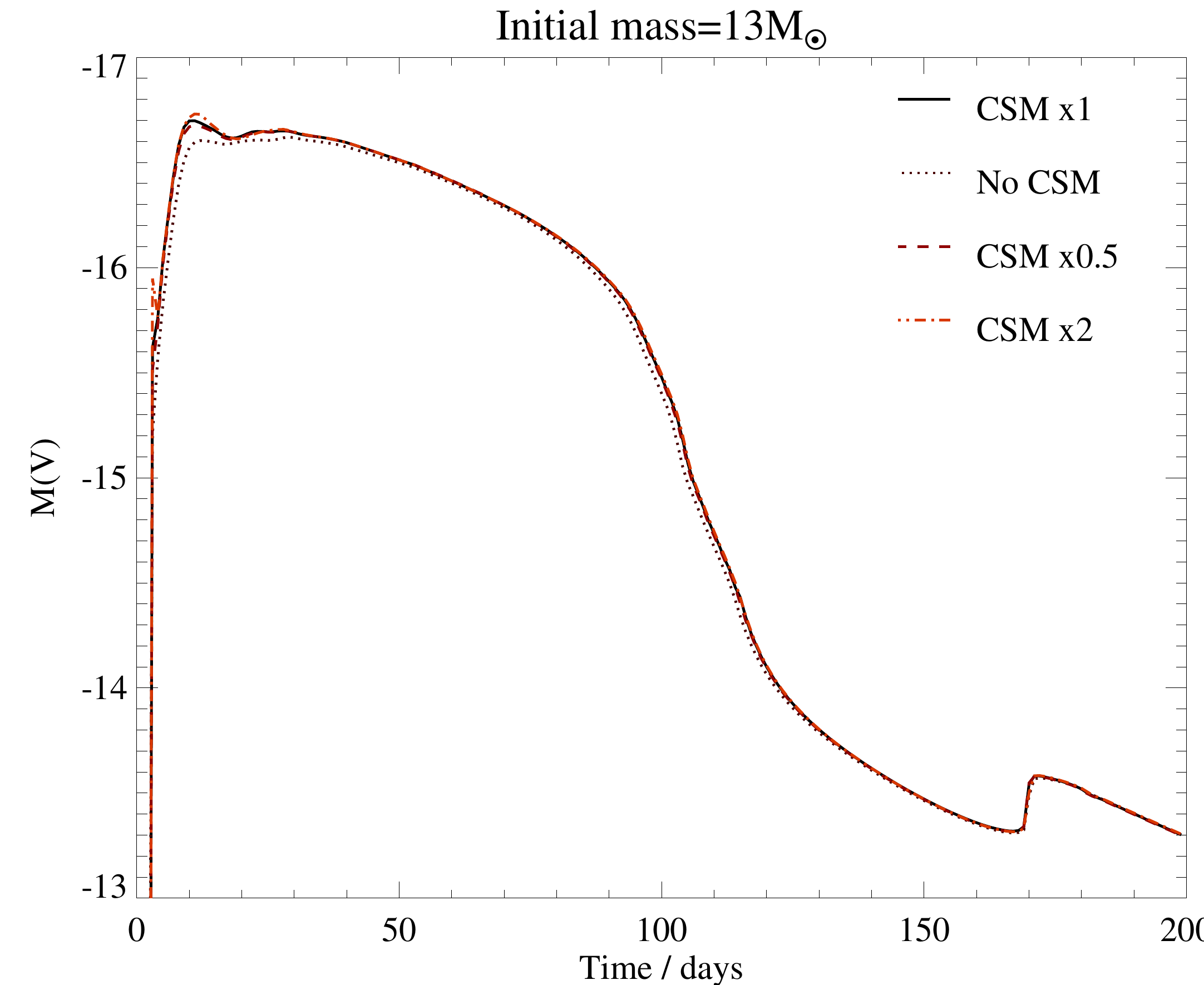}
\includegraphics[width=0.65\columnwidth]{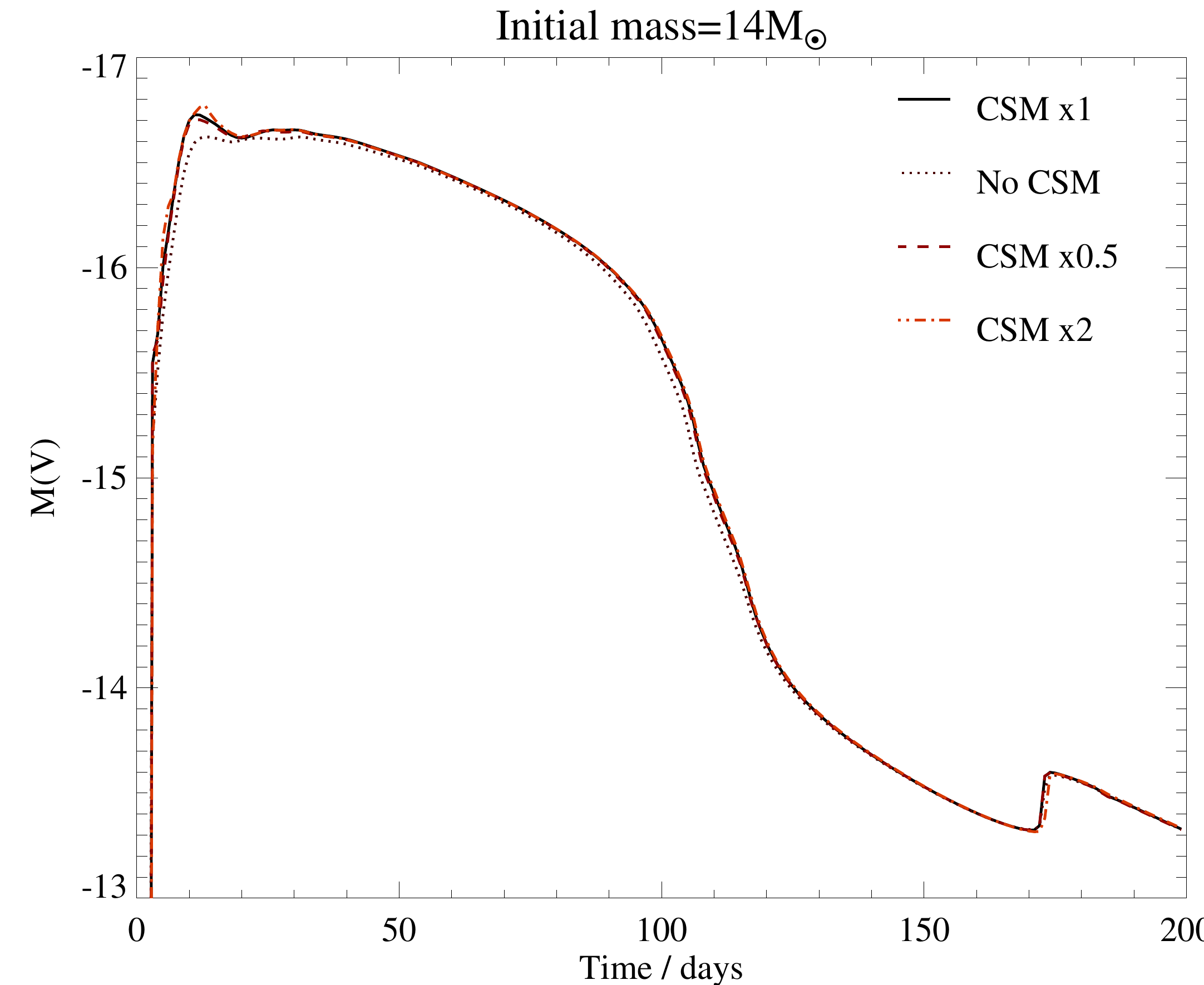}\\
\includegraphics[width=0.65\columnwidth]{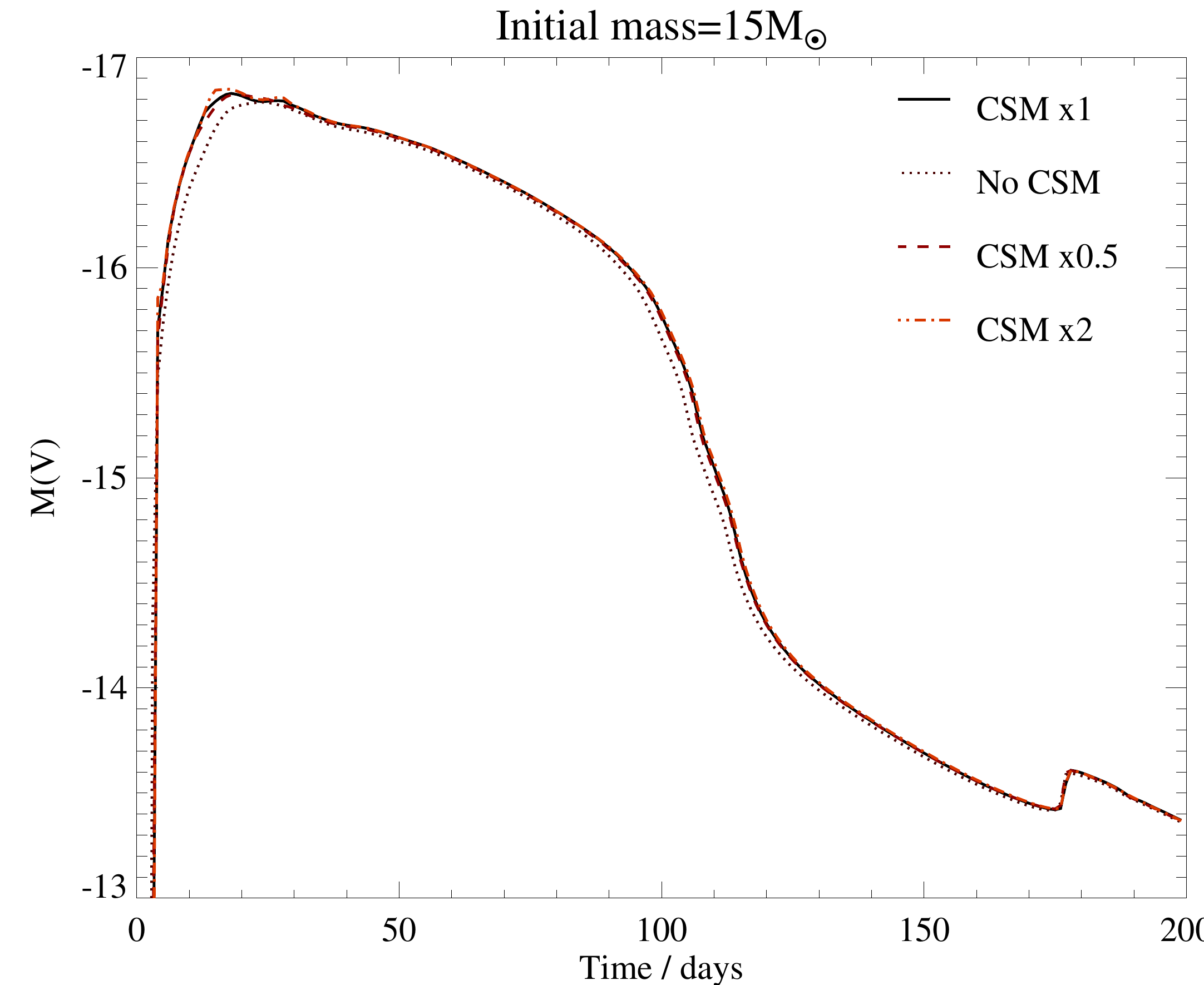}
\includegraphics[width=0.65\columnwidth]{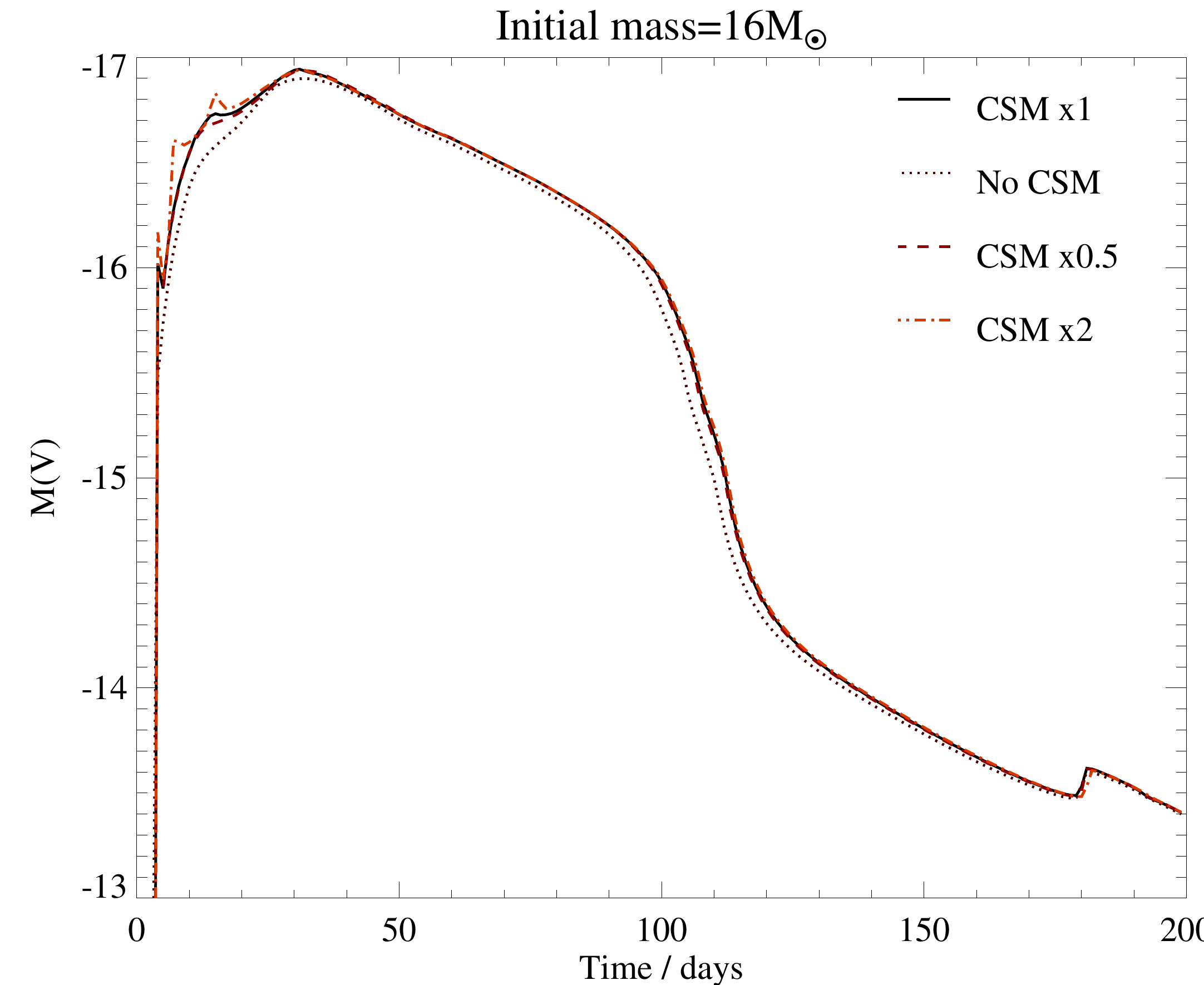}
\includegraphics[width=0.65\columnwidth]{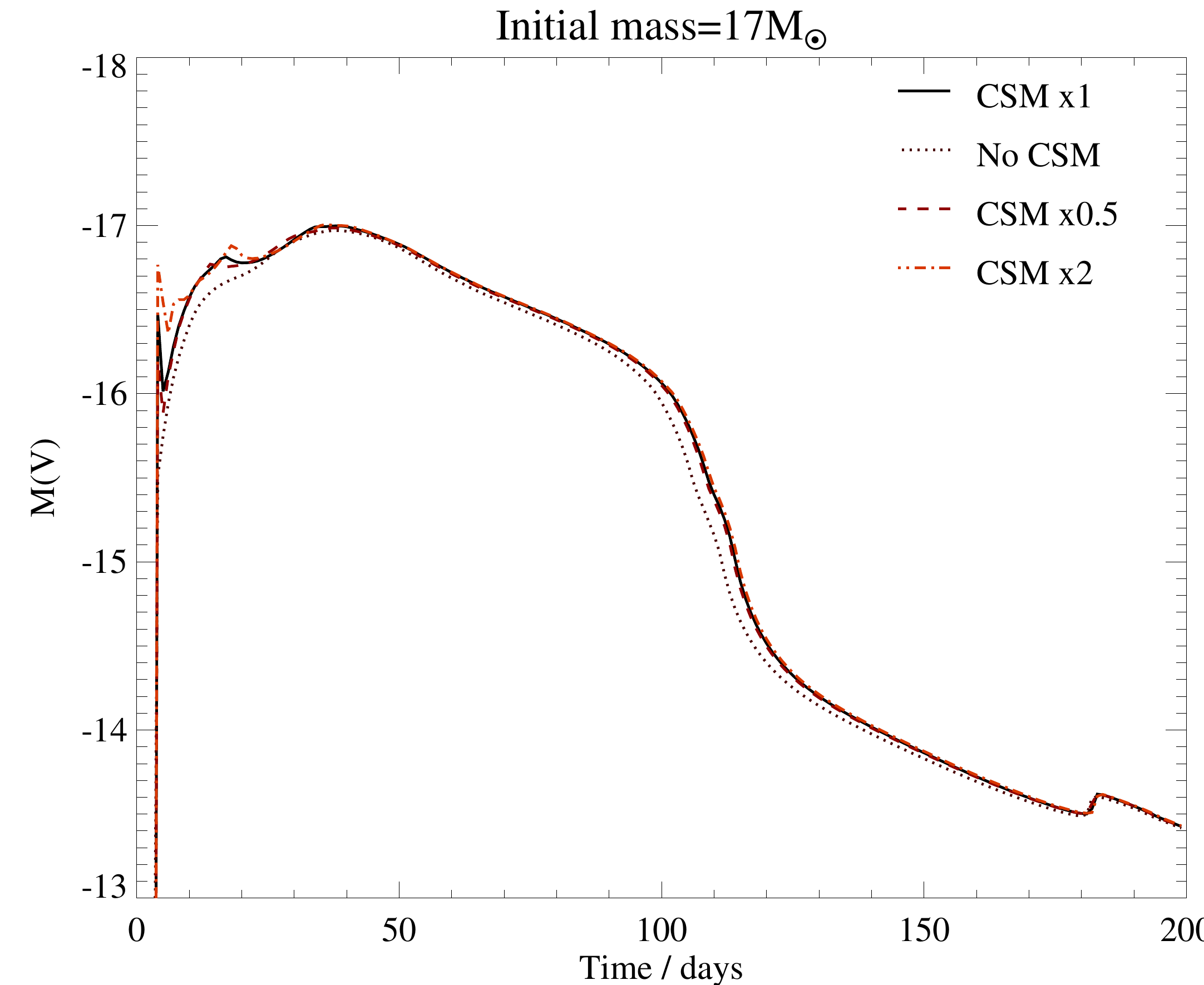}\\\includegraphics[width=0.65\columnwidth]{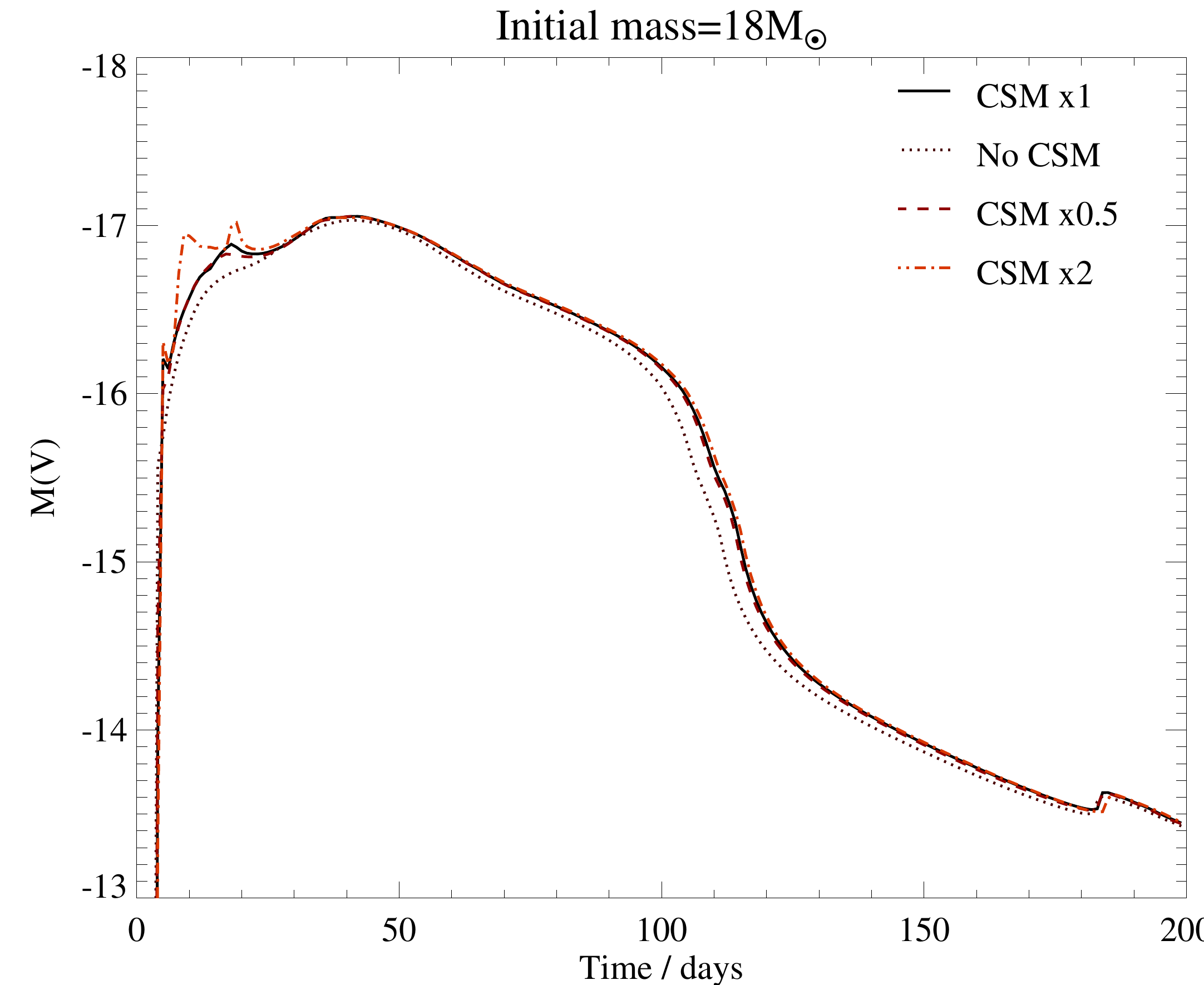}
\includegraphics[width=0.65\columnwidth]{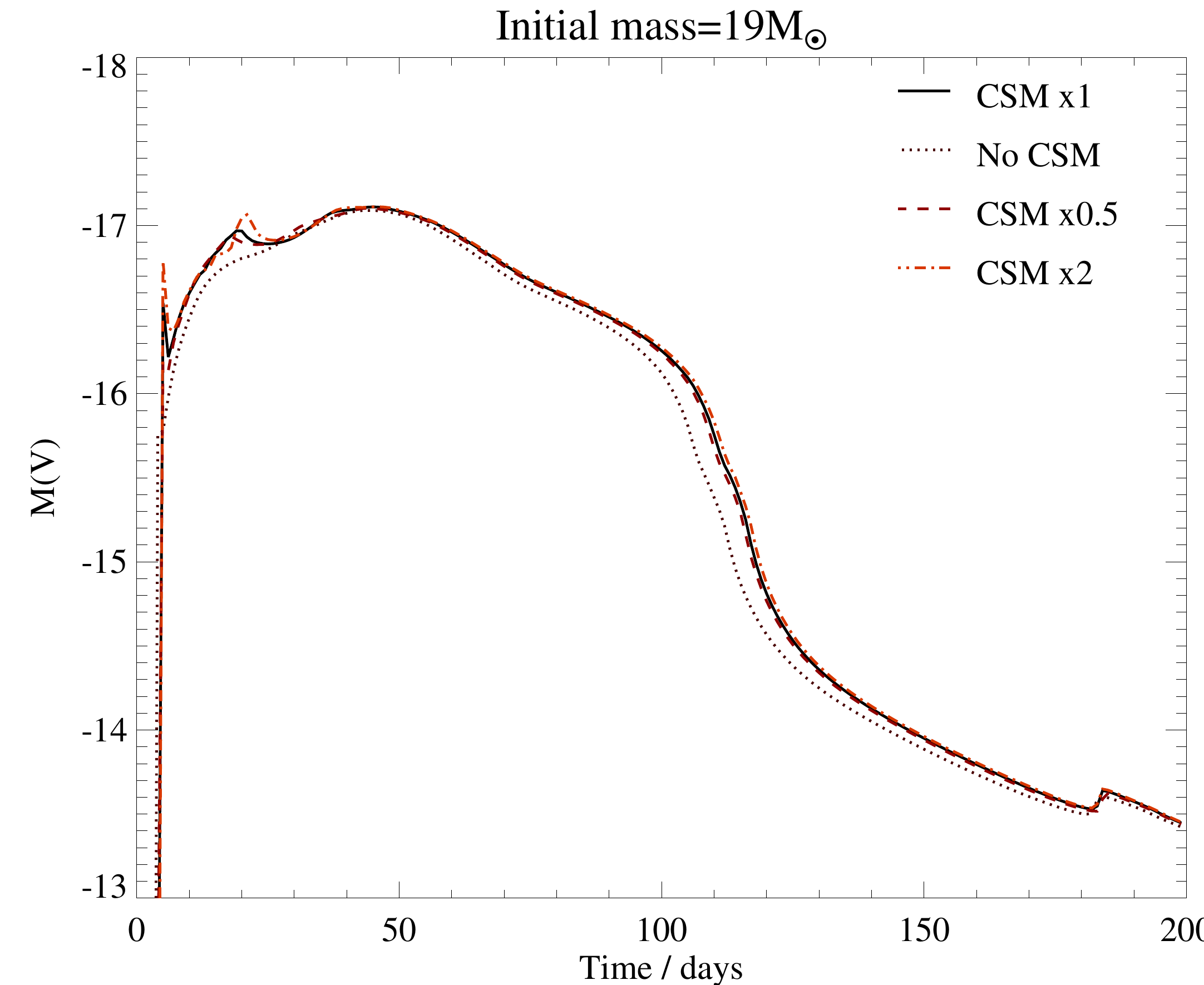}
\includegraphics[width=0.65\columnwidth]{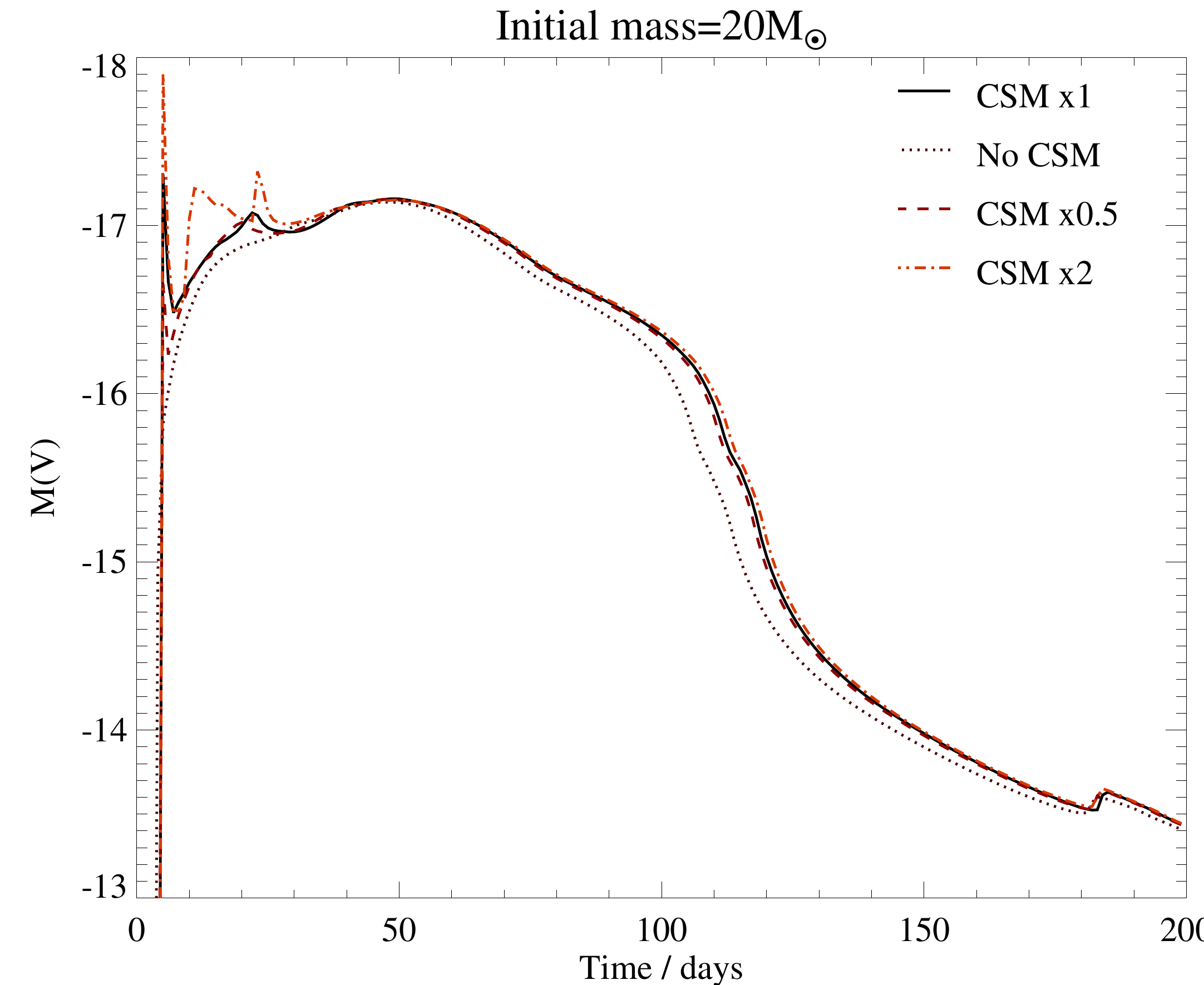}\\
\caption{As in Figure \ref{fig:csmtest2} but now showing stars in the initial mass range $6<$M/M$_\odot<20$, for which changing the circumstellar medium has little affect.}
\label{fig:csmtest1}
\end{center}
\end{figure*}

Here we show the evolution of absolute V band magnitude for our lightcurve models, as in Figure 1, but to aid clarity each panel only shows one initial mass progenitor star, with either 5 lines indicating the effect of changing the $\beta$ acceleration parameter of the stellar wind (Figure \ref{fig:csmtest3}), or 4 lines showing the effect of different assumed circumstellar medium densities (Figures \ref{fig:csmtest2} and \ref{fig:csmtest1}).

\end{appendix}

\end{document}